\documentclass[psfig]{mn2e}
\usepackage{graphicx}
\usepackage{caption}
\usepackage{subcaption}

\begin{document}

\title[Spectroscopy of the Radio Galaxy PKS 1934-63]
 {MUSE 3D Spectroscopy and Kinematics of the gigahertz peaked spectrum Radio Galaxy PKS 1934-63:  Interaction, Recently Triggered AGN and Star Formation}
\author[N. Roche, A. Humphrey, P. Lagos,  P. Papaderos, M. Silva, L.S.M. Cardoso, J.M. Gomes]
 {Nathan Roche$^{1}$ 
 \thanks{nathanroche@mac.com},
 Andrew Humphrey$^{1}$,  Patricio Lagos$^{1}$, Polychronis Papaderos$^{1}$,\\\\
 {\LARGE \rm  Marckelson Silva$^{1,2}$, Leandro S. M. Cardoso$^{1,2}$, Jean Michel Gomes$^{1}$}\\\\
$^1$ Instituto de Astrof\'isica e Ci\^encias do Espa\c co, 
Universidade do Porto, CAUP, Rua das Estrelas, 4150-762 Porto, Portugal.\\
$^2$ Departamento de F\'isica e Astronomia, Faculdade de Ci\^encias, Universidade do Porto, Rua do Campo Alegre 687, 4169-007 Porto, Portugal.\\}

\bibliographystyle{unsrt} \bibliographystyle{unsrt}

\date{1 May 2016}

\pagerange{\pageref{firstpage}--\pageref{lastpage}} \pubyear{2016}

\maketitle
 
\label{firstpage}

\begin{abstract} 
We observe the radio galaxy PKS 1934-63 (at $z=0.1825$) using MUSE (Multi Unit Spectroscopic Explorer) on the Very Large Telescope (VLT). The radio source is   GigaHertz Peaked Spectrum and compact (0.13 kpc), implying an early stage of evolution ($\leq 10^4$ yr). Our data show an interacting pair of galaxies, projected separation 9.1 kpc, velocity difference $\Delta(v)=216$ km $\rm s^{-1}$. The larger galaxy is a  $\rm M_{*}\simeq 10^{11}M_{\odot}$ spheroidal with the  emission-line spectrum of a high-excitation young radio AGN, e.g. strong [OI]6300 and [OIII]5007. Emission-line ratios indicate a large contribution to the line luminosity from high-velocity shocks ($\simeq 550$ km $\rm s^{-1}$) . The companion is a non-AGN disk galaxy, with extended $\rm H\alpha$ emission from which its star-formation rate is estimated as  $\rm 0.61~M_{\odot}yr^{-1}$. 

Both galaxies show rotational velocity gradients in $\rm H\alpha$ and other lines, with the interaction being prograde-prograde. The SE-NW velocity gradient of the AGN host is misaligned from the E-W radio axis, but aligned with a previously discovered central ultraviolet source, and a factor 2 greater in amplitude in $\rm H\alpha$ than in other (forbidden) lines (e.g. [OIII]5007). This could be produced by a fast rotating (100--150 km $\rm s^{-1}$) disk  with circumnuclear star-formation. We also identify a broad component of [OIII]5007 emission, blueshifted with a velocity gradient aligned with the radio jets, and associated with outflow. However, the broad component of [OI]6300 is redshifted.  In spectral fits, both galaxies have old  stellar populations plus  $\sim 0.1\%$ of  very young stars,  consistent with the galaxies undergoing first perigalacticon, triggering infall and star-formation from $\sim 40$ Myr ago followed by the radio outburst.
    \end{abstract}
   

\section{Introduction}
Gigahertz Peaked Spectrum (GPS) radio galaxies afford an excellent opportunity to study the birth and evolution of powerful radio sources, and to obtain insight into radio-mode feedback, as the radio source interacts with the ambient interstellar medium of the host galaxy. 

In this paper we present our new integral field spectroscopic study of PKS 1934-63, an interacting pair of galaxies host to a powerful GPS radio source.
The radio source was detected in the Parkes 2700 MHz survey (Shimmins 1971) and Wall \& Cannon (1973) identified the optical counterpart on a Schmidt (Mount Stromlo) plate, classifying the galaxy as a $18^m$ Seyfert.  Peterson and Bolton (1972), at the same telescope, measured the redshift as 0.182.

We first describe the radio properties.
Tzioumis et al. (1989) with Very Long Baseline Interferometry (VLBI, Australia to South Africa) at 2.3 GHz, resolved the source into two peaks separated by  $42.0\pm0.2$ milliarcsec (0.13 kpc) on a position angle (hereafter PA; defined with North as zero and E as $90^{\circ}$) of $-90.5^{\circ}\pm1$. PKS 1934-63 is classed as a GPS source as its radio  flux density is maximum (16.4 Jy) near 1.4 GHz with a low frequency cut off; the spectral energy distribution (SED) can be represented as a double power law $S_{\nu}\propto \nu^{0.78}$ at $\nu< 1.4$ GHz, then  $S_{\nu}\propto \nu^{-0.73}$ (de Vries, Barthel \& O'Dea 1997). 

GPS radio sources have small diameters, $D<1$ kpc, which might indicate (e.g. 
O'Dea, Baum and  Stanghellini 1991) that they have become confined (trapped) within the host galaxy nucleus by a very dense (clumpy and dusty) interstellar medium, or simply that they are observed very soon after the radio outbursts is triggered (e.g. Morganti et al. 1997). It is now considered they are young sources, forming part of an evolutionary sequence (An \& Baan 2012; Kunert-Bajraszewska 2015), which begins with  High Frequency Peaked sources, with few parsec diameters and the low-frequency turnover at several GHz. These rapidly expand and increase in radio power to form GPS sources like the very luminous PKS 1934-63. After some $\sim 10^5$ years the radio source extends  $D \sim 1$--3 kpc and is then described as compact steep spectrum (CSS), and the expansion continues while radio luminosity remains near constant at its maximum. The more powerful radio Active Galactic Nuclei (AGNs), like this one (high ionization galaxies) continue evolving to become giant FR-II (Fanaroff \& Riley 1974) type radio galaxies and eventually fade, while the less luminous (classed as low ionization galaxies) follow a parallel track to become extended FR-Is.

While the expansion velocity and age of this source appear still to be uncertain (compare Ojha et al. 2004 and Tzioumis et al. 2010), in the An \& Baan (2012) model the hotspot separation velocity for a source of this size (if it is destined to become a FR-II) must be $\geq 0.05c$, implying the radio burst began no more than $10^4$ yr ago.
The observed  flux corresponds to a luminosity $P_{1.4{\rm GHz}}\simeq 10^{34.2}$ erg $\rm s^{-1}Hz^{-1}$, placing the galaxy neatly on the upper sequence of the size-luminosity plot of An \& Baan (2012). 

In the optical wavelengths, Heckman (1980) described the galaxy's spectrum as transitional LINER-Seyfert, with line ratios indicating a high electron density  and temperature ($T_e>18000$ K from [OIII]).
Heckman et al. (1986) in $V$ and $R$ imaging found two galaxy nuclei separated by 3--4 arcsec, and `fans or tail like features' with a surface brightness $V\simeq 23$--23.5 mag $\rm arcsec^{-2}$, and concluded the galaxy is undergoing a collision or merger.
Fosbury et al. (1987) took a spectrum with the long-slit on a PA of $284^{\circ}$ ($-76^{\circ}$), crossing the two nuclei, confirmed the radio source is in the brighter eastern nucleus and found a strong velocity gradient, with the fainter West component more redshifted.

  Holt, Tadhunter and Morganti (2008) examined the kinematics of the [OIII]5007 emission line. 
Labiano et al. (2008) with Hubble Space Telescope Advanced Camera for Surveys (HST-ACS)  imaging of some compact/GPS radio galaxies in F330W (near-ultraviolet) detected an interesting sub-arcsec extended source at the centre of PKS 1934-63.

Kawakatu, Nagao and Woo (2009) examine the optical emission lines of compact (young) narrow-line, powerful radio galaxies at $z<1$ (including PKS 1934-63) and find  systematically much higher $\rm [OI]6300/[OIII]5007$ and $\rm [OIII]4363/[OIII]5007$ ratios than for radio-quiet Seyfert 2s.
However, there is a wide range of  line fluxes and ratios in young radio galaxies and Son et al. (2012) divide their sample into high and low excitation galaxies (HEG/LEG), attributing the difference to variations in spectral energy distribution and accretion rate (Eddington ratio).

Inskip et al. (2010) image many radio galaxies in the $K$-band (finding the majority to be massive ellipticals and about half are interacting or disturbed), and fit both PKS 1934-63 components with S\'ersic profiles, the residuals showing `faint tidal features' and a `third component' to the south.
Ramos Almeida et al. (2011) in deep $r$-band imaging with the Gemini Multi Object Spectrograph (GMOS), find most PKS radio galaxies have evidence of interaction and describe PKS 1934-63 as two bright nuclei separated by 9 kpc with `two tidal tails extending towards the SW from the radio galaxy and to the north from its companion' with surface brightness 23.1 and 23.4 mag $\rm arcsec^{-2}$.

 Dicken et al. (2012) took {\it Spitzer} mid-infrared spectroscopy of some $z<0.7$ radio galaxies to look for polycyclic aromatic hydrocarbon (PAH) emission lines, a (dust-insensitive) signature of star-formation. For PKS 1934-63 detection was significant ($1.3\pm 0.1\times 10^{-14}$ erg $\rm cm^{-2}s^{-1}$ for the $\rm 11.3 \mu m$ line) but described as low equivalent width. 
  Ramos Almeida et al. (2013)  measure a low clustering amplitude of $\rm B_{gq}^{av}=119\pm 163$, meaning this galaxy is not in any cluster (in comparison, for Abell class 0 $\rm B_{gq}>400$).

Integral field spectroscopy (IFS) can reveal more about the radio galaxies and their structure and interactions, separate AGN-dominated and star-forming regions  and show their kinematics, detect localised shocks and outflows and for young sources like this, show how they are triggered.

In May 2014 the European Southern Observatory announced a call for short observing proposals for the Science Verification run of an integral field spectrograph, the Multi Unit Spectroscopic Explorer (MUSE), newly installed on the Very large Telescope (VLT). 
We drew up a proposal for the radio galaxy PKS 1934-63, chosen from Ramos Almeida et al. (2011), on the basis of its multiple morphology, tidal features, reported evidence of extended emission and star-formation, its being well placed for this observing period, and the interesting IFS findings from other PKS  galaxies (e.g. Inskip et al. 2007, 2008; Shih, Stockton \& Kewley 2013).

From IPAC NED, the J2000 co-ordinates of PKS 1934-63 are 
RA 19:39:25.0261 and Dec -63:42:45.625.
 For $\rm H_0=70$ km $\rm s^{-1}Mpc^{-1}$, $\Omega_M=0.27$, $\Omega_{\Lambda}=0.73$ (assumed throughout) and our measured redshift of $z=0.1825$ the  distance modulus is 39.744 mag and the angular diameter distance 635.6 Mpc, so that 1 arcsec subtends 3.081 kpc, and the lookback time is 2.254 Gyr. 
 
In Section 2 we describe the observations and data processing, in section 3 present spectra and an analysis of the line measurements, In section 4 we show images of the radio galaxy and analyze the morphology. In section 5 we study the kinematics in $\rm H\alpha$, [OIII] and several other lines. In section 6 we fit models to the spectra to estimate the star-formation history and in sections 7 and 8 discuss and summarise these findings on the evolution of the radio source and galaxies.

\section{Observations and Data Reduction}
Our observations were made using the MUSE spectrograph (e.g. Bacon et al. 2015), which is installed on the Nasmyth B focus of Yepun, the 8.2m aperture VLT UT4 telescope. It is here operated in natural seeing (adaptive optics are not available at this time) and the wide field mode, in which  MUSE covers  a square arc min field of view, sampled by a system of 24 spectrographs. It gives essentially seeing-limited spatial resolution (here $\rm FWHM\simeq 0.7$ arcsec), and coverage 4800--$9300\rm \AA$, with spectral resolution from 1770 at $4800\rm\AA$ up to 3590 at $9300\rm\AA$, ie.  about $2.6\rm\AA$. The
 MUSE data cubes consist of $300\times 300$ spatial pixels of $0.2\times 0.2$ arcsec, with each of these sampled at a $1.25\rm\AA$ pixel size in the spectral dimension.

Our MUSE data were taken as Program 60.A-9335(A), Roche et al., `Spectroscopy of the Merging Radio Galaxy PKS 1934-63', during the Science Verification run (1.5 hours were allocated), and in 2 Observing Blocks, the first on 22 June 2014 with two exposures of 600 seconds, and the second a full hour on 23 June 2014 with 4 exposures of 
710 seconds, making in total 4040s on the target galaxy. Pointings centred on the AGN co-ordinates with small dithers of approx 2 arcsec between exposures (but no rotations).

For data reduction we used the first v1.0 public version of the MUSE data Reduction software, together with the MUSE Data Reduction Cookbook (VLT-MAN-ESO-14670-6186) written by J. Vernet and R. Bacon.
The first step was to generate the master bias, and then the master flat by combining lamp flat-fields. Then a wavelength calibration was automatically fitted using a set of 3 arc frames, one each for the Neon, Xenon and HgCd lamps. Then the raw sky-flat frames were combined into master sky-flat. The next step, a procedure muse\_scibasic, uses the above to debias, flat-field and wavelength-calibrate, and was run on each science exposure in turn and also on a 30 sec exposure of the standard star GD153-N used for flux-calibration.

Another step muse\_standard uses the processed standard-star exposure and a tabulated spectrum to set up the flux calibration. The final stage of the basic reduction, muse\_scipost, processed and combined the 6 scarce frames into a single flux-- and wavelength-calibrated data cube, with rejection of cosmic rays, corrections for the atmospheric extinction at Paranal and telluric absorption bands, registration of the exposures and sky line subtraction. This also generates 2D images of the field-of-view, summed over wavelength for any chosen or the full $\lambda$ range. With this step the galaxies and spectra  are finally revealed to the eye.

  However, so great are the memory requirements that our computer was unable to process the whole data cube at once, and we had to divide the MUSE wavelength range into 3 intervals and  run scipost for each separately, giving the reduced data cube in 3 slices of 4800--$6400\rm\AA$, 6400--$8000\rm\AA$ and 8000--$9300\rm\AA$, approximately $V$, $R$ and $I$-bands, each with a broad-band image (in the subsequent analysis we re-combine the spectra extracted from the 3 slices).
 
\section {Analysis of Aperture Spectra}
\subsection{Extracting the Spectra}
Fig 1 shows our MUSE image summed over the whole $R$-band (6400--$8000\rm \AA$). The radio galaxy is clearly seen as a double object near the centre, the largest and brightest galaxy in the field of view. The image also contains several bright stars and a number of fainter galaxies. 

\begin{figure}
 \includegraphics[width=1.03\hsize,angle=0]{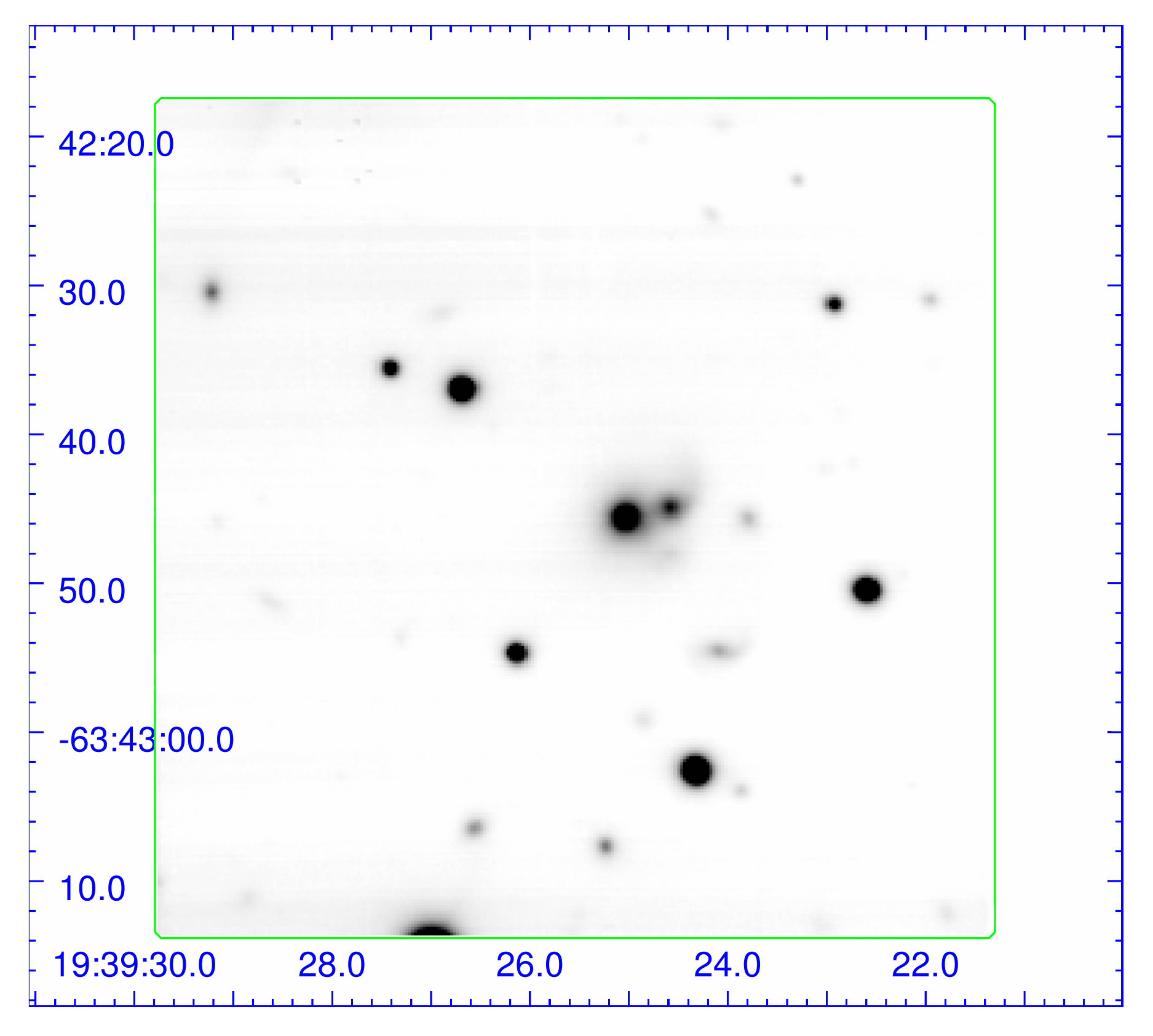}
\caption{MUSE $\rm 6400<\lambda<8000\AA$ ($R$-band) image of the field of view, slightly trimmed to $56\times56$ arcsec, with PKS 1934-63 prominent as bright double galaxy slightly right of the centre -- the AGN host is the larger, East component. Axes show RA (hh:mm:ss) and Declination (degrees). North is top, East is left, as in all plots in this paper. Arcsinh intensity scaling.}
 \end{figure}
 \begin{figure}
 \includegraphics[width=1.03\hsize,angle=0]{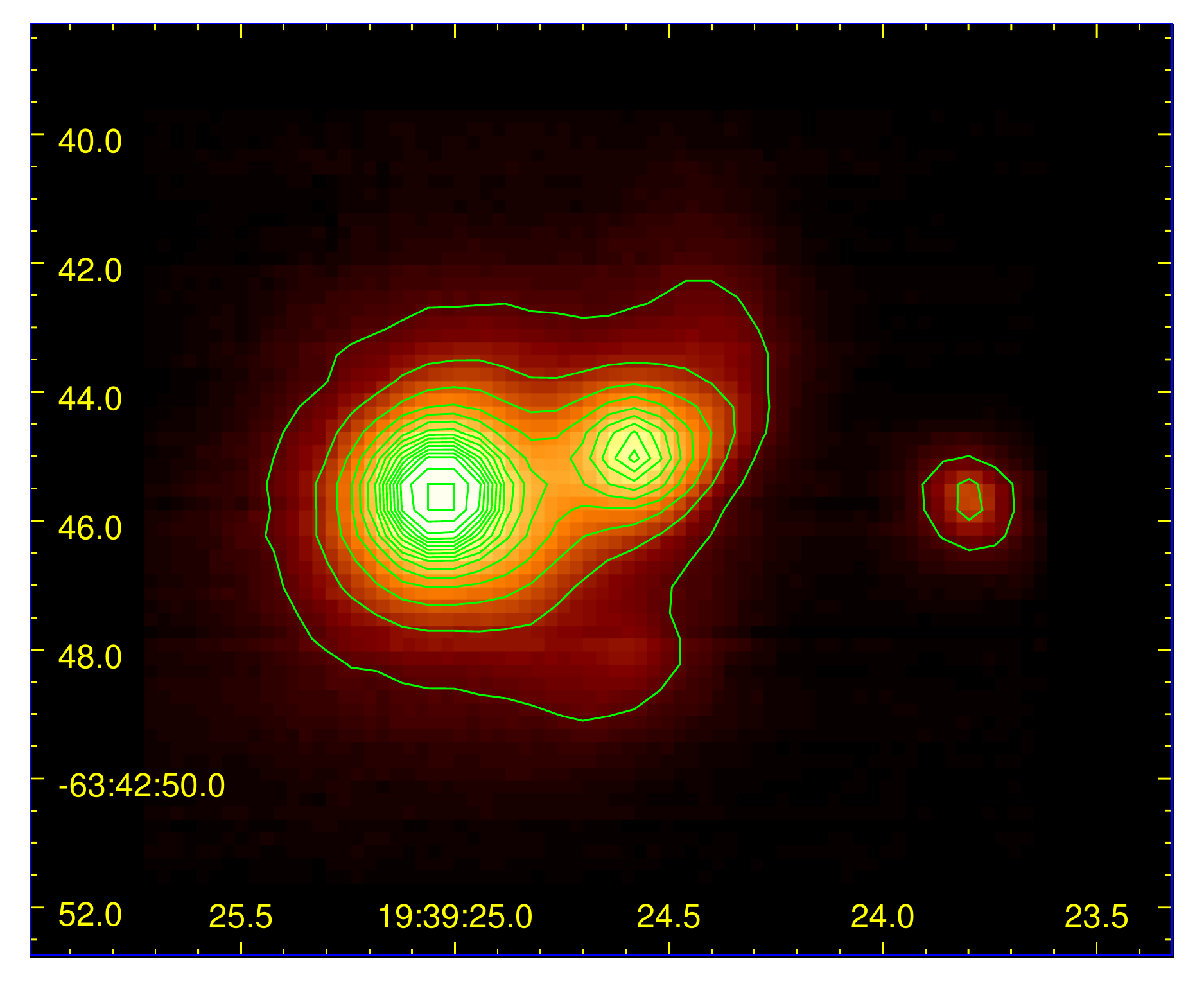}
\caption{Close-up of PKS 1934-63 in the $R$-band, with its two unequal nuclei, highlighted by contours, linearly spaced in intensity at multiples of $1.25\times 10^{-18}$ erg $\rm cm^{-2} s^{-1}\AA^{-1}arcsec^{-2}$.  Image is $15\times12$ arcsec, RA and Dec axes are shown. The faint (apparent) `extension' to the SW of the radio galaxy, and the small separate galaxy to the right are background galaxies at higher redshifts (see Fig 5 and Appendix)}.
 \end{figure}

The spectra in the calibrated images and datacubes are in units of $\rm 10^{-20}$ ergs $\rm cm^{-2}s^{-1}\AA^{-1}$. For the purposes of image analysis we define an AB magnitude system for the three wavelength intervals,
$m_{5400}$, $m_{7200}$ and $m_{8650}$ ($V$, $R$ and $I$) with zero-points on the basis of the central wavelength, 28.851, 28.305, and 27.907 respectively. These do not include the effect of Galactic dust reddening at this relatively low Galactic latitude $-29^{\circ}$. For the spectra we correct for this using {\small IRAF} `deredden', with $A_v=0.234$ as tabulated for these co-ordinates by NED. For the broad-bands we derive corrections integrating the dereddened and uncorrected primary galaxy spectra over the respective wavelength ranges; 0.228, 0.164 and 0.118 mag for $VRI$ (and subtract these from the zero-points).

Sources were detected using {\it s-extractor} (Bertin \& Arnouts 1996) 
with a detection threshold of $2\sigma$ in 5 pixels (where the sky noise is  0.443504 $\times 10^{-20}$ erg $\rm cm^{-2}s^{-1}\AA^{-1} pixel^{-1}$),. This gave the greatest number of detections, 44, for the $R$ image  and we then use the $R$ image to catalog the sources, and run in double-image mode to obtain photometry at the same positions in $V$ and $I$. Six detections are classed as bright stars ($17.7<m_{7200}<20.7$). Of the galaxies, the primary (radio source) and companion galaxy are the brightest with $r=5$ pixel aperture (total) magnitudes of $m_{7200}=18.62\pm0.02$(17.64) and $19.80\pm0.02$(18.92).  Their aperture colours are  $m_{5400}-m_{7200}=0.79$ and 0.72, and $m_{8650}-m_{7200}=0.13$ and 0.34 . 
The two galaxy centroids are separated by $14.8\pm0.2$ pixels,  2.96 arcsec ($9.12\pm 0.13$ kpc), on a PA $-77.0^{\circ}$.

  Spectra for the two galaxies were extracted by a summation of the MUSE data cube spectra for all pixels in radius $r=5$ pixel (1.0 arcsec) circular apertures centred on the {\it s-extractor} positions. The spectra were still visibly contaminated by sky lines, so in order to remedy this, MUSE spectra were summed from two regions east and one west of the radio galaxy (containing no visible sources and totalling 1500 pixels), to produce a sky-residual spectrum which  was subtracted from all the galaxy spectra, greatly improving their appearance.
  
\begin{figure}
 \includegraphics[width=0.75\hsize,angle=-90]{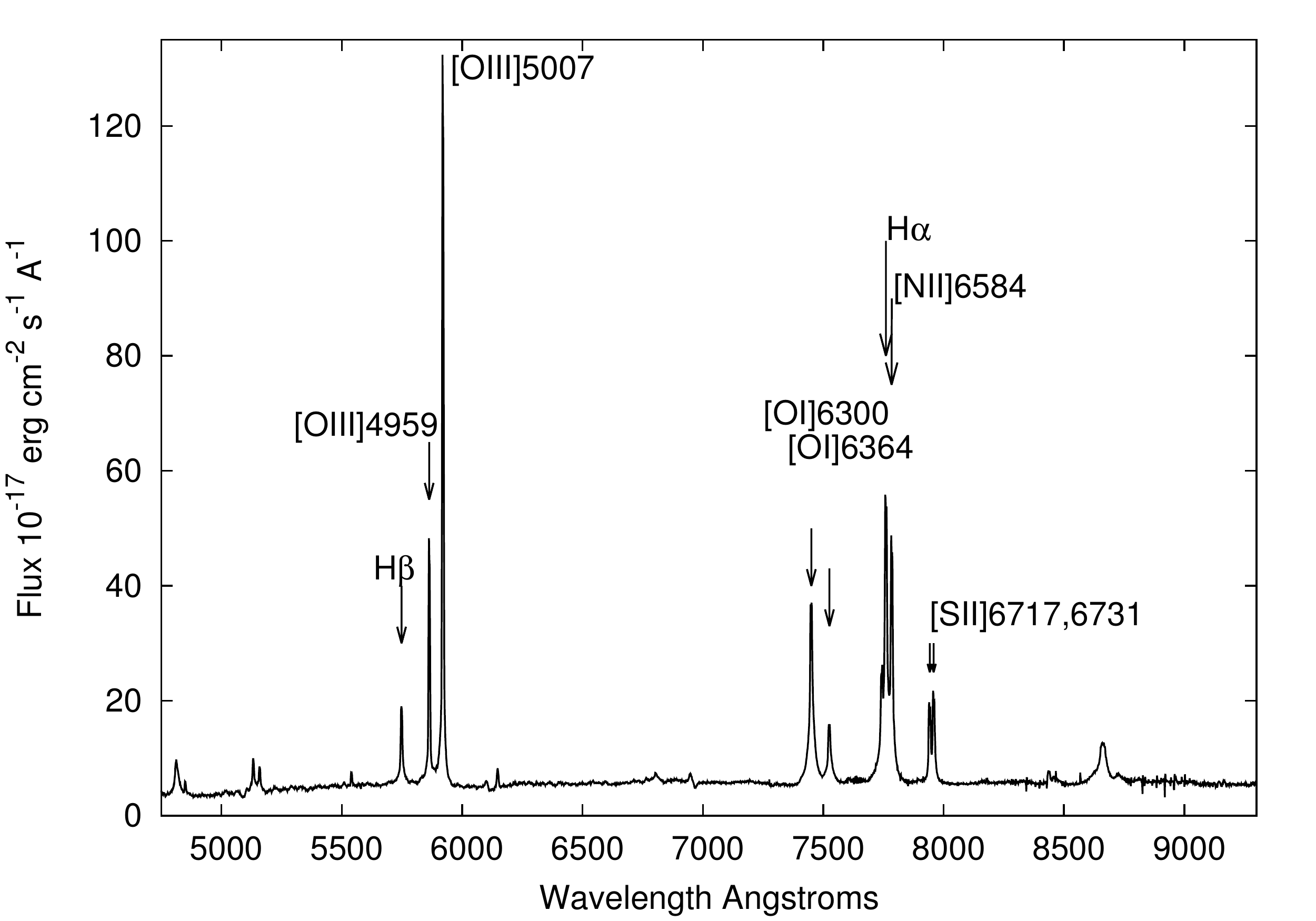}
  \includegraphics[width=0.75\hsize,angle=-90]{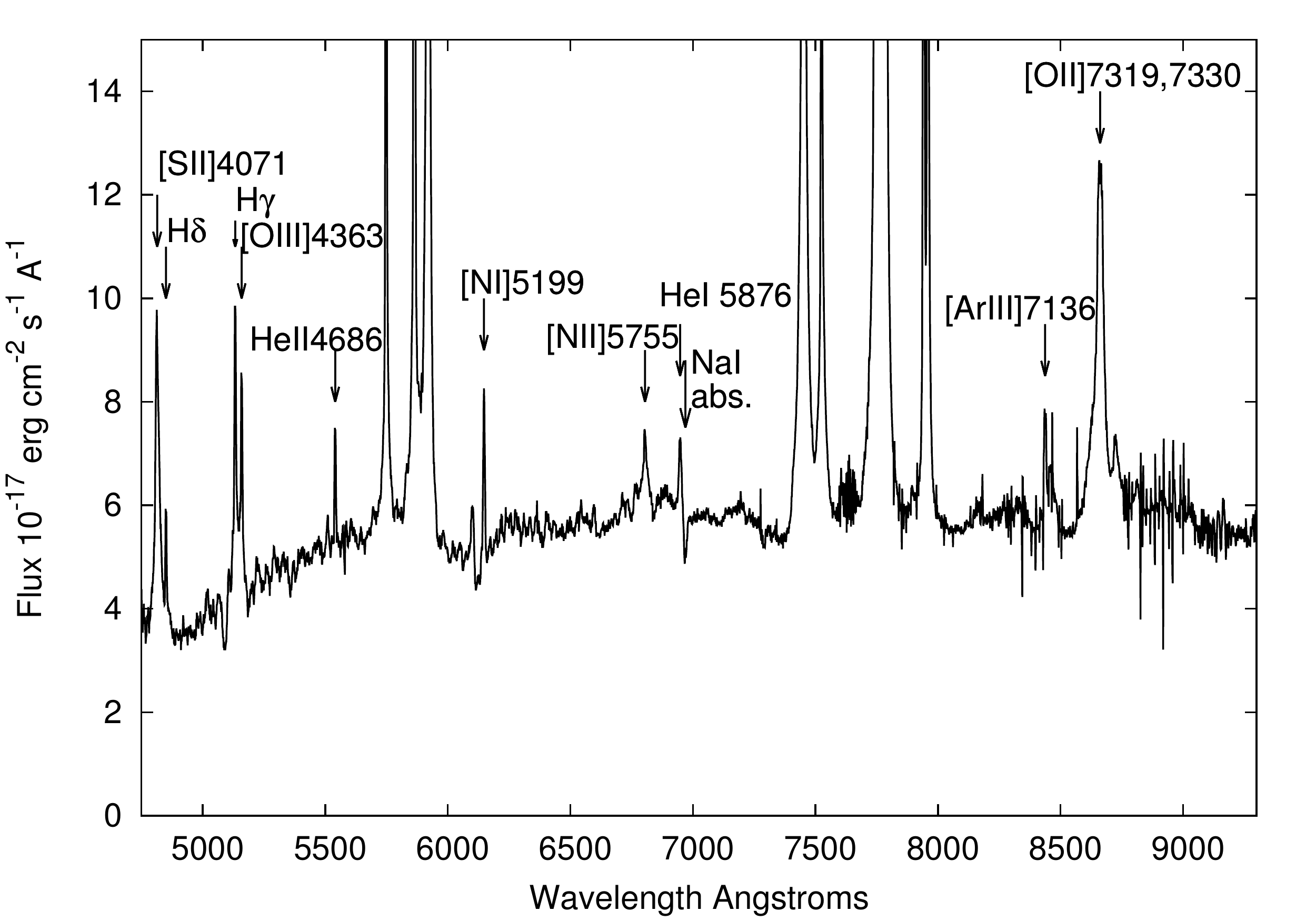}
\caption{Spectrum of the primary galaxy (containing radio AGN), extracted in a $r=5$ pixel (1.0 arcsec radius) aperture (containing 80 pixels), shown with two scalings to depict both the strongest emission lines and fainter features.}
 \end{figure}
\begin{figure}
 \includegraphics[width=0.75\hsize,angle=-90]{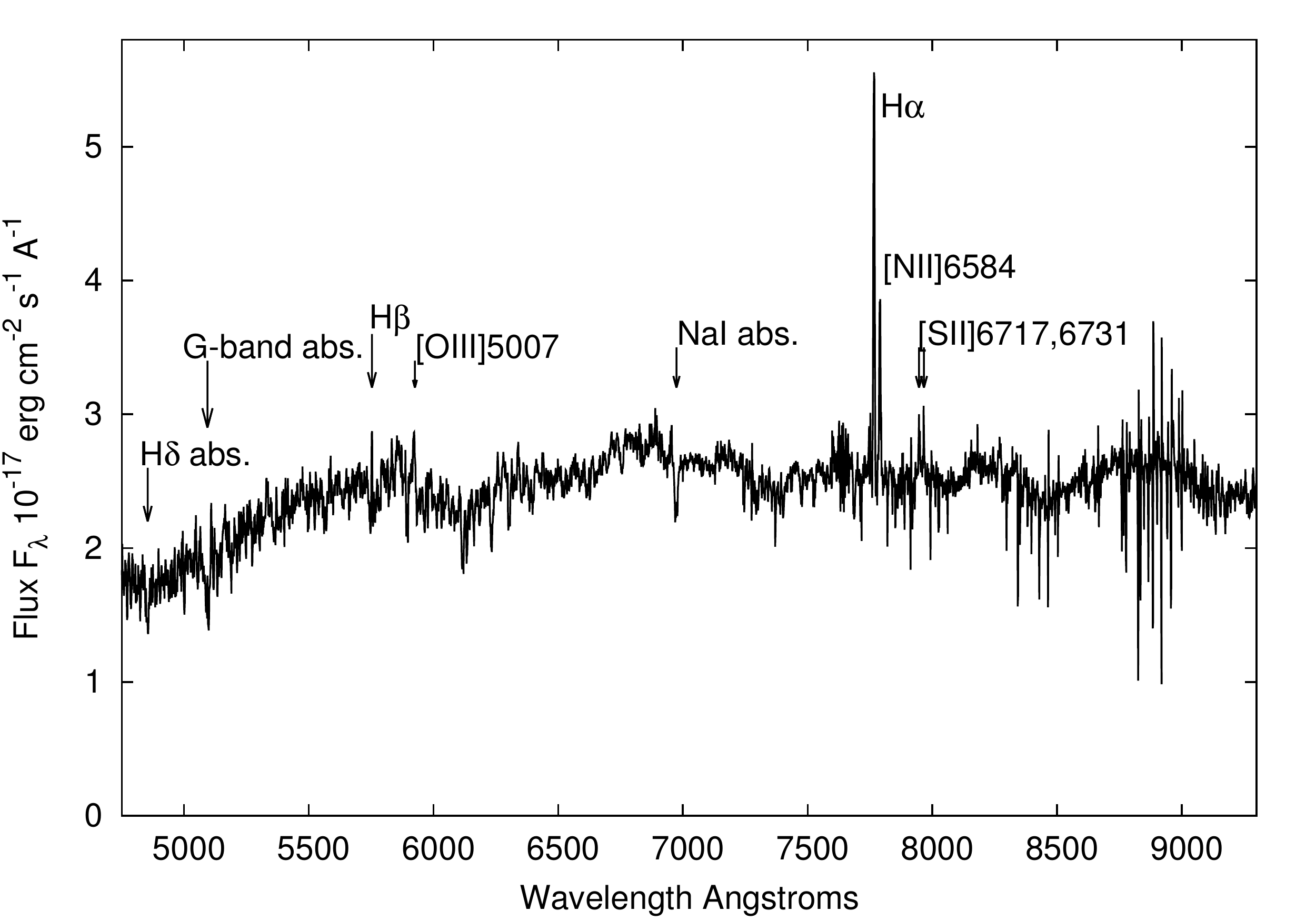}
\caption{Spectrum of the secondary (companion) galaxy to the west of the AGN, extracted in a $r=5$ pixel (1.0 arcsec radius) aperture.}
 \end{figure}
 
The radio galaxy spectrum (Fig 3) shows at least 20 emission lines, many are characteristic of AGN (i.e. high ionization) and the brightest is [OIII]5007. For $\rm H\alpha$ we measure a wavelength   $\rm 7760.5\AA$, giving redshift $z=0.1825$ and observer-frame equivalent width $\rm EW_{obs}(H\alpha)\simeq113\AA$. The AGN emission lines appear winged with flat or slightly double peaks. We measure a $\rm H\alpha$ FWHM of $12.58\pm 0.03\rm\AA$; subtracting in quadrature the instrumental $2.6\rm\AA$ gives $12.31\rm\AA$, multiplying by $c/\lambda$ this corresponds to 476 km $\rm s^{-1}$ (or a velocity dispersion $\sigma\simeq 203$ km $\rm s^{-1}$). 
 
 Fig 4 shows the spectrum of the companion galaxy to the west, summed over the same sized aperture, which appears to be non-AGN and star-forming, with $\rm H\alpha$ and [NII]6584 the most prominent features.  The $\rm H\alpha$ at $\rm 7766.1\AA$ gives a redshift of 0.18335 and a velocity offset $\Delta(v)=+216$ km $\rm s^{-1}$ from the AGN. We find a near-Gaussian line profile with $ \rm H\alpha$ FWHM $6.19\pm0.14\rm\AA$, intrinsically $5.62\rm \AA$ or 217 km $\rm s^{-1}$. The  $\Delta(v)$ is consistent with an interacting pair, with a lower limit (depending on the inclination) on the dynamical mass $\rm M_{dyn}\geq (\Delta(v))^2(9.12 kpc)/G = 1.0\times 10^{11}M_{\odot}$.

 Both galaxy spectra show $\rm NaI_{5890}^{5896}$ absorption lines with moderate EWs of about $2\rm \AA$ and wavelengths in the galaxy rest-frames of $5893.4\rm\AA$ and $5892.5\rm\AA$. These lines are consistent with stellar absorption, with neither galaxy showing enhanced strength and blueshifted NaI lines that  might be be a signature of a neutral outflow, as is seen in many starburst galaxies (e.g. Heckman et al. 2000).

Inskip et al. (2010) noted `a third faint object lying just to the south of the interacting companion galaxy', which was also described by Ramos Almeida et al. (2011) as a second tidal tail.
In our data, this object is a slight intensity peak about 3 arcsec south of the companion nucleus and about 2 magnitudes fainter.  Extracting its spectrum in a $r=4.0$ pixel aperture (Fig 5), we find it is actually a background galaxy at $z=0.6461$, with emission lines  of [OII]3727 ($\rm EW_{obs}=22\AA$), $\rm H\beta$ and [OIII]5007. 
\begin{figure}
 \includegraphics[width=0.73\hsize,angle=-90]{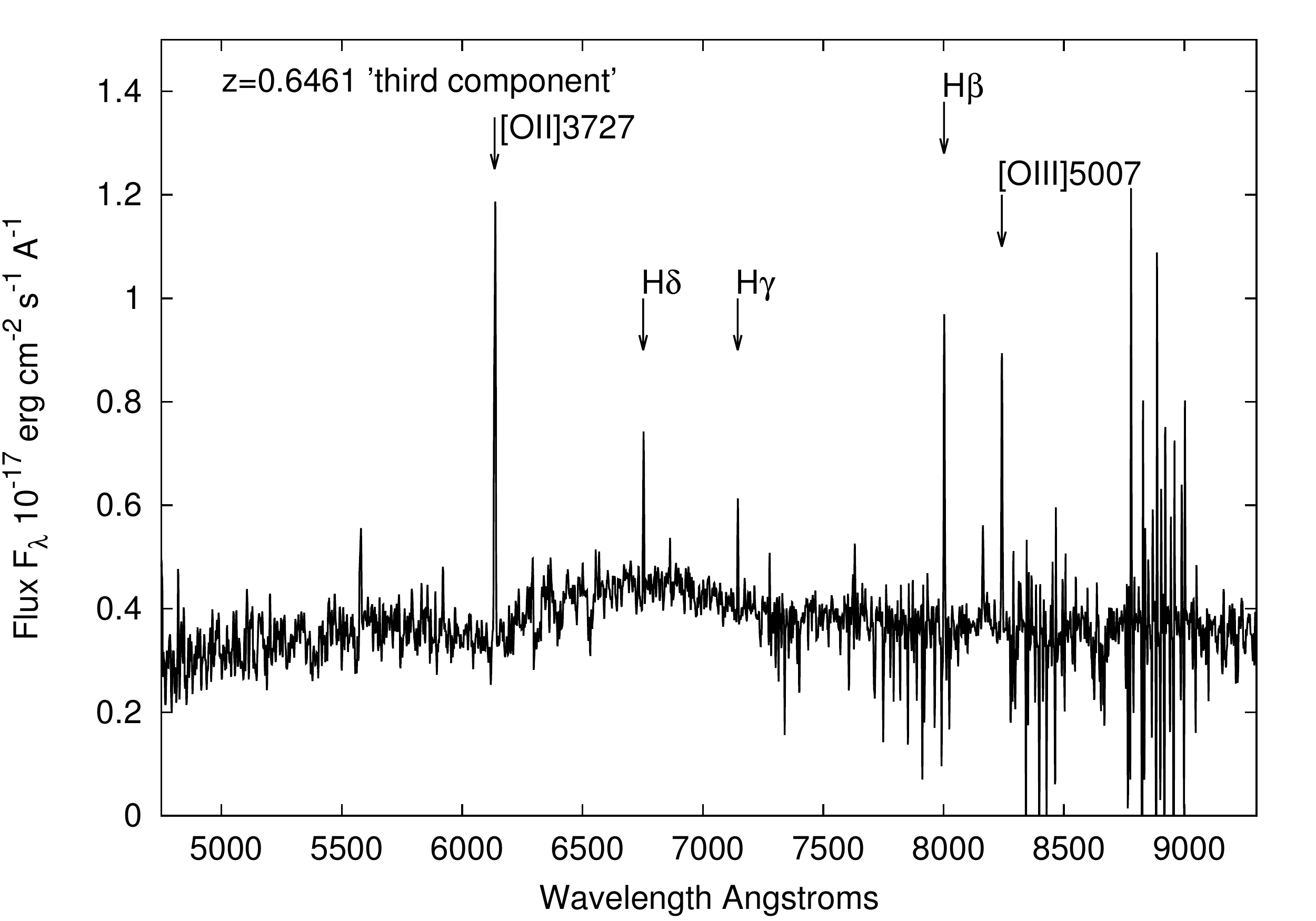}
\caption{Spectrum of the `third component' i.e. the low surface brightness object 3 arcsec south of the companion galaxy, which is not a tidal tail but a background galaxy at $z=0.6461$ (spectrum is extracted in $r=4$ pixel aperture and slightly smoothed).}
 \end{figure}
 We examine the spectra of other galaxy detections in the field and find redshifts for a further 10 galaxies, all but one more remote than PKS 1934-63; these spectra are given in the Appendix. 

Table 1 gives emission-line fluxes measured with {\small IRAF} `splot'  or `fitprof' for the AGN and companion, with Gaussian profiles and in some cases double Gaussians (to better include the flux from broad components).    
   \begin{table*}
   \caption{List of emission and absorption lines detected in the $r=5$ pixel aperture spectra for the PKS 1934 primary/AGN and companion galaxy, in units of  $10^{-17}$ erg $\rm cm^{-2}s^{-1}$. Approximate dust corrections are applied based on the observed Balmer decrement in each galaxy, giving $E(B-V)=0.408$ (AGN) and 0.357, and the Calzetti (2000) extinction law. From the faintest detected lines in the companion galaxy, the limit for line detection is close to $10^{-17}$ erg $\rm cm^{-2}s^{-1}$.}

\begin{tabular}{lcccccc}
\hline
Line & \multispan{3} Primary & \multispan{3} Companion  \\
       & $\lambda_{obs}$~\AA & Emission Flux & $\rm Flux_{corr}$ & $\lambda_{obs}$~\AA & Em. Flux &
 $\rm Flux_{corr}$\\
  \hline
$\rm  [SII]_{4069}^{4076}$ &  4814.9 & $101.6\pm 0.7$ & 781. & - & - & - \\                        
 $\rm H\delta$ & 4850.0 & $12.2\pm 0.4$ &  92.6 & abs & abs & abs \\                                 
 $\rm H\gamma$ & 5132.8 & $61.5\pm 0.7$ & 422. & 5136.1 & $1.95\pm 0.21$ & 10.6 \\     
 $\rm [OIII]4363$ & 5158.2 & $46.9\pm0.7$ & 319. & - & - \\                                                       
 HeII 4686 & 5540.4 & $15.7\pm 0.3$ & 94.1 & - & - & - \\                                                          
 $\rm H\beta$ & 5748.6 & $154.0\pm 1.3$ & 867. & 5752.9 &  $5.55\pm 0.17$ & 25.2 \\    
 $\rm [OIII]4959$ & 5863.5 & $364.8\pm 1.4$ & 1984. & 5865.8 & $1.84\pm 0.22$ &  8.10 \\                                               
 $\rm [OIII]5007$ & 5919.9 & $1130.9\pm 1.6$ & 6052. & 5921.7 & $4.56\pm0.31$ & 19.8 \\   
 $\rm [NI]5199$ & 6147.7 & $23.0\pm0.3$ & 116.  & - & - & - \\                         
 $\rm  [NII]5755$ & 6805.3 & $32.7\pm 0.5$ & 139. & - & - & - \\           
 HeI 5876 & 6947.9 & $16.6\pm 0.3$ & 68.3 & - & - & - \\                         
 $\rm [OI]6300$ & 7450.6 & $615.3\pm 2.0$ & 2272. & - & - & - \\     
 $\rm [OI]6364$ & 7525.7 & $183.1\pm 1.2$ & 671. & - & - & - \\      
 $\rm [NII]6548$ & 7742.5 & $205.4\pm 1.7$ & 719. & 7748.4 &  $1.83\pm 0.16$ & 5.47 \\    
 $\rm H\alpha$ & 7760.5 & $770.0\pm 2.8$ & 2687. & 7766.0 & $24.12\pm0.21$ & 72.0 \\   
 $\rm [NII]6584$ & 7784.2 & $584.2\pm 2.2$ & 2028. & 7789.7 & $8.34\pm 0.22$ & 24.8 \\   
 $\rm [SII]6717$ & 7941.9 & $152.1\pm 0.6$ & 525. & 7947.5 & $3.62\pm 0.26$ & 10.5 \\    
 $\rm [SII]6731$ & 7958.8 & $170.7\pm 1.0$ & 572. & 7965.7 & $2.17\pm 0.20$ & 6.26 \\  
 $\rm [ArIII]7136$ & 8437.6 &  $26.3\pm 0.7$ & 80.8 & - & - & - \\
 $\rm [OII]_{7319}^{7330}$ & 8660.8 & $215.0\pm 1.7$ & 636. &  - & - & - \\
 \hline
 \end{tabular}

  \begin{tabular}{lcccc}
  \hline
Line & \multispan{2} Primary & \multispan{2} Companion  \\
       & $\lambda_{obs}$~\AA & Absorption EW & $\lambda_{obs}$~\AA & Absorption EW \\
  $\rm H\delta$ &    emission & emission & 4853.8 & $1.84\pm0.22$ \\
   $\rm NaI_{5890}^{5896}$    & 6969.0 & $1.98\pm 0.04$ & 6972.9 & $2.01\pm 0.12$ \\
  \hline      
 \end{tabular}
     \end{table*}

\subsection{Detailed Analysis of AGN Emission Lines}

For the primary galaxy, flux ratios $\rm [NII]/H\alpha=0.76$, $\rm [SII]/H\alpha=0.42$ and $\rm [OIII]/H\beta=7.34$ are in the AGN region on the well-known BPT  (Baldwin, Philips \& Terlevich 1981) diagram.  
Due to the high $\rm[OIII]/H\beta$, PKS 1934-63 is classed as a `high excitation galaxy'  (HEG) with an `excitation index' (EI) of 1.05 in the scheme of Son et al. (2012), and the luminosities (from our spectra) $\rm L_{[OIII]}=10^{42.03}$ and $\rm L_{H\alpha}=10^{41.86}$ erg $\rm s^{-1}$ are sufficient for the HEG class.  The luminosity and ratios place the galaxy on the upper evolution sequence in the models of An \& Baan 2012 and Kunert-Bajraszewska (2015), destined to become a FR-II. The $\rm [OI]/H\alpha$ ratio of 0.80 is above most other HEGs and places the galaxy near the Seyfert-LINER borderline. The ratios $\rm[OI]6300/[OIII]5007=0.544(\pm 0.004)$ and $\rm[OIII]4363/[OIII]5007=0.042$ exceed those of almost all Seyfert 2s in Kawakato, Nagao \& Woo (2009), but are typical for narrow-line radio-loud AGN selected as young i.e. compact ($D<1$ kpc) or medium (1--10 kpc) symmetric objects.

The $\rm H\alpha/H\beta$ ratio (Balmer decrement), sensitive to dust extinction, is measured as 5.00, high compared to 2.86 expected for dust-free star-forming galaxies. The intrinsic ratio may be slightly higher for AGN e.g $\sim 3.1$ assumed by Groves, Brinchmann and Walcher (2012), while our best-fit model described below predicts 2.95. Even for the highest of these values, on the basis of the Calzetti extinction law (Calzetti et al. 2000) the extinction is a factor of  $(5.0/3.1)^{2.615}=3.49$ in $\rm H\alpha$, or a reddening $\rm E(B-V)=0.408$ mag.

These flux ratios can also be used to estimate electron density and temperature. Using {\it stsdas.nebular.temden} in {\small IRAF}, the density diagnostic $\rm [SII]6717/[SII]6731=0.891$ in  combination with $\rm T_e$-sensitive $\rm[OIII]4959+5007/[OIII]4363=31.9$ gives the solution $\rm n_e=1321~cm^{-3}$
and $\rm T_e=23420$ K. Another temperature indicator $\rm [NII](6548+6584)/[NII]5755=24.1$ gives  a similar $\rm n_e=1299~cm^{-3}$
and $\rm T_e=22801$ K. This source has a generally high temperature and density as originally reported by Heckman (1980).
Shock ionization from the interaction of AGN jets with gas clouds or the host galaxy itself may help to explain young radio galaxy line ratios (e.g. Inskip et al. 2006; Shih et al. 2013),  and also the high electron temperatures.

\begin{figure*}
\begin{subfigure}{0.5\textwidth}

\centering
 \includegraphics[width=1.11\hsize,angle=0]{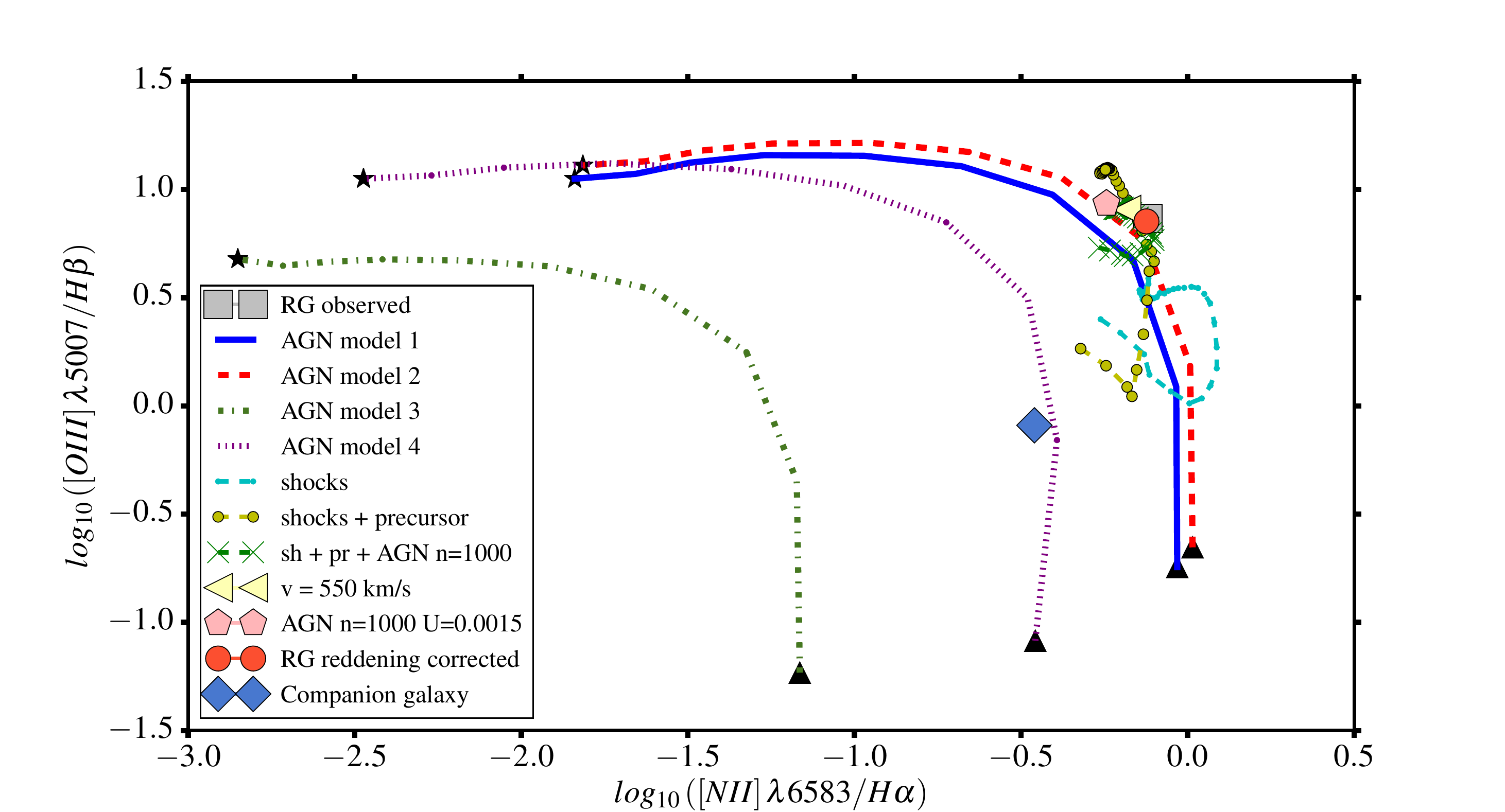}
 \end{subfigure}%
 \begin{subfigure}{0.5\textwidth}
 \centering
  \includegraphics[width=1.11\hsize,angle=0]{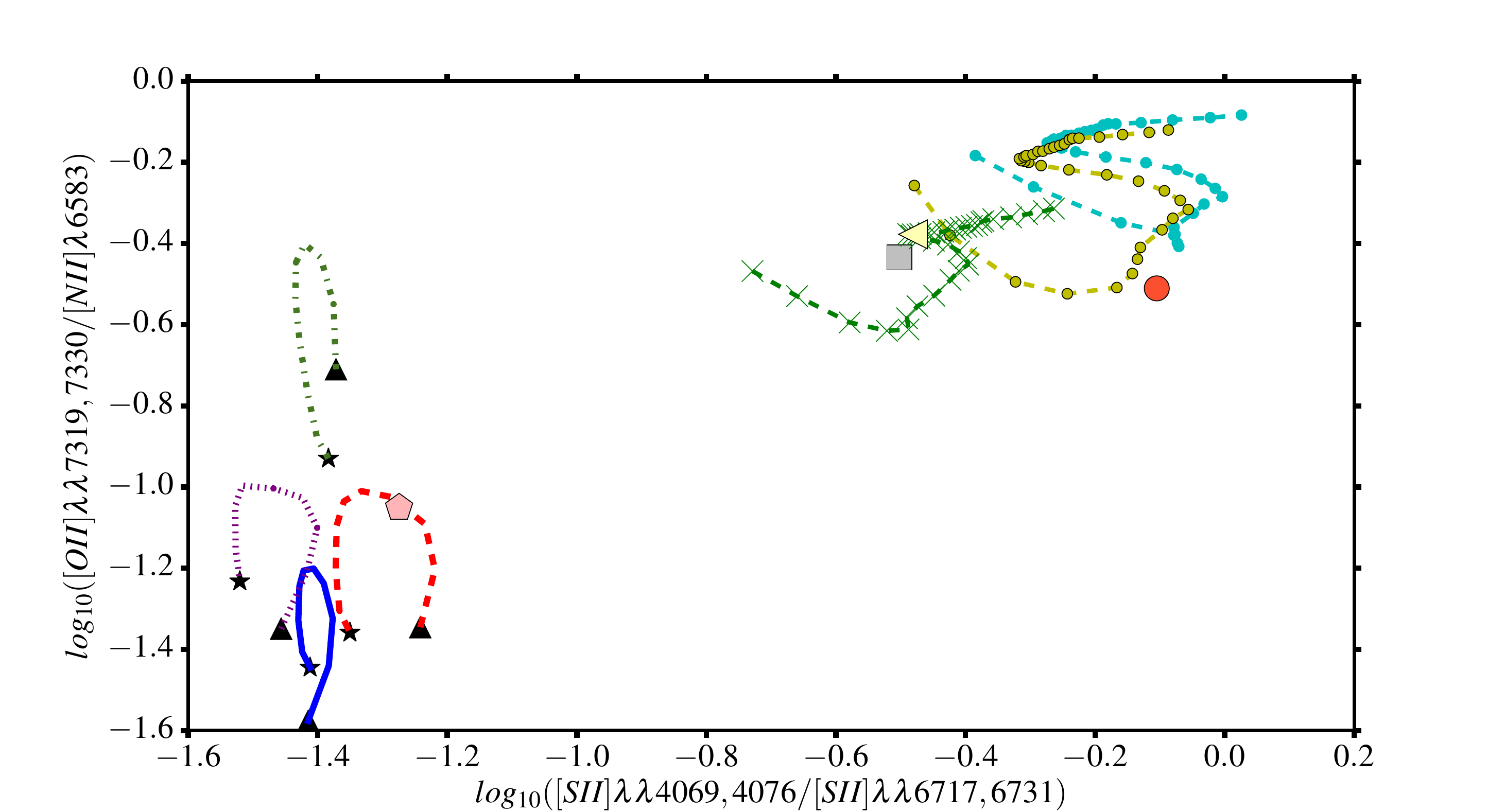}
  \end{subfigure}
  \begin{subfigure}{0.5\textwidth}
  \centering
   \includegraphics[width=1.11\hsize,angle=0]{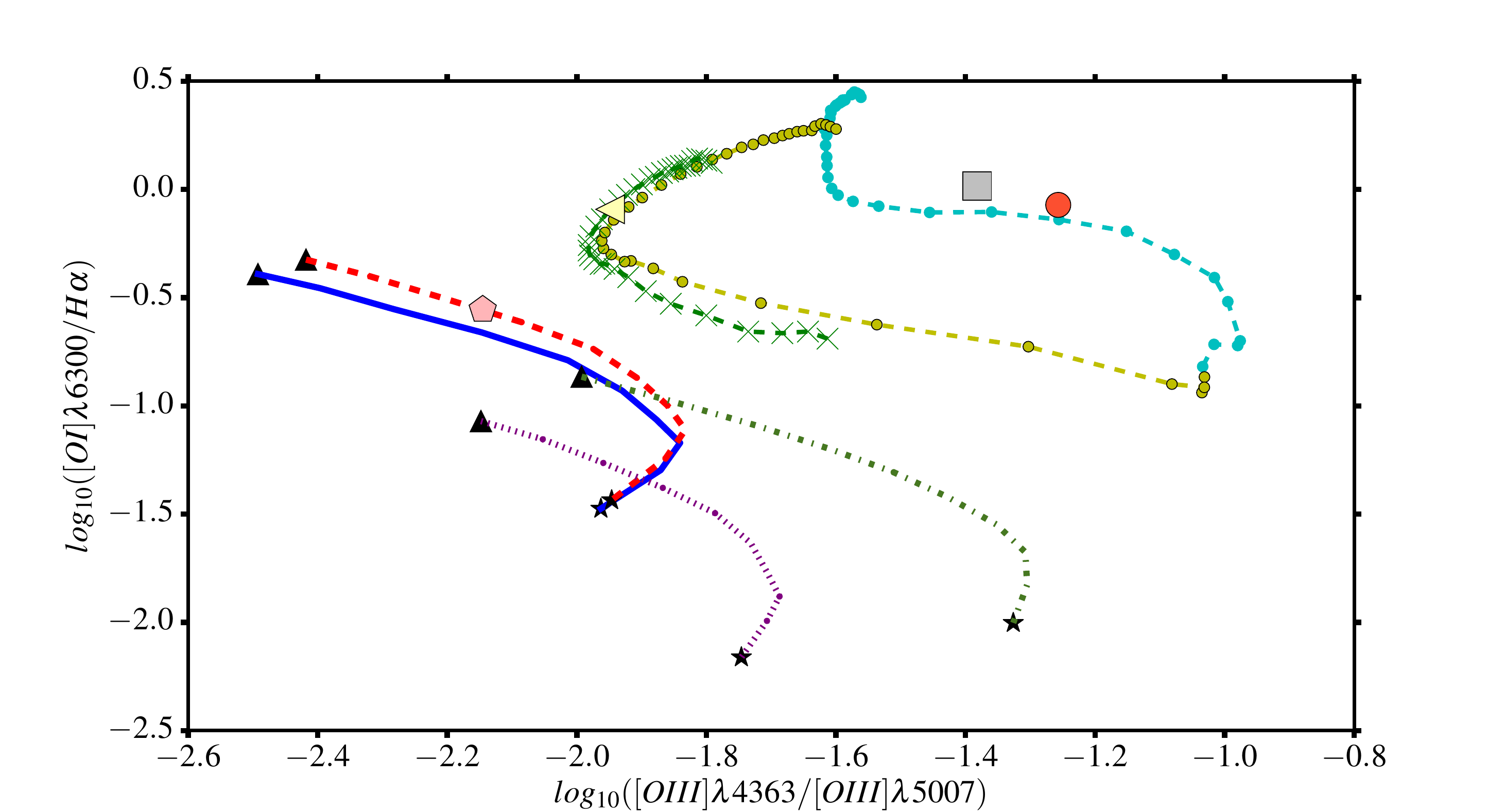}
    \end{subfigure}%
 \begin{subfigure}{0.5\textwidth}
 \centering
    \includegraphics[width=1.11\hsize,angle=0]{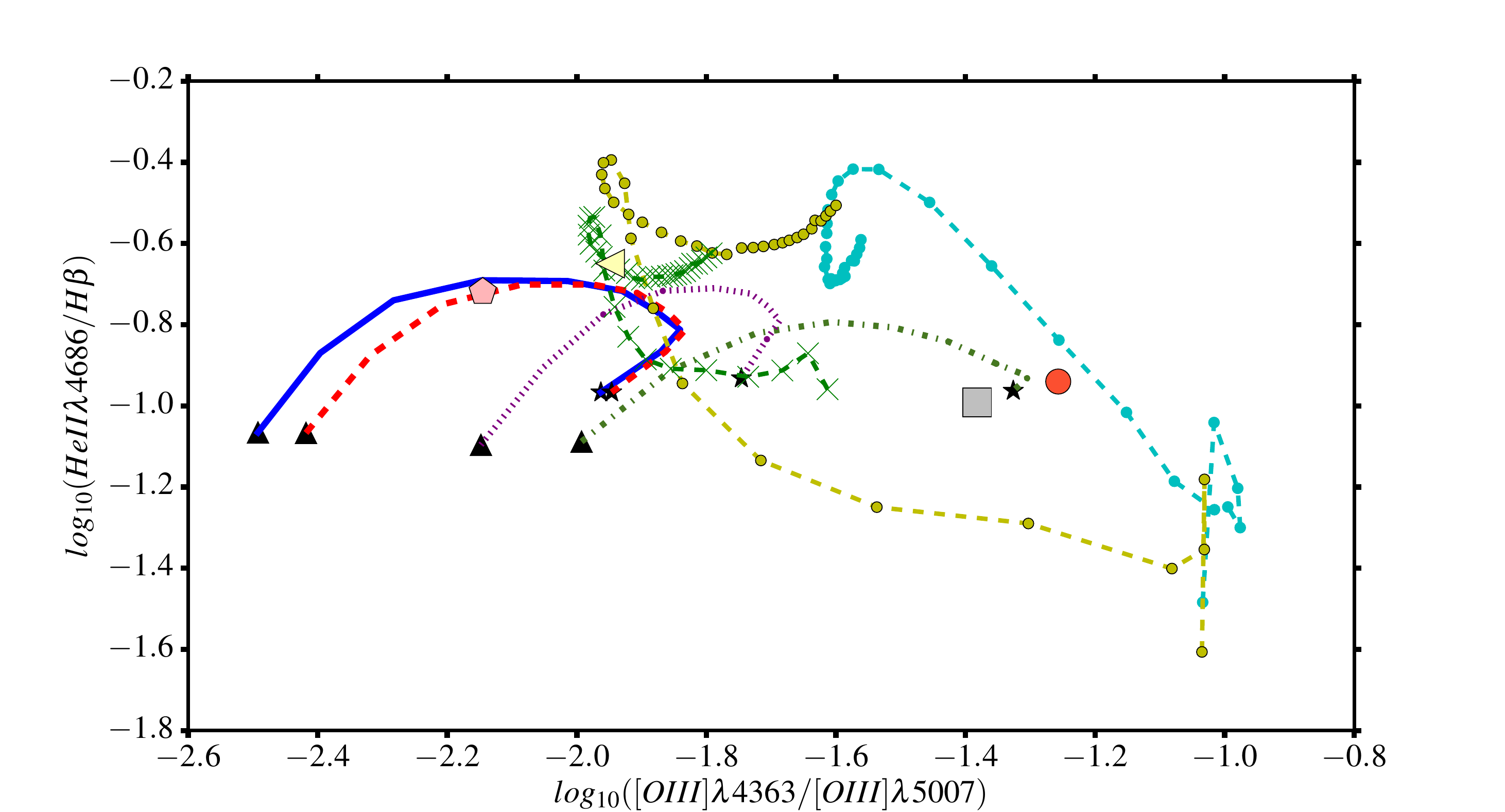}
    \end{subfigure}
\caption{Plots of line ratios showing the observed and dust-corrected ratios from our aperture spectrum for the primary galaxy, compared to (i) four pure AGN photoionization models described in the text, showing a sequence of ionization parameter U increasing from the triangle to the star symbol; a symbol indicates the closest model (model 2) with $\rm n_e=1000~cm^{-3}$ and $U=0.0015$ (ii) a pure shocks and a shocks-plus-precursor model from Allen et al. (2008), showing a sequence of velocity $v_{shock}$ (iii) a 50:25:25 composite of AGN (model 2  with fixed $U=0.0015$ ), precursor and shock models, and showing a sequence of $v_{shock}$. A yellow-triangle symbol indicates the chosen `best'  composite model with $v_{shock}=550$ km $\rm s^{-1}$. The first plot also shows the secondary galaxy which has the different line ratio typical of a star-forming disk galaxy. }
 \end{figure*}

To investigate this we compare with a set of models. On Fig 6 are plotted the primary's emission-line ratios, observed and with an approximate dust-reddening correction based on the Balmer decrement and Calzetti law. In addition, we have computed sequences of pure AGN photoionization is computed using Mappings Ie (Binette, Dopita \&Tuohy 1985; Binette et al. 2012).

(1) A sequence of   ionization parameter U increasing from 0.0001 to 2.0 (from triangle to star symbol), with power-law SED
$f_{\nu}\propto \nu^{ -1.5}$, solar abundances, $n_H = 100$
cm$^{-3}$, and assuming a Maxwell-Boltzmann distribution of electron
energies. The high energy cut-off for the ionizing continuum is 50 keV; see Humphrey et al. (2008). These parameters have been shown to be able to generally reproduce the optical emission line ratios of powerful, low redshift radio galaxies (e.g. Robinson et al. 1987).

(2) As for (1), except with $\rm n_H = 1000$ cm$^{-3}$, as suggested by the [SII]6717/[SII]6731 ratio.

(3) As for (1), except that the gas metallicity is reduced to 0.1 solar.
Other metals are scaled with oxygen, except for nitrogen, which is scaled quadratically such that
N/O $\propto$ O/H while $\rm O/H > 0.4$ solar, and linearly (N/H $\propto$
O/H) at O/H $\le$ 0.4 solar (e.g. Henry, Edmunds and K\"oppen 2000; Humphrey et al. 2008). 

(4) As for (1), except that the electron energy distribution was set to a
non-equilibrium, $\kappa$-distribution with $\kappa=10$ 
(Binette et al. 2012; Nicholls, Dopita \& Sutherland 2012; Nicholls et al. 2013; Humphrey and Binette 2014). 

Also plotted is a shock model sequence from Allen et al. (2008), with a high electron density of $\rm n_e=1000$ $\rm cm^{-3}$, solar abundances and magnetic equipartition with $B=100\mu G$; the sequence plotted is of shock velocity from $v=100$ to 1000 km $\rm s^{-1}$. Also plotted is the corresponding model with the inclusion of the ionized precursor. These two models are described as representing `the physical extremes of having little or no gas ahead of the shock (shock-only) and having an extensive, radiation bounded, precursor region ahead of the shock', thus it could be realistic to use a mixture of the two.

On the first plot, the BPT diagram  of $\rm [NII]/H\alpha$ against $\rm [OIII]/H\beta$, the first two AGN models run close to the observed ratios (which here are insensitive to dust), with the the $ \rm n_e=1000$ model (model 2) slightly closer, and so does the shocks-plus-precursor model. For the AGN models the ionization parameter is quite tightly constrained to $U\simeq 0.0015$.  On the second and third plots, none of the pure AGN models can simultaneously fit the high ratios of  $\rm [OII]_{7319}^{7330}/[NII]6584$, $\rm [SII]_{4069}^{4076}/[SII]_{6717}^{6731}$, $\rm [OI]6300/H\alpha$ and [OIII]4363/[OIII]5007, all of which are more consistent with shock excitation, with high velocities ( $\geq 500$ km $\rm s^{-1}$). It appears that $\rm [OII]_{7319}^{7330}/[NII]6584$, $\rm [SII]_{4069}^{4076}/[SII]_{6717}^{6731}$ and [OIII]4363/[OIII]5007
are all useful diagnostics for distinguishing high-velocity shocks from AGN photoionization, and the
 $\rm [OI]6300/H\alpha$ ratio is a sensitive indicator of shock velocity (although with a moderate dependence on electron density and the magnetic field).

On the fourth plot, all the models of AGN and high-velocity shocks models can fit the $\rm HeII 4686/H\beta$ ratio  of 0.10, which gives no discrimination between them but does establish that these are the dominant sources of Balmer-line emission, rather than star-formation, for which typically $\rm HeII 4686/H\beta\simeq 0.01$ (Shirazi \& Brinchmann 2012).

The low-metallicity AGN model does fit some line ratios more closely -- especially it enhances [OIII]4363 by  its higher electron temperature, and the $\kappa$ (kappa) model (model 4) has a similar effect due to its different distribution of electron energies. However, both of these models, especially the low metallicity, greatly underpredict the [NII]6584 and [OI]6300 fluxes, and so are disfavoured for this galaxy. 
 The emission line profiles seem too narrow to be produced entirely by high-velocity shocks, but their winged form (both narrow and broad components) together with the flux ratios points to  a composite AGN/shocks model. We try combining the best-fit AGN model  (model 2) with $\rm n=1000~cm^{-3}$ and $U$ fixed at 0.0015,  with the pure shocks and the shocks plus precursor models in the ratio 0.5:0.25:0.25 (in $\rm H\beta$ flux). On the plots we show this combined model as a sequence of increasing shock velocity;  the best-fit to the $\rm [OI]6300/H\alpha$  ratio (0.80) has $v_{shock}\simeq 550$ km $\rm s^{-1}$ (this model indicated by the triangular symbol).

In summary, standard AGN models can explain the position of this radio galaxy on the BPT diagram, and the ratio $\rm HeII 4686/H\beta$, but they do not account for the high values of the other line ratios considered here. The observed line ratios are much closer to our composite model with an approximately equal mixture of AGN emission and high-velocity shocks. These conclusions apply whether or not the  (uncertain) dust reddening corrections are applied.

\subsection{Spectrum of Second Galaxy}

The companion galaxy spectrum has fewer visible lines and the $\rm EW_{obs}$ of $\rm H\alpha$ is only $\rm 9.4\rm\AA$. Some emission-line fluxes, especially $\rm H\beta$,  may be underestimated  because of stellar absorption. To correct for this, the emission-line fluxes given in the table for the companion galaxy, are remeasured after subtraction of a fitted stellar model (from {\it Starlight} see Section 6) from the observed spectrum, which has the effect of increasing $\rm H\alpha$ by about $9\%$, $\rm H\beta$ even more.
With these corrected fluxes the $\rm [NII]/H\alpha$ ratio is $0.35\pm 0.01$,  and the $\rm [OIII]/H\beta$ ratio $0.82\pm0.08$, consistent with a star-forming galaxy. The Balmer decrement $\rm H\alpha/H\beta$ is high at 4.35, which corresponds to a dust extinction of a factor $\simeq (4.35/2.86)^{2.615}=2.99$  in $\rm H\alpha$ or $E(B-V)= 0.357$ mag, almost equal to the primary.
From the $\rm [NII]/H\alpha$ (N2) ratio, metallicity can be estimated as $\rm 12+log(O/H)\simeq 8.53$--8.64 with the lower calibration of Marino et al. (2013) and the higher of Pettini and Pagel (2004, PP04). The O3N2 index of 0.370 (again see Marino et al. 2013)  gives $\rm 12+log(O/H)\simeq 8.45$--8.61, but with greater uncertainties.

At this distance the observed $\rm H\alpha$ flux in the aperture gives  $\rm L_{H\alpha}=10^{40.36}$ erg $\rm s^{-1}$, corresponding to a star-formation rate (SFR)  $\rm 4.4\times10^{-42}L(H\alpha)M_{\odot}yr^{-1}$ (Sobral et al. 2014, for Chabrier IMF), only $\rm 0.10~ M_{\odot}yr^{-1}$, but correcting for internal dust and for emission outside the aperture (estimated below) will increase this.

\section{Emission-line Imaging and Galaxy Morphology}
A narrow-band $\rm H\alpha$ image (Fig 7) is extracted by summing the data cube over $7750\leq \lambda\leq 7774\rm \AA$ (20 pixels, containing the line from both galaxies) and subtracting the continuum image at $7840\leq \lambda \leq 7900\rm \AA$.
The primary/AGN galaxy appears round and stellar and is by far the brightest source, but there is clearly $\rm H\alpha$ emission from the second galaxy, extending to at least 28 pixels (17 kpc) west of the AGN, with a more complex structure, almost V-shaped with an elongated central region and the tail or arm extending NW. The $\rm H\alpha$ image does not show the
`third component', because of the mismatched redshift, nor any other sources in the $56\times 56$ arcsec (172 kpc) area, meaning there are no neighbouring emission-line galaxies (or extended nebulae)  within $\Delta(z)\pm0.0032$ or $\pm 967$ km $\rm s^{-1}$ (however, there is an absorption-line galaxy, see Appendix). 

With {\it s-extractor} run on the narrowband $\rm H\alpha$ image, only the AGN and companion are detected, with ($r=5$ pix) aperture fluxes $7.265\pm 0.009\times 10^{-15}$ and $2.186\pm0.042\times 10^{-16}$ erg $\rm cm^{-2}s^{-1}$,  similar to those from fitting lines in the spectra. The `Kron' total fluxes are $8.5133\pm 0.012\times 10^{-15}$ and $5.224\pm0.132\times 10^{-16}$ erg $\rm cm^{-2}s^{-1}$.

\begin{figure}
 \includegraphics[width=1.03\hsize,angle=0]{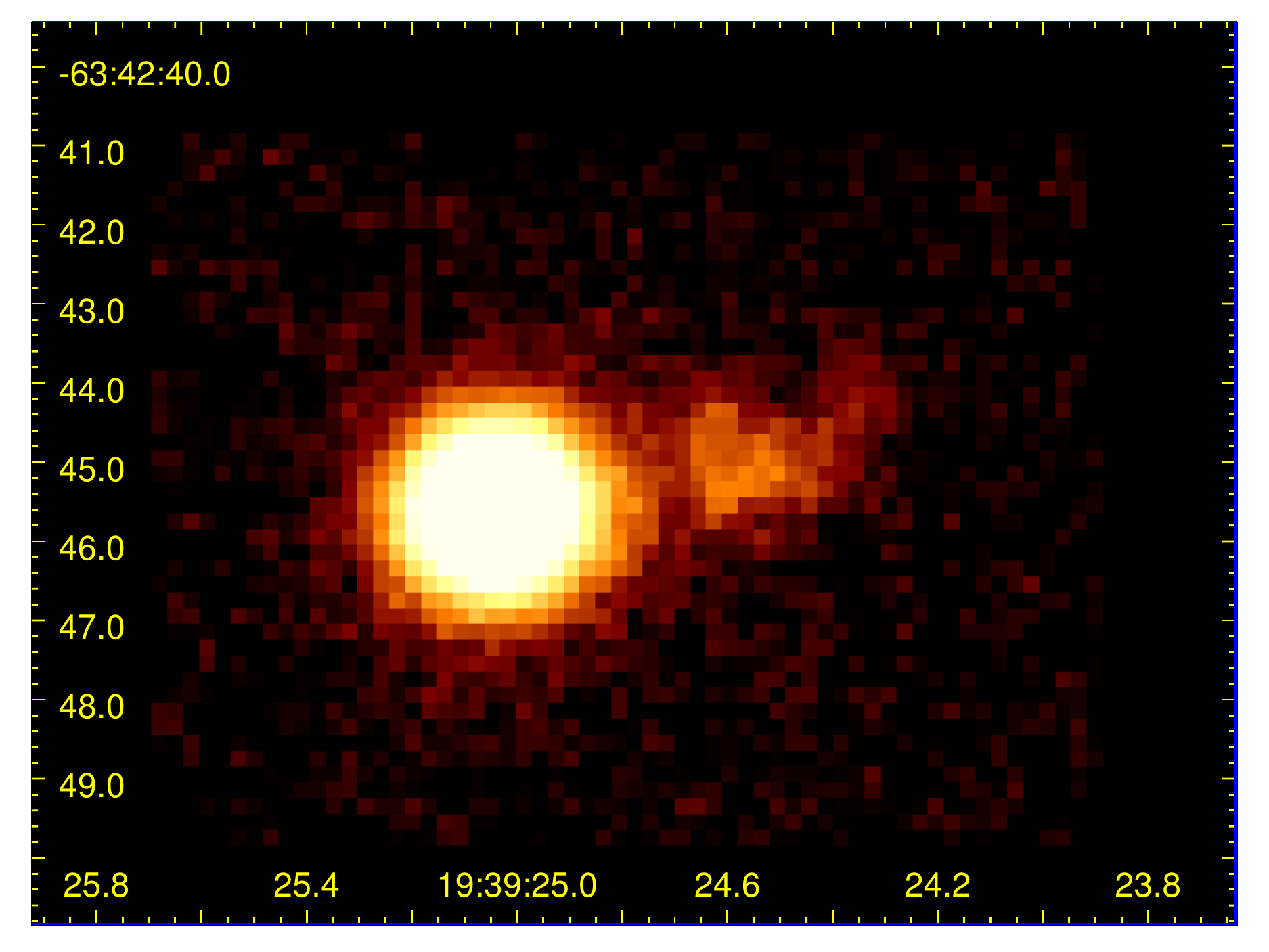}
\caption{Continuum-subtracted $\rm H\alpha$ image of PKS 1934-63, both galaxies visible, the bright AGN and the more irregular companion galaxy to the right (west). $12\times 9$ arcsec area, asinh scaling, RA and Dec axes.}
 \end{figure}
 \begin{figure}
 \includegraphics[width=1.03\hsize,angle=0]{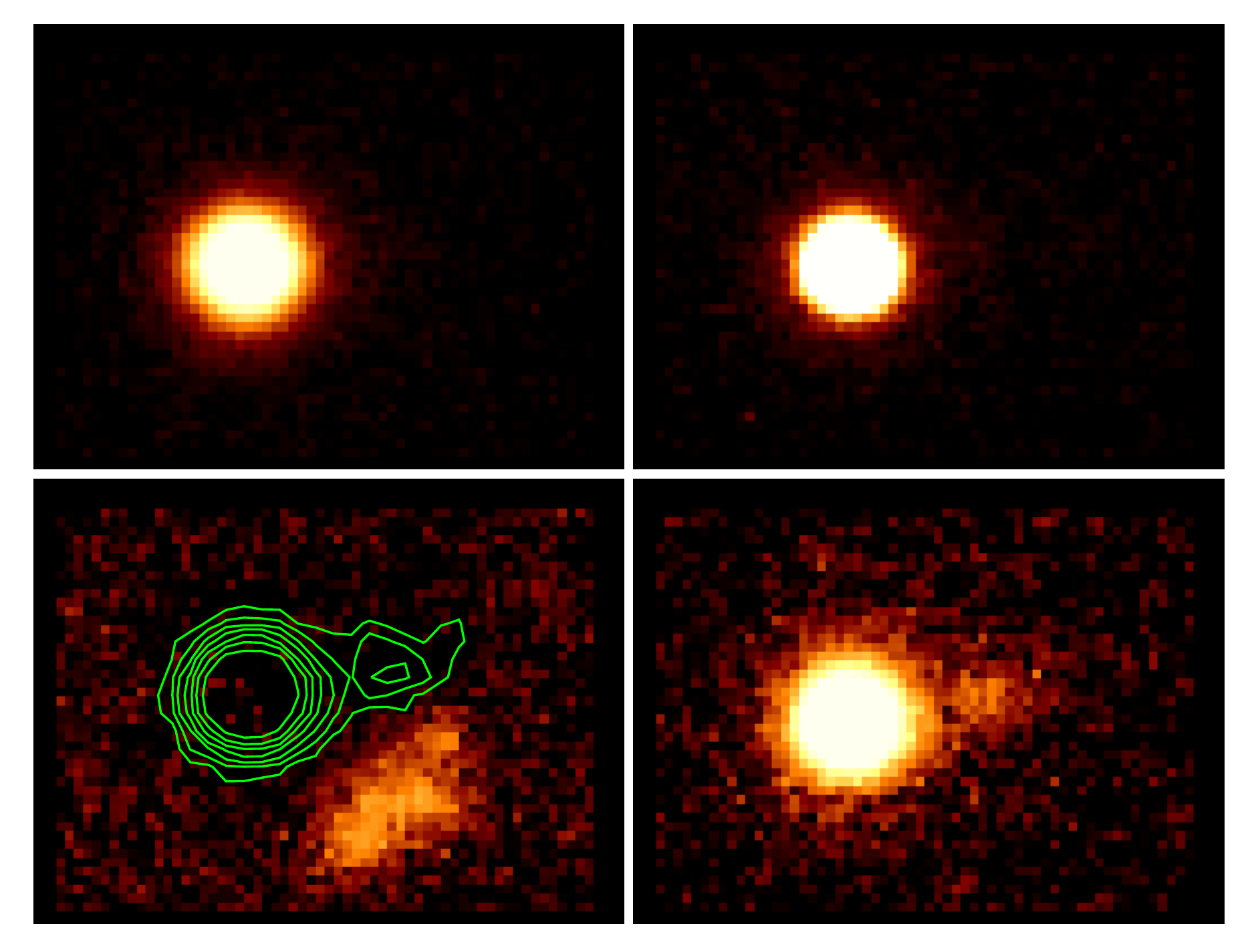}
\caption{Continuum-subtracted images of PKS 1934-63 in [OI]6300 (top left) and [OIII]5007 (top right).  $12\times 9$ arcsec area. Only the AGN in the primary galaxy is visible, giving a star-like appearance. Below (left), a continuum-subtracted 6131--$\rm 6142\AA$ image revealing the more distant `third component' galaxy in [OII]3727 (green contours show the position of PKS 1934-63 as seen in $\rm H\alpha$). Below right is the [NII]6584 image which does show the companion, a little less strongly than in $\rm H\alpha$.}
 \end{figure}
 
 On continuum-subtracted images for the [OIII]5007 and [OI]6300 emission lines (Fig 8), the contribution from star-formation will be far less significant, and indeed only the AGN host is visible. Also we show a continuum-subtracted 6131--$\rm 6142\AA$ image which reveals the faint but large  `third component' galaxy to the SW and $z=0.6461$ in its strong [OII]3727 emission.
 
 To further investigate morphology and the question of extended emission, we first fit profiles with {\small IRAF} {\it stsdas.analysis.isophote.ellipse}, centering on the two nuclei (and for comparison, one of the bright stars).
The broad-band profiles of the primary (Fig 9) have detectable flux out to more than 30 pixels (18 kpc), with half the ({\it s-extractor} total) flux enclosed within (half-light radii)  $r_{hl}\simeq 6.4$ pixels in $R$ and 6.8 pixels in $I$. In contrast, the $\rm H\alpha$, [OI] and [OIII] profiles appear only slightly more extended than a point source (star). 

For the companion galaxy, the $\rm H\alpha$ and continuum have similarly extended profiles, with the former somewhat  flatter near the centre (Fig 10). At $r>1.5$ arcsec (the midpoint of the pair) the  $\rm H\alpha$ profile turns up because of the emission from the far more luminous AGN. We  tried removing the AGN from the $\rm H\alpha$ image by subtracting  the model profile (suitably scaled) fitted by IRAF to the [OI]6300 image (where only the primary is seen). The AGN-subtracted $\rm H\alpha$ profile then decreases near-exponentially with no turn-up, and out to 15 pixels  sums to give a total $\rm H\alpha$ flux for the companion galaxy of $4.47\times 10^{-16}$ erg $\rm cm^{-2}s^{-1}$; by comparison to narrowband image summed over just the $r=5$ pixel aperture, this gives an aperture  correction as a factor 4.47/2.186=2.045 ($\pm 0.1$ approx.).
The total $\rm H\alpha$ flux from the primary, integrating to the largest radii (20 pixels) and subtracting off the companion's contribution, sums to $8.7\times 10^{-15} $ erg $\rm cm^{-2}s^{-1}$
($\rm L_{H\alpha}=10^{41.915}$ erg $\rm s^{-1}$).

\begin{figure}
 \includegraphics[width=0.75\hsize,angle=-90]{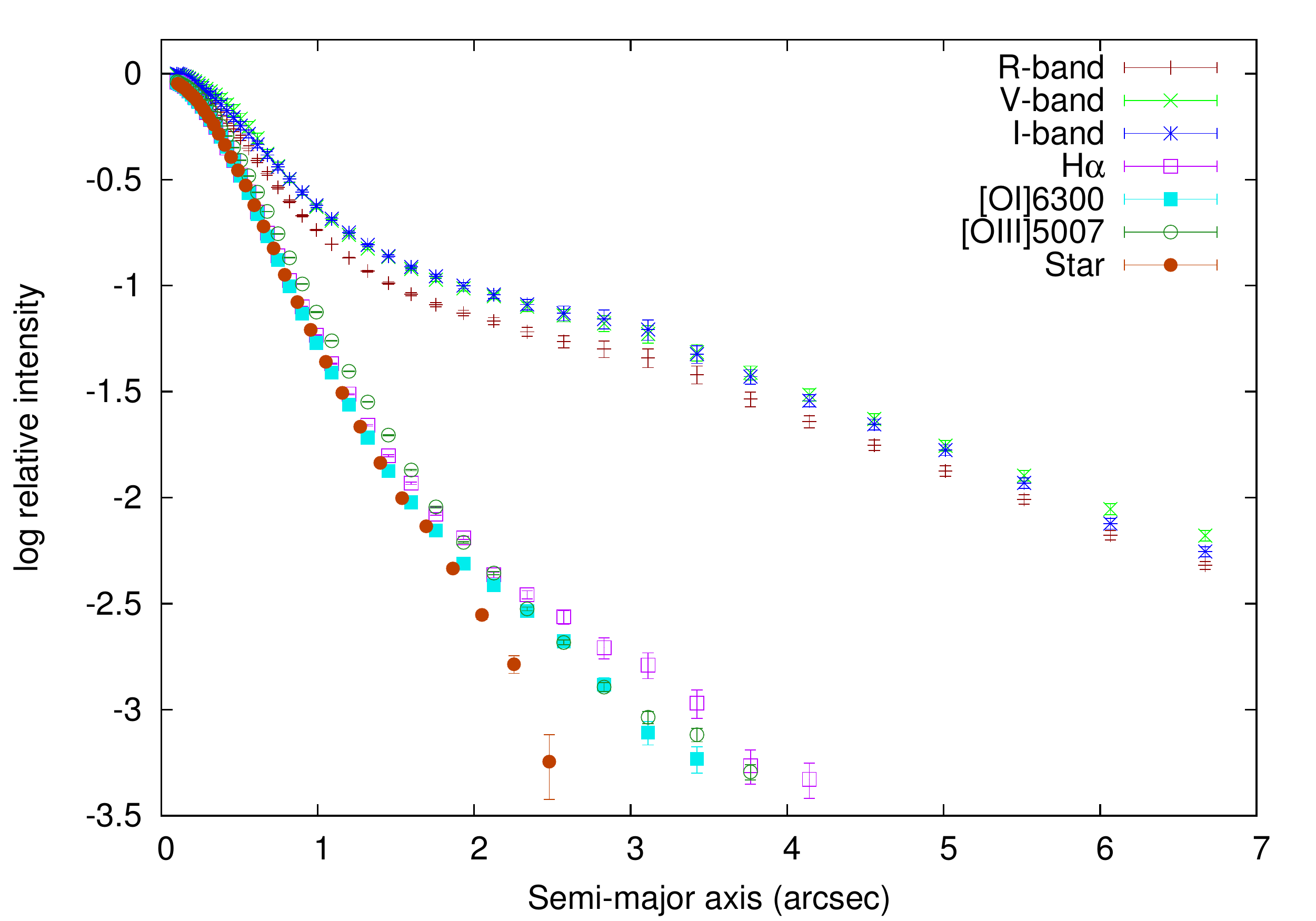}
\caption{Flux density profiles centred on the primary galaxy, in broad and narrow (emission-line) bands, scaled to the same central intensity, and compared with a star on the same MUSE image. The companion galaxy at $r\simeq3$ arcsec is not masked-out here and affects some profiles.}
 \end{figure}
\begin{figure}
 \includegraphics[width=0.75\hsize,angle=-90]{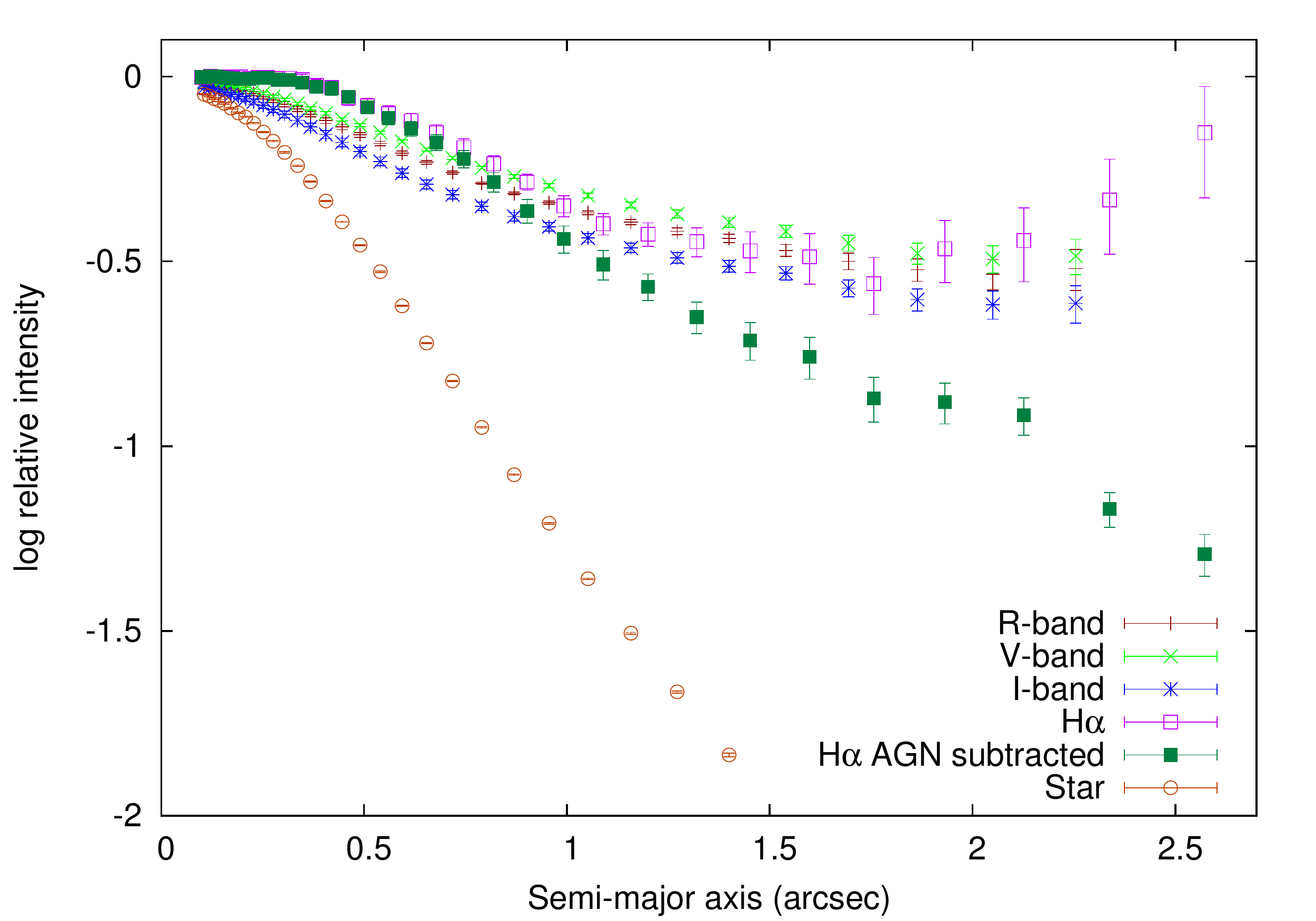}
\caption{Flux density profiles centred on the companion galaxy, in broad bands, $\rm H\alpha$, and in $\rm H\alpha$ with the AGN subtracted from the image (using a model profile, fitted to the[OI] image, scaled to the same central intensity). Compared with the same stellar profile.}
 \end{figure}
  \begin{table*}
  \caption{{\it Galfit} model fits to the primary (g1) and companion (g2) giving effective radius on semi-major axis $r_{eff}$, S\'ersic index $n$, axis ratio b/a, position angle and either $R$ mag (for I \& II) or flux in units $10^{-17}$ erg $\rm cm^{-2}s^{-1}$. $r_{eff}$ is half-light radius (for exponential profiles the scale length is converted as $r_{hl}=1.678~ r_{exp}$).
Fit I is a S\'ersic profile fit to both galaxies on the $R$-band image (6400--$\rm 8000\AA$). In Fit II, g1 is fit with two components, a point source (denoted `p')  and a S\'ersic profile (this gives a larger $r_{hl}$ and is  significantly closer than Fit I). Fits III and IV are our two most successful fits to the narrow-band $\rm H\alpha$ image. Fit III uses a combined point source and exponential for g1 and a S\'ersic model for g2, and in Fit IV the extended component of g1 is fitted with a de Vaucouleurs profile ($n=4$). V, VI and VIII are fits to the [OIII] [OI] \& [NII] narrowband images with exponentials plus a point source (the companion galaxy is not included in the [OI] fit). Error bars are as given by {\it Galfit} for parameters which are allowed to vary freely.}

\begin{tabular}{lcccccc}
\hline
Fit & \multispan{2} $r_{eff}$ & S\'ersic & b/a & PA & {\it mag} or flux \\
    & pixels & kpc & $n$ &   & deg.   & in units $10^{-17}$\\
\hline                   
I g1 & $5.59\pm 0.04$ & $3.44\pm 0.03$ & $3.96\pm 0.05$ & $0.93\pm 0.01$ & $-22.5\pm 4.0$ & {\it 17.64}\\
\smallskip
($R$)g2 & $6.18\pm 0.07$ & $3.81\pm 0.04$ & $1.62\pm 0.03$ & $0.66\pm 0.01$ & $-41.4\pm 1.3$ & {\it 18.92} \\
\hline
II:g1 & $9.99\pm 0.16$ & $6.16\pm 0.10$ & 4.0 & $0.91\pm 0.01$ & $-19.9\pm 3.1$ & {\it 17.64} \\
 ($R$)g1              & p & p & p & - & - &  ${\it 19.61\pm 0.02}$ \\
                   \smallskip
g2 & $6.27\pm 0.06$ & $3.87\pm 0.04$ & $1.69\pm 0.03$ & $0.66 \pm0.01$ & $-46.0\pm 1.2$ & {\it 18.92} \\
\hline
III:g1 & $2.38\pm 0.08$ & $1.47\pm 0.05$ & 1.0 & $0.69 \pm0.02$ & $-46.0\pm 2.6$ & $120\pm 5$\\
($\rm H\alpha)$g1                  & p & p & p & - & - & $746\pm 7 $\\
                     \smallskip
g2 & $7.10\pm 0.36$ & $4.37\pm 0.22$ & $0.92\pm 0.07$ & $0.49\pm 0.02$ & $-75.2\pm 2.0$ & $53.6\pm 1.5$\\
\hline
IV g1 & $1.43\pm 0.10$ & $0.88\pm 0.06$ & 4.0 & $0.73\pm 0.02$ & $-50.2\pm 3.9$ & $173\pm 7$\\
 ($\rm H\alpha)$ g1            & p & p& p & - & - & $708\pm7$ \\
                \smallskip
g2 & $6.41\pm 0.30$ & $3.95\pm 0.19$ & $0.79\pm 0.07$ & $0.48\pm 0.02$ & $-74.9\pm 2.0$ & $46.7\pm 1.3$\\
\hline
V g1 & $1.29\pm 0.03$ &   $0.80\pm 0.03$    &  1.0 & $0.49\pm 0.04$ & $84.6\pm 1.9$ & $239\pm 11$ \\
([OIII]) g1 & p & p & p & - & - &  $1000\pm 10$ \\
g2 & $27.0\pm 3.0$ &  $16.6\pm1.8$      & 1.0 & $0.59\pm 0.05$ & $-82.7\pm 7.6$  & $50.3\pm 3.5$ \\
\hline
VI g1 & $9.20\pm 0.60$  & $5.67\pm 0.37$    &  1.0 &  $0.77\pm0.04$ & $-63.1\pm7.8$ & $36.7\pm 1.1$  \\
([OI])  g1 & p & p & p & - & - & $540\pm 6$ \\
\hline
VII g1 &$6.33\pm 0.27$ & $3.89\pm 0.17$ & 1.0 & $0.66\pm 0.02$ & $-22.0\pm 2.8$ & $70.8\pm 2.0$ \\
([NII]) g1 & p & p & p & - & - & $624\pm 6$ \\
g2 & $6.91\pm 0.55$ & $4.26\pm 0.34$ & 1.0 & $0.53\pm 0.04$ & $-72.0\pm 4.5$ & $26.5\pm1.6 $\\
\hline
\end{tabular}
\end{table*}

 We simultaneously fit the profiles of the two galaxies using {\it Galfit} (Peng et al. 2010). First,  a comparison star in the broad $R$-band is fit with a Moffat profile, giving $\rm FWHM= 3.66$ pixels (0.73 arcsec) and $n=2.53$, and this fit was used as the point-spread function that {\it Galfit} convolves with model profiles (in $\rm H\alpha$ the star gives  $\rm FWHM=3.55$ pixels and $n=2.51$)
 
Then, we fit the $R$-band image with single S\'ersic profiles (simple ellipsoidals) for each galaxy (Fit I in Table 2) and found for the primary a de Vaucouleurs ($n\simeq 4$) profile of relatively small half-light radius ($r_{hl}$), but the residual image showed a central peak surrounded by negative values. Another fit is made (Fit II) by including for the primary an additional point-source component, with flux allowed to vary; it was fitted at $14\%$ of flux with the 
 $r_{eff}$ of the extended S\'ersic (index fixed at $n=4$) component increased to $r_{hl}=9.99$ pixels, 6.16 kpc. This is visibly a better fit with a significantly lower $\chi^2$, so this $r_{hl}$ will be more correct.
 
 The companion galaxy is fitted with a more disk-like S\'ersic index of 1.69, $r_{hl}= 6.27$ pixels or 3.87 kpc, and the residual image (Fig 11) shows the pattern of a barred spiral, with  (as noted by Ramos Almeida et al. 2011), a tidal tail or arm extending west and curving north.
The $\rm H\alpha$ brightest region corresponds to the bar and the inner western arm, forming a 'V' shape. These faint bar and arm features have a surface brightness $23.7\pm 0.5$ $R$ mag $\rm arcsec^{-2}$ compared to the secondary's central 20.4 $R$ mag $\rm arcsec^{-2}$.

 Inskip et al. (2010) previously ran {\it Galfit} on $K$-band imaging of this galaxy and similarly fitted the primary and companion with respective S\'ersic indices 4 and 2, and $r_{eff}$ of 4.38 kpc (smaller than ours) and 5.99 (larger) kpc.
 
 In emission line images, the primary galaxy appears almost stellar, but there are strong indications of an extended component. To obtain good fits with {\it Galfit} requires  two components;  (fits III to VII in Table 2) each time the point-source dominates the flux, but with $\sim 8$--$25\%$ extended emission. Its best-fit morphology varied for different lines: in [OIII] the  extended component is bright but very compact (sub-kpc) and may be aligned with the radio axis, in $\rm H\alpha$ it is strong, more extended, and elongated on PA$\sim -45^{\circ}$, in [OI] and [NII] most extended but with a low surface brightness and a smaller flux fraction.
   The companion galaxy in $\rm H\alpha$ is fit with a similar $r_{hl}$ as it has in the $R$-band, with an even lower $n$ index (near pure exponential), and similarly in the narrowband [NIII]6584. 
 


 \begin{figure}
 \includegraphics[width=1.05\hsize,angle=0]{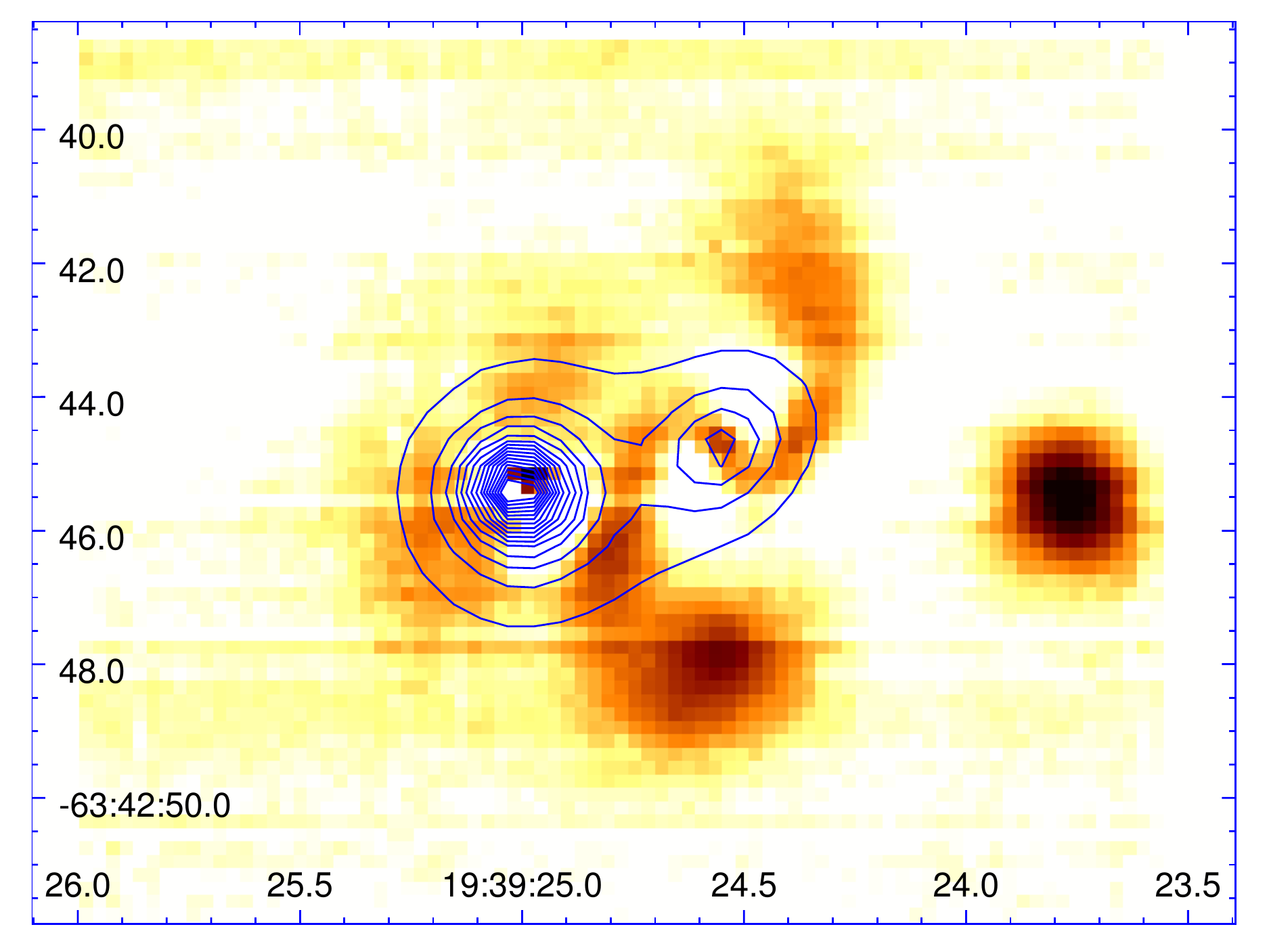}

\caption{Residuals of the $R$-band image to the S\'ersic (plus point source) model fits (contours), showing patterns (like reverse `S') suggestive of a barred spiral for the companion galaxy. Showing $16\times 13$ arcsec area with RA/Dec axes.  The `blob' to the far south of the spiral is not part of the arm but the `third component' background galaxy.}
 \end{figure}

\section{Kinematics of the two Galaxies}
\subsection{Velocity Maps in $\rm H\alpha$ and Other Lines}
In this section we focus on the area with detected $\rm H\alpha$ emission from  PKS 1934-63, a box about $8\times5$ arcsec. Spectra for these 1000 pixels are extracted, sky-subtracted and corrected for Galactic reddening. Fig 12 shows the spectra for the $\rm H\alpha$ and [NII]6548,6584 wavelength region summed in one $\rm arcsec^2$ cells (each $5\times 5$ pixels) in a grid covering this area, with the primary/AGN left of centre and the companion galaxy on the right. In addition to the $\Delta(v)=216$ km $\rm s^{-1}$ offset between the two galaxies these emission lines reveal a significant velocity gradient within each. 


  \begin{figure}
 \includegraphics[width=0.75\hsize,angle=-90]{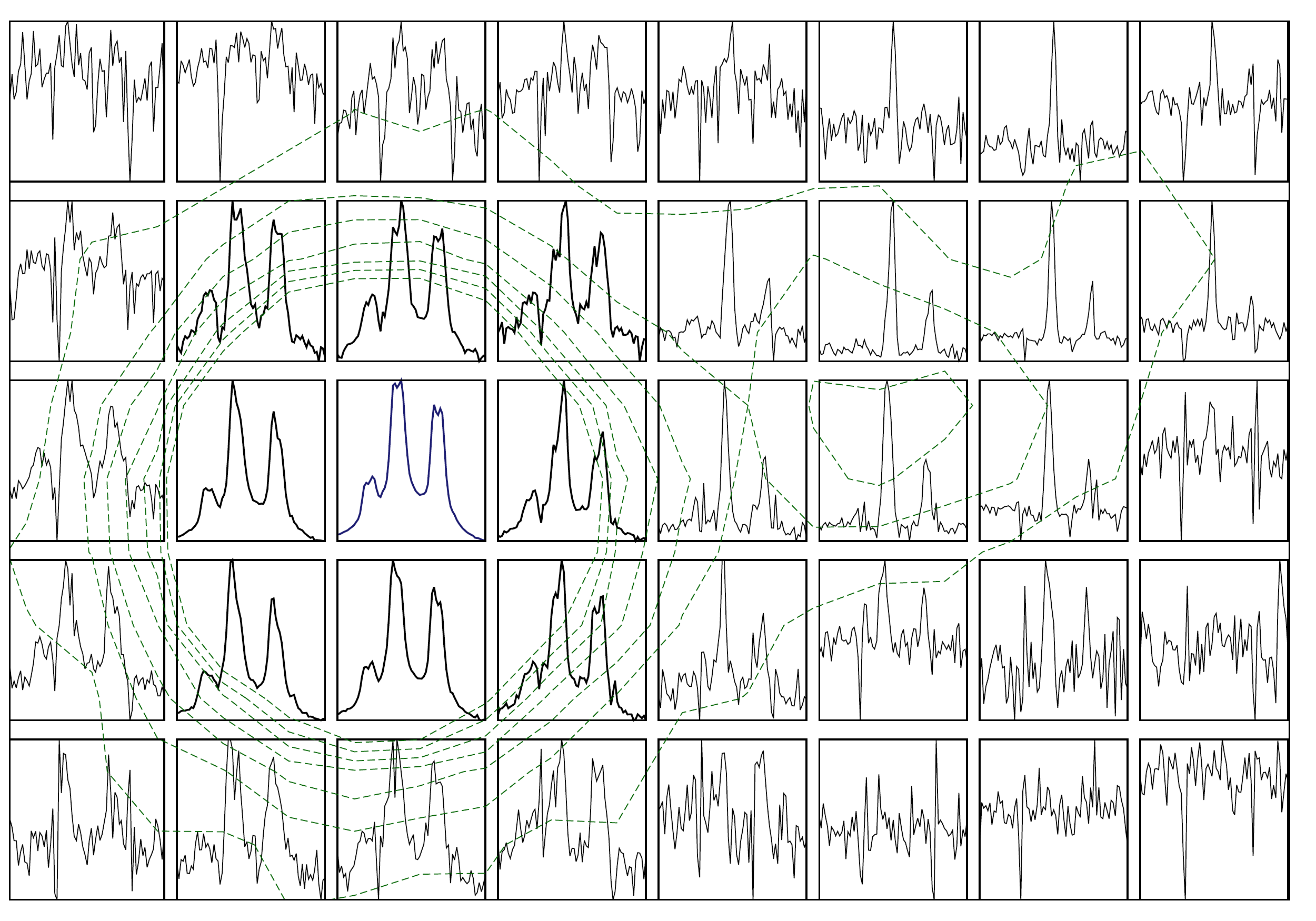}
\caption{The $\rm H\alpha$ and [NII] region of the spectrum (7724--$7814\rm\AA$), summed in one square arcsec ($5\times 5$ pixels) cells, in a grid of $8\times 5$ arcsec covering the primary and companion (the whole region of $\rm H\alpha$ emission). The AGN is within cell (3,3), denoted bold/blue. Overplotted is a  $\rm H\alpha$ contour map of the same area to show the galaxy positions (contour spacing $3.125\times10^{-17}$ erg $\rm cm^{-2}s^{-1}arcsec^{-2}$, for clarity only the first 7 shown).}
 \end{figure}
  \begin{figure}
 \includegraphics[width=0.75\hsize,angle=-90]{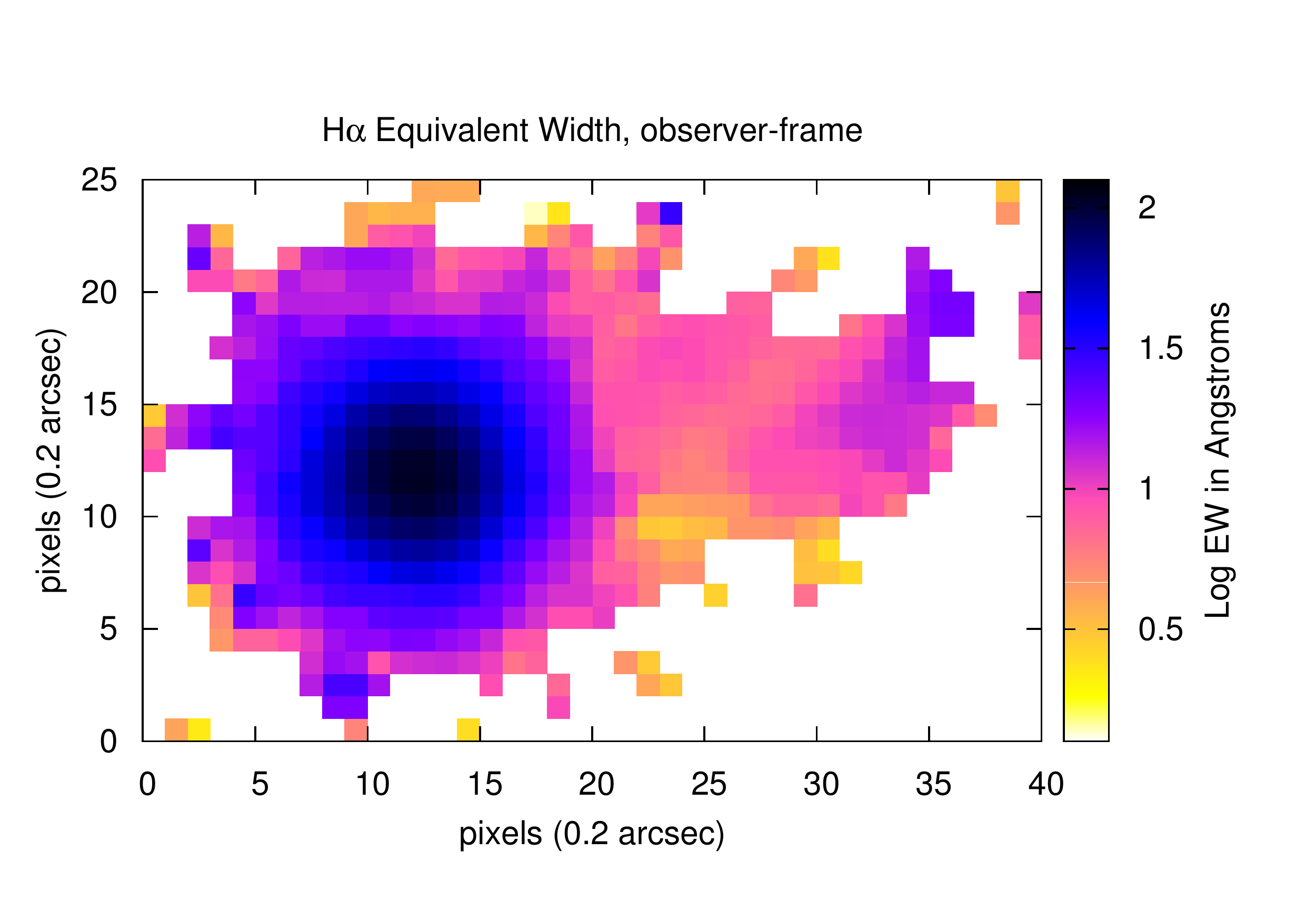}
\caption{$\rm H\alpha$ observer-frame equivalent width in Angstroms (from Gaussian fits), on log scale showing the gradient in the companion galaxy.}
\end{figure}

To map the kinematics,  Gaussian profiles are fit to the $\rm H\alpha$ and [NII]6584 lines of each individual pixel within this area, using {\small IRAF} `fitprof'. As it soon became apparent that different lines did not always show the same velocity gradients, we fit independent wavelength  to $\rm H\alpha$ and [NII]. 
 The output wavelengths were then all converted into line-of-sight velocity shifts  $\Delta(v)$ relative to the approximate AGN zeropoint set at $\rm \lambda(H\alpha)=7760.5\AA$. 
 
Firstly, Fig 13 shows the $\rm H\alpha$ equivalent width given by the Gaussian fit, showing the line is well detected over most of the `box'.  Equivalent width peaks centrally at $110\rm \AA$ in the primary; in the companion it is more uniform with the maximum not in the centre but in the NW arm, where it reaches $\sim20\rm \AA$ in a few pixels.

\begin{figure}

  \includegraphics[width=0.8\hsize,angle=-90]{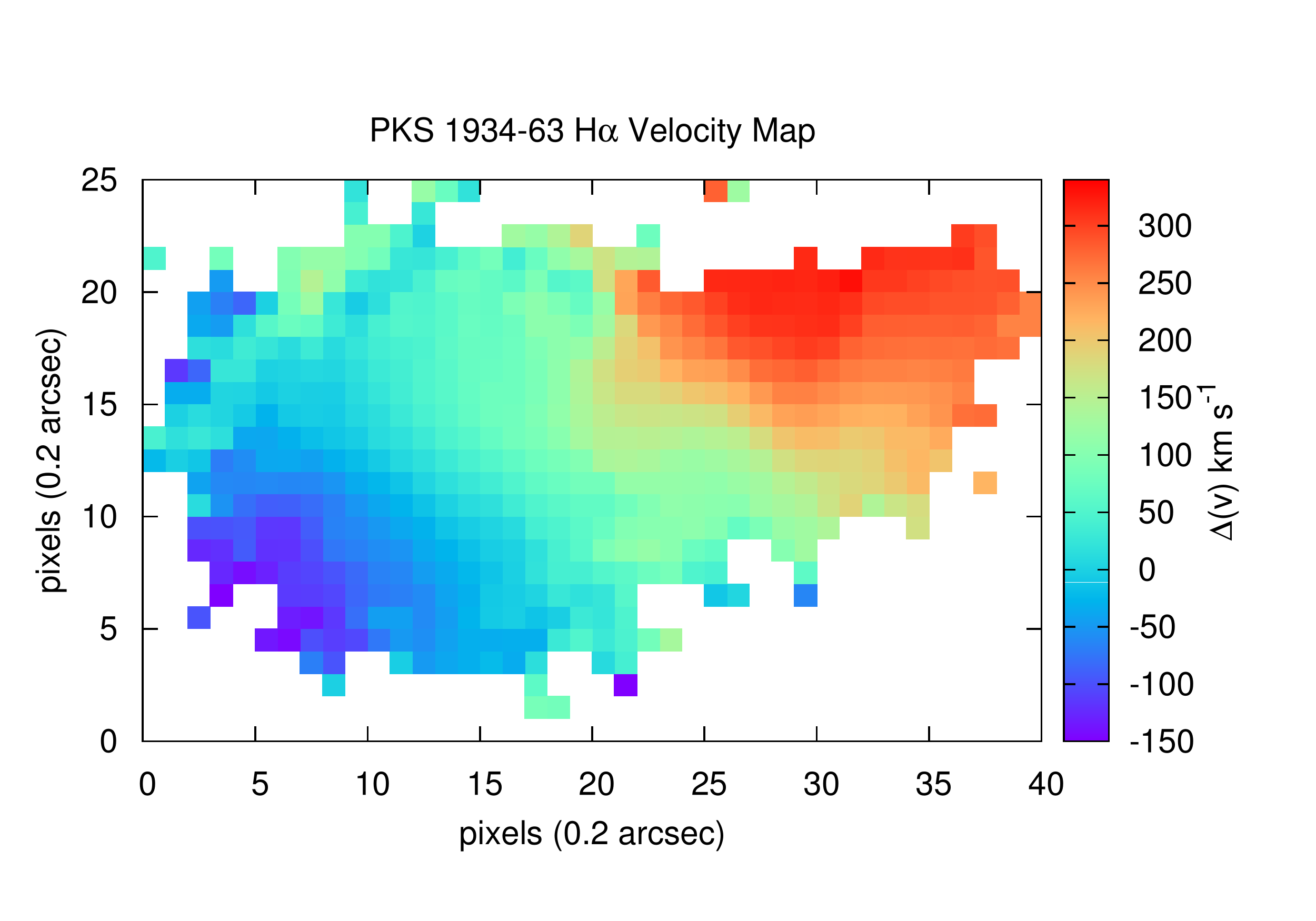}

 \includegraphics[width=0.8\hsize,angle=-90]{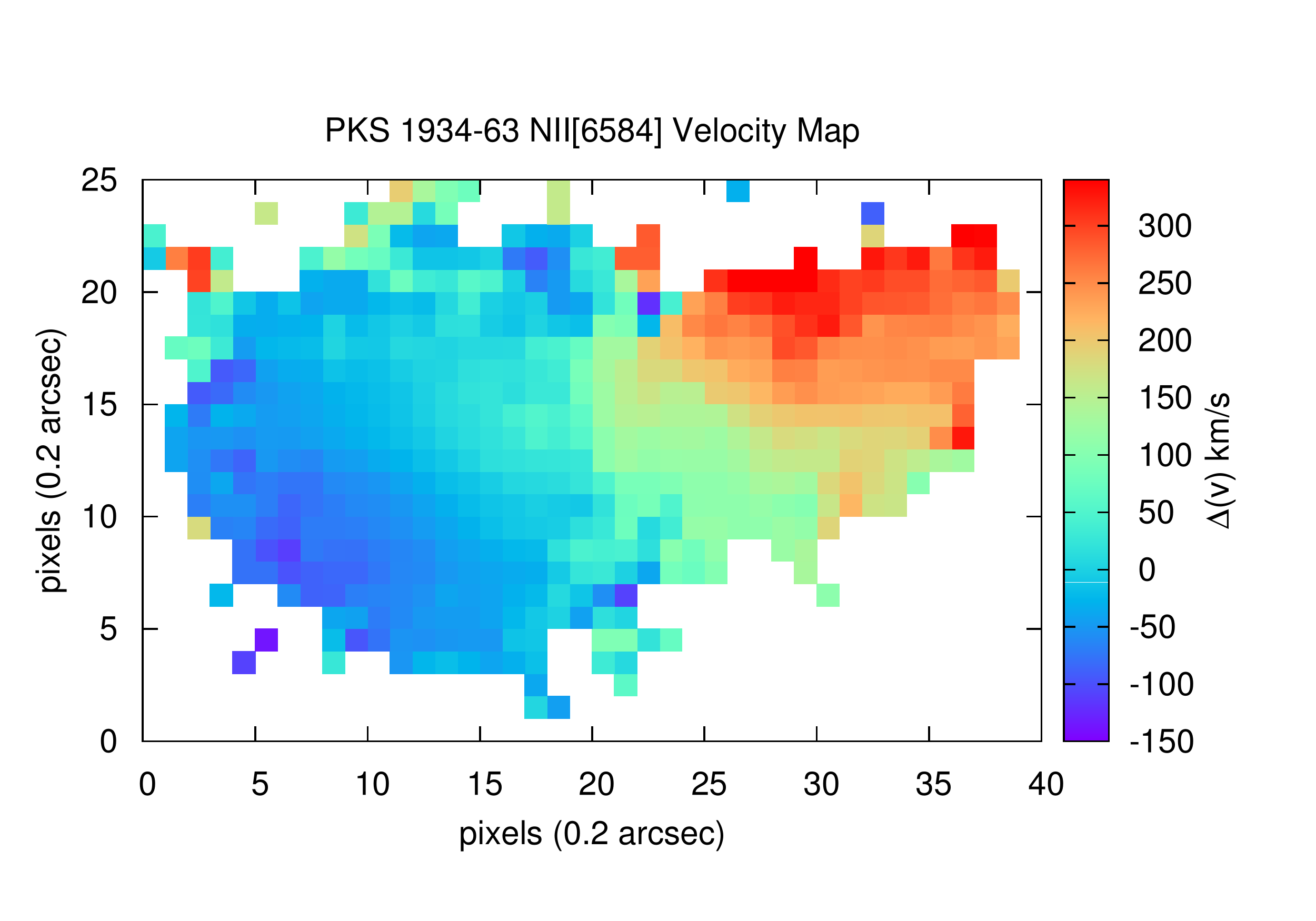}
\caption{Line of sight velocity, $\Delta(v)$ map of PKS 1934-63, both AGN and companion galaxy, obtained by fitting to the $\rm H\alpha$ (above) and [NII]6584 (below) emission lines  in the spectra of the individual pixels.}
\end{figure}

 Fig 14 shows $\Delta(v)$ maps for $\rm H\alpha$ and for [NII]6584. The two lines give a similar picture of a strong, continuous and monotonic velocity gradient across the system from SE to NW, composed of the two near-parallel gradients within the two galaxies (apparently from the rotation of each galaxy) and the  inter-nuclear vector which has $\Delta(v)$ in a similar direction (closer to E-W). This means the interaction can be described as prograde-prograde. However, in the primary galaxy,  $\rm H\alpha$ shows a stronger velocity gradient than [NII], as we investigate further.

 \begin{figure}
 \vskip -0.6cm
 \includegraphics[width=0.8\hsize,angle=-90]{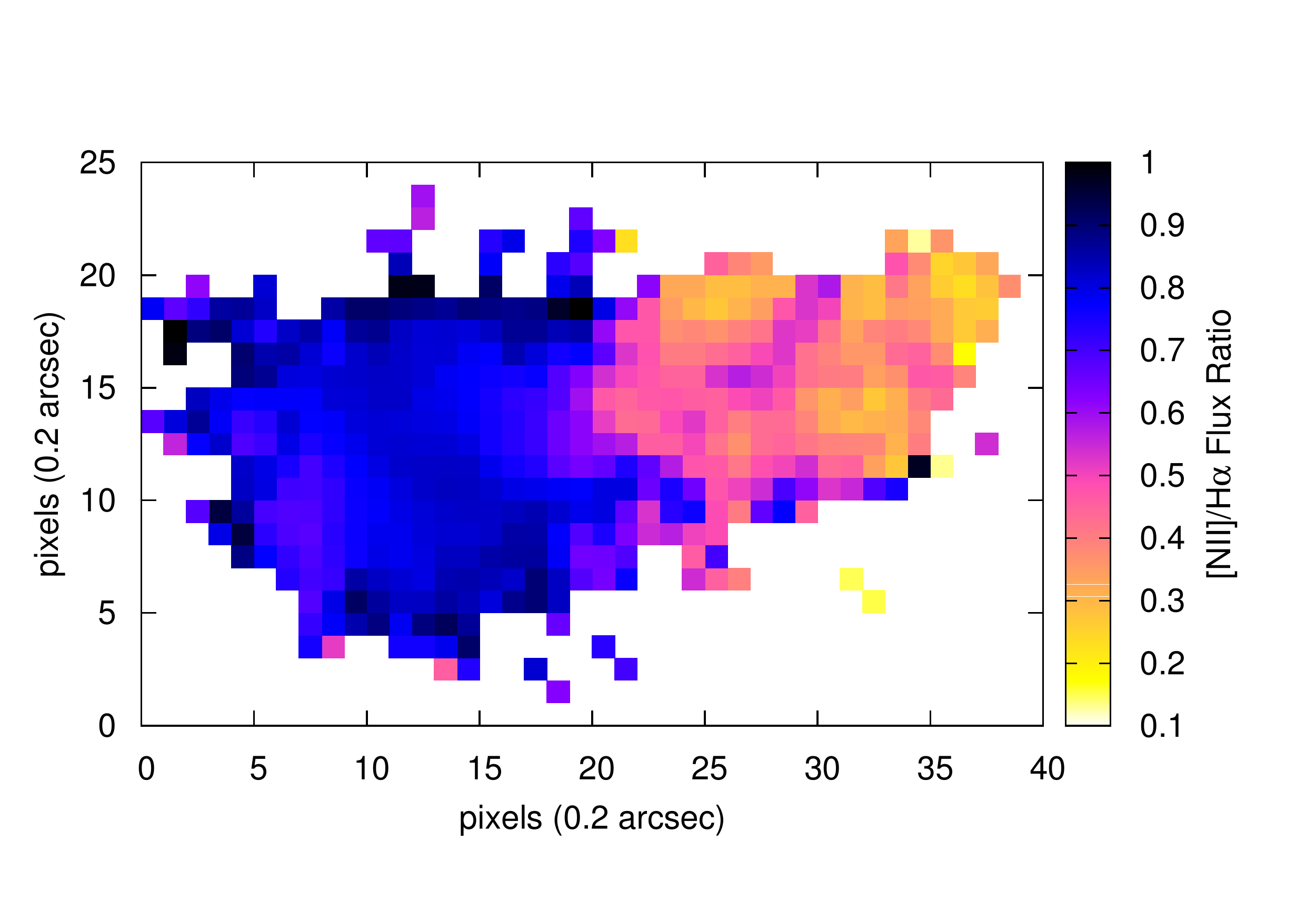}
\caption{Map of the flux ratio of [NII]6584 to $\rm H\alpha$, from the  fits to emission lines in the spectra of the individual pixels}
\end{figure}
  \begin{figure}

\includegraphics[width=0.8\hsize,angle=90]{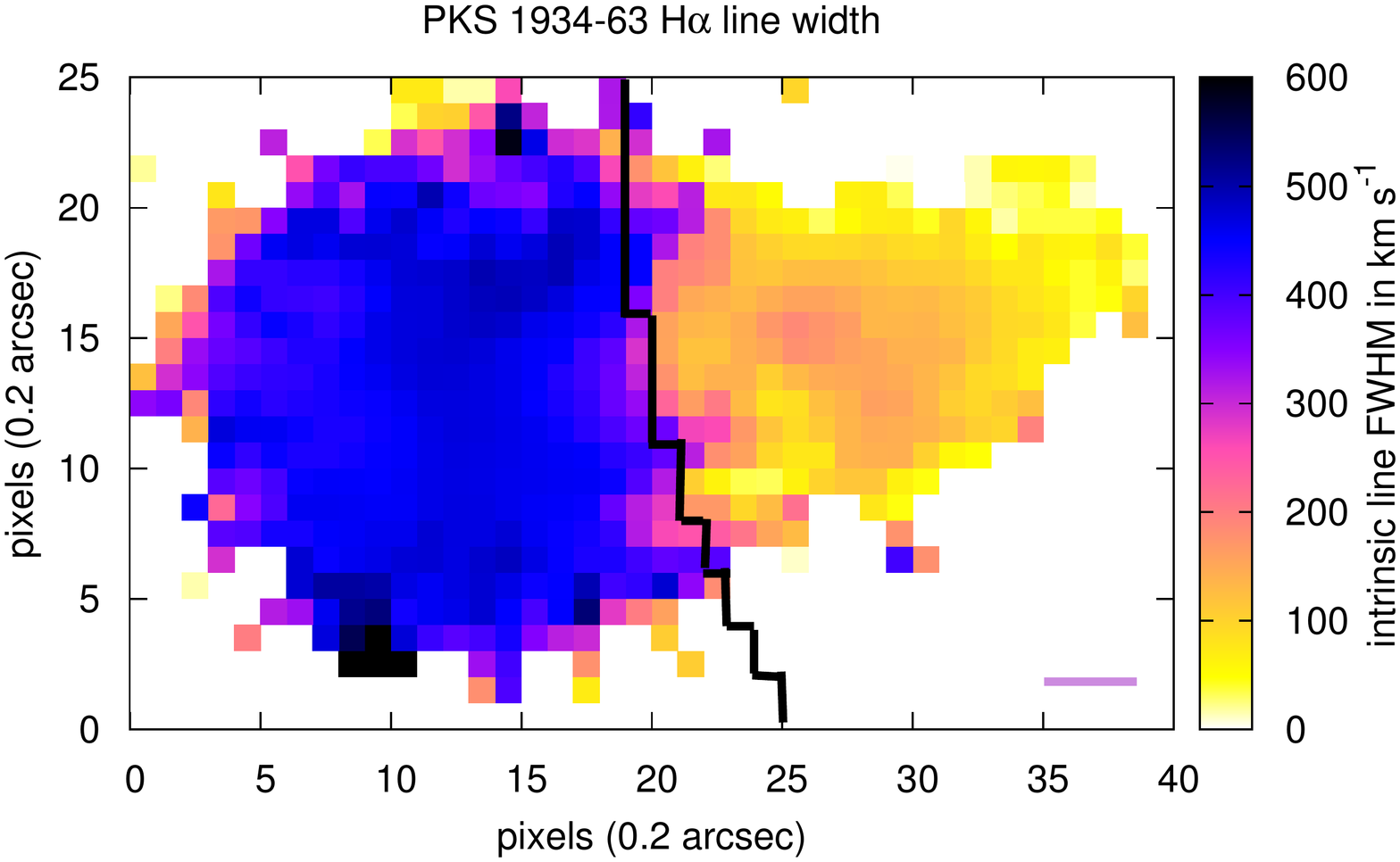}
\caption{ $\rm H\alpha$ line FWHM map, obtained from Gaussian fits to the emission line in the spectra of individual pixels. To convert from Angstroms to velocity, the instrumental FWHM of $2.6\rm \AA$ is subtracted in quadrature and then $\rm 1\AA$ corresponds to 38.6 km $\rm s^{-1}$. The black line is the divide used to separate the pixels of the primary and secondary galaxies for the kinematic analysis.The magenta bar lower-right shows the FWHM of  spatial resolution.}
\end{figure}

 \begin{figure}
 \includegraphics[width=0.7\hsize,angle=-90]{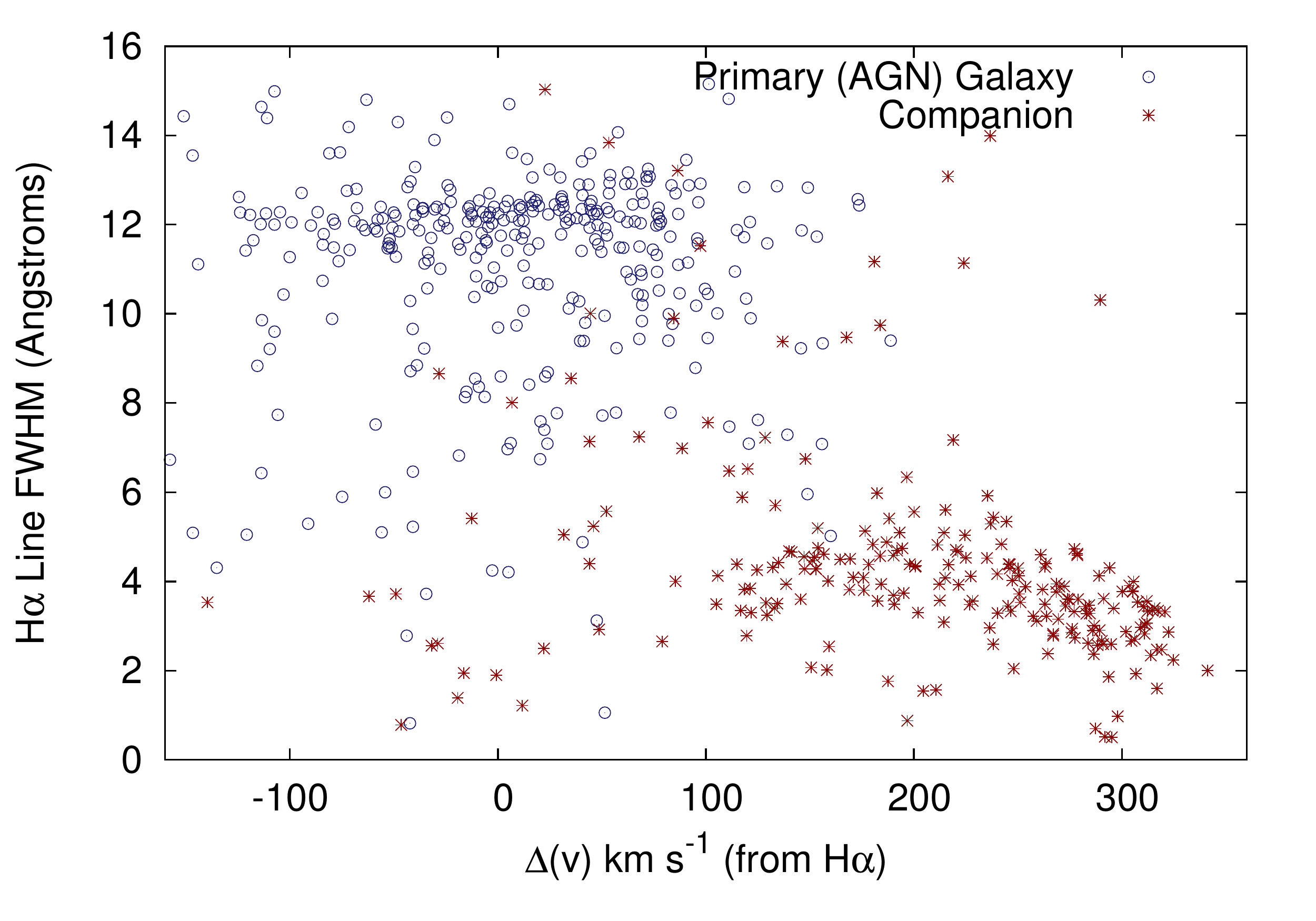}
\caption{The $\rm H\alpha$ line FWHM as measured in Angstroms (instrumental $\mathrm{FWHM}\simeq 2.6\rm \AA$ is not subtracted here) against velocity shift (also measured from the $\rm H\alpha$ line) showing the sharp difference between the two galaxies. }

\end{figure}

 Maps of the $\rm [NII]6584/H\alpha$  ratio (Fig 15) and the line width as a  velocity (Fig 16)  show a sharp difference between the two galaxies with higher values in the primary/AGN. Using these two maps we define an approximate divide between pixels belonging to the eastern or western galaxy, at ${r_1\over r_3}=1.2$ where $r_1$, $r_2$ are radial distances from the centres of the primary and secondary. 

Fig 17 shows single-pixel $\rm H\alpha$  FWHM  against  $\rm H\alpha$ $\Delta(v)$, divided by galaxy. The two galaxies overlap only slightly in $\Delta(v)$ and 
in line FWHM are widely separated,  most pixels in the primary having 10--$14\rm \AA$ compared to 3--$\rm 6\AA$ in the companion.

Line kinematic fits are performed in the same way for other emission lines, [OIII]5007, [OI]6300, $\rm [SII]_{6717}^{6731}$ and $\rm H\beta$ (Fig 18), but the strength of these is only sufficient to  obtain $\Delta(v)$ maps for the primary galaxy. All show similar, apparently rotational, velocity patterns to those in $\rm H\alpha$ and [NII], with a velocity gradient oriented in the same SE-NW direction.

\begin{figure}
 \includegraphics[width=0.75\hsize,angle=-90]{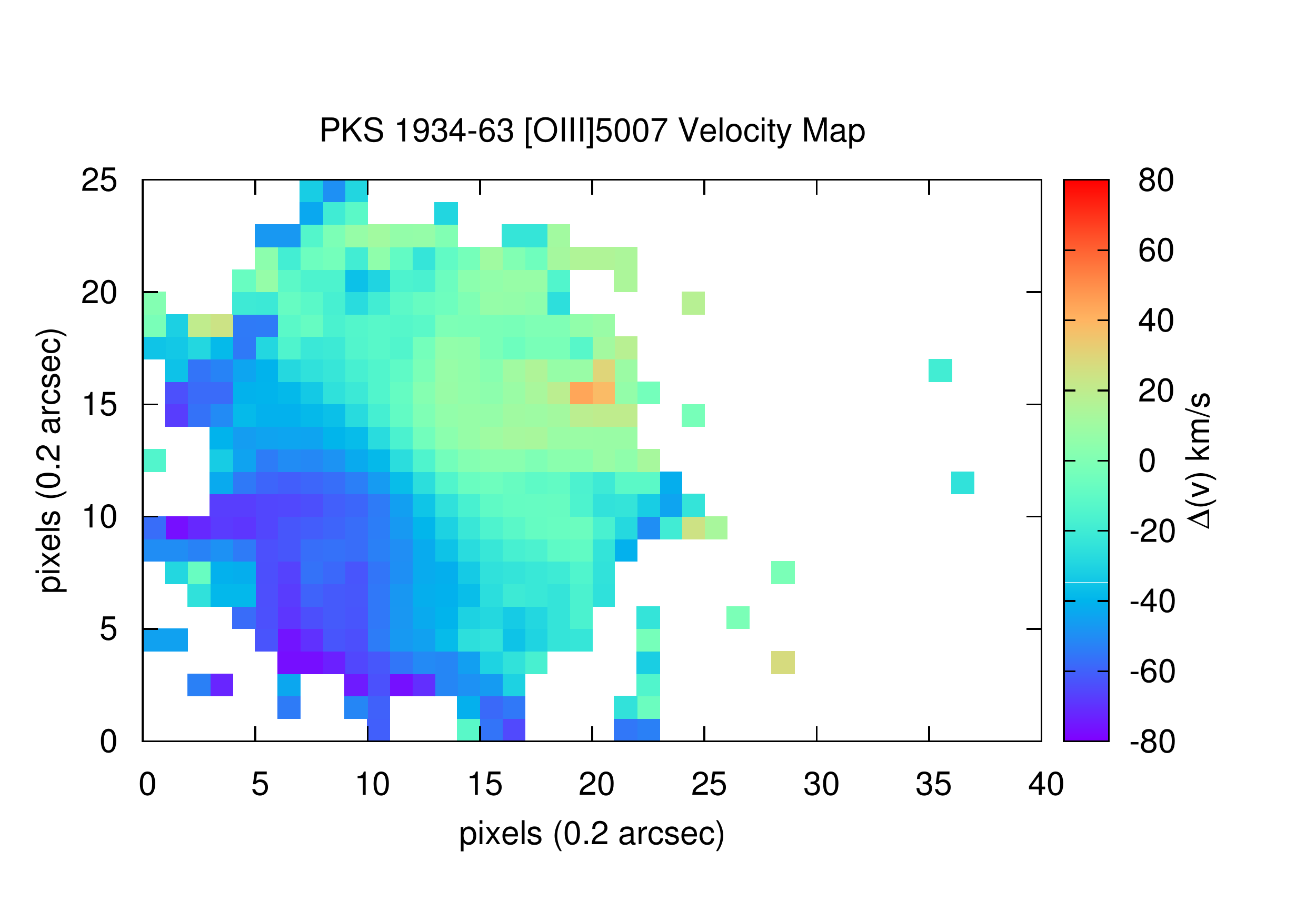}
 \vskip -1.0cm
\includegraphics[width=0.75\hsize,angle=-90]{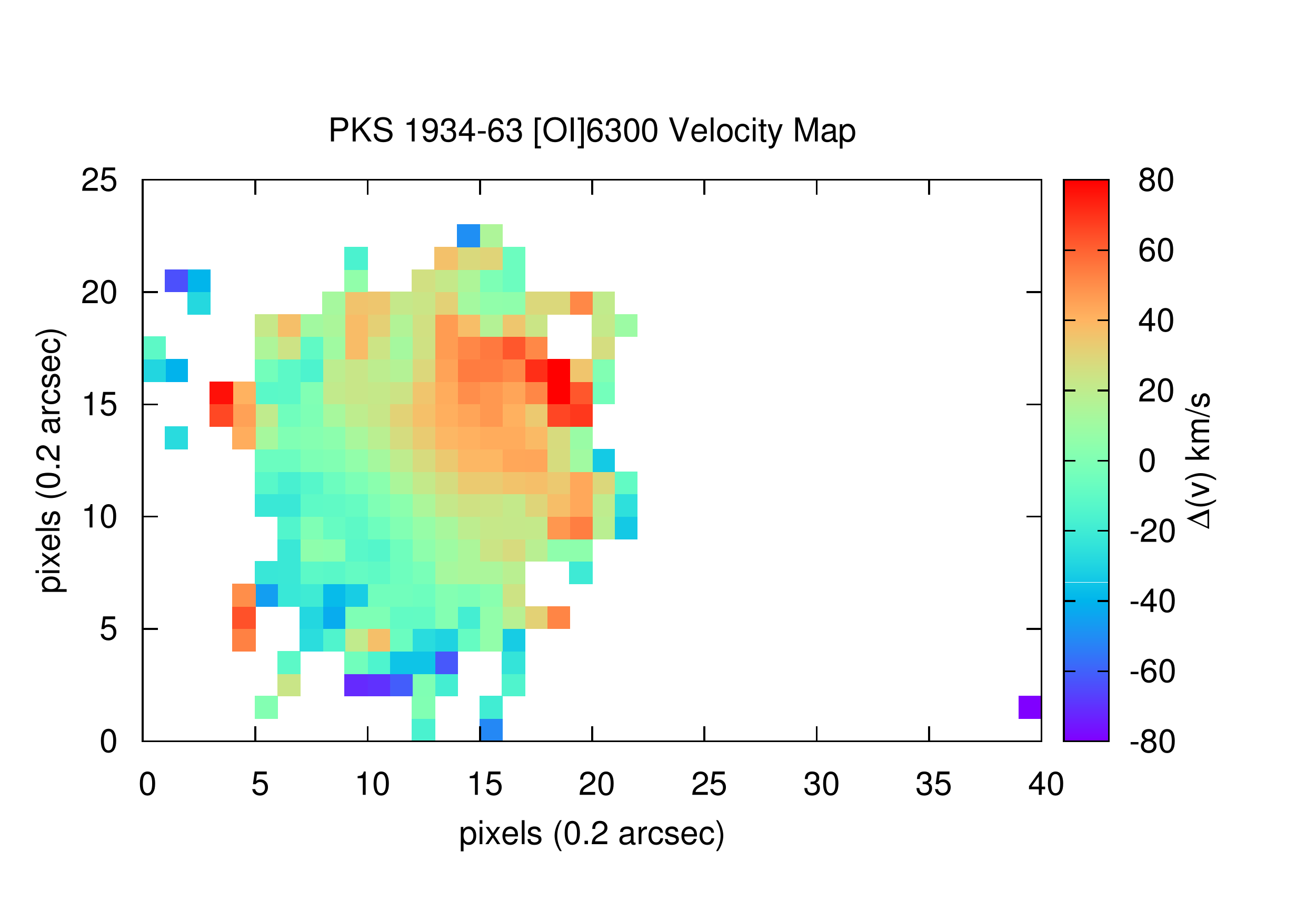}
\vskip -1.0cm
\includegraphics[width=0.75\hsize,angle=-90]{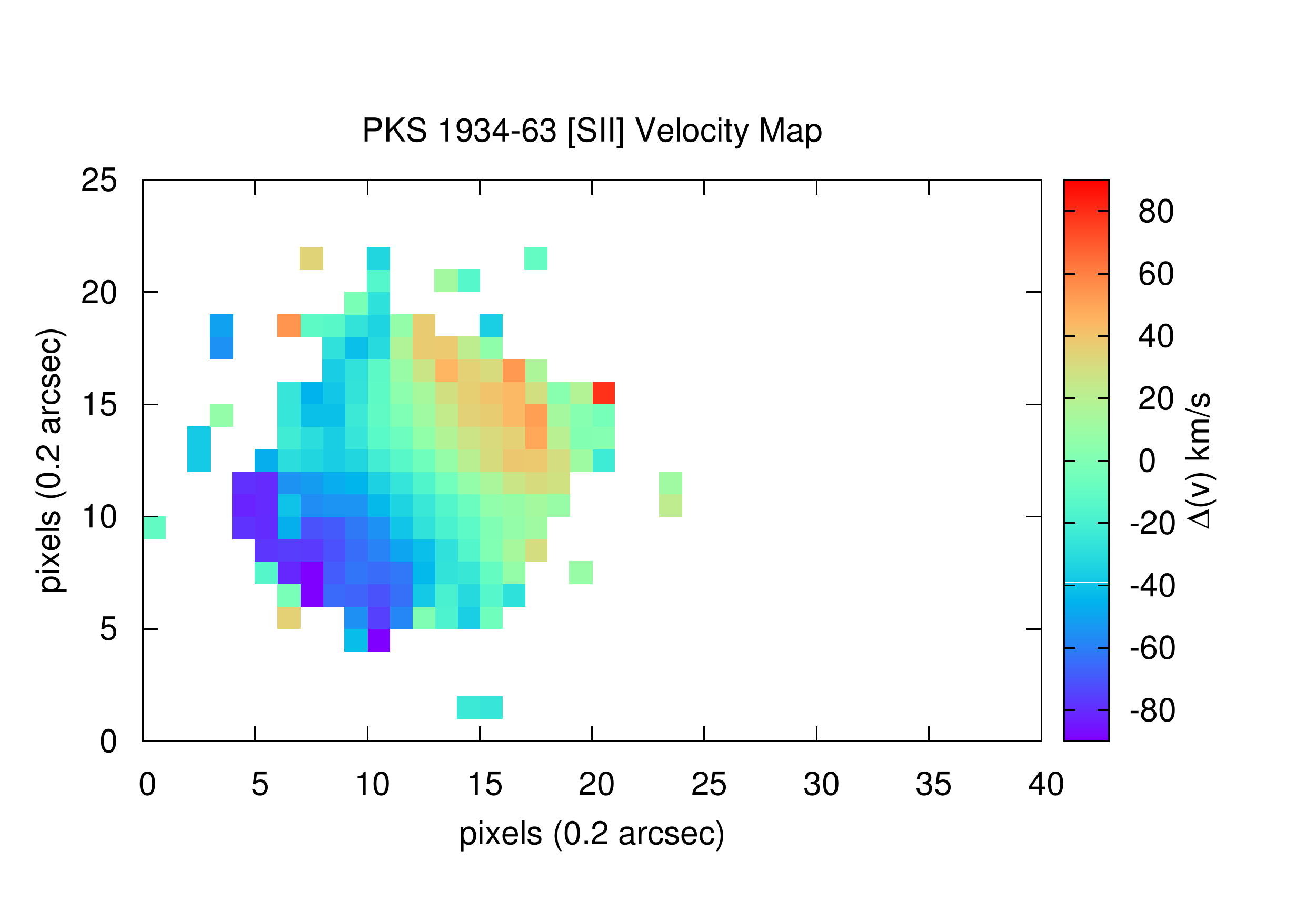}
\vskip -1.0cm
\includegraphics[width=0.75\hsize,angle=-90]{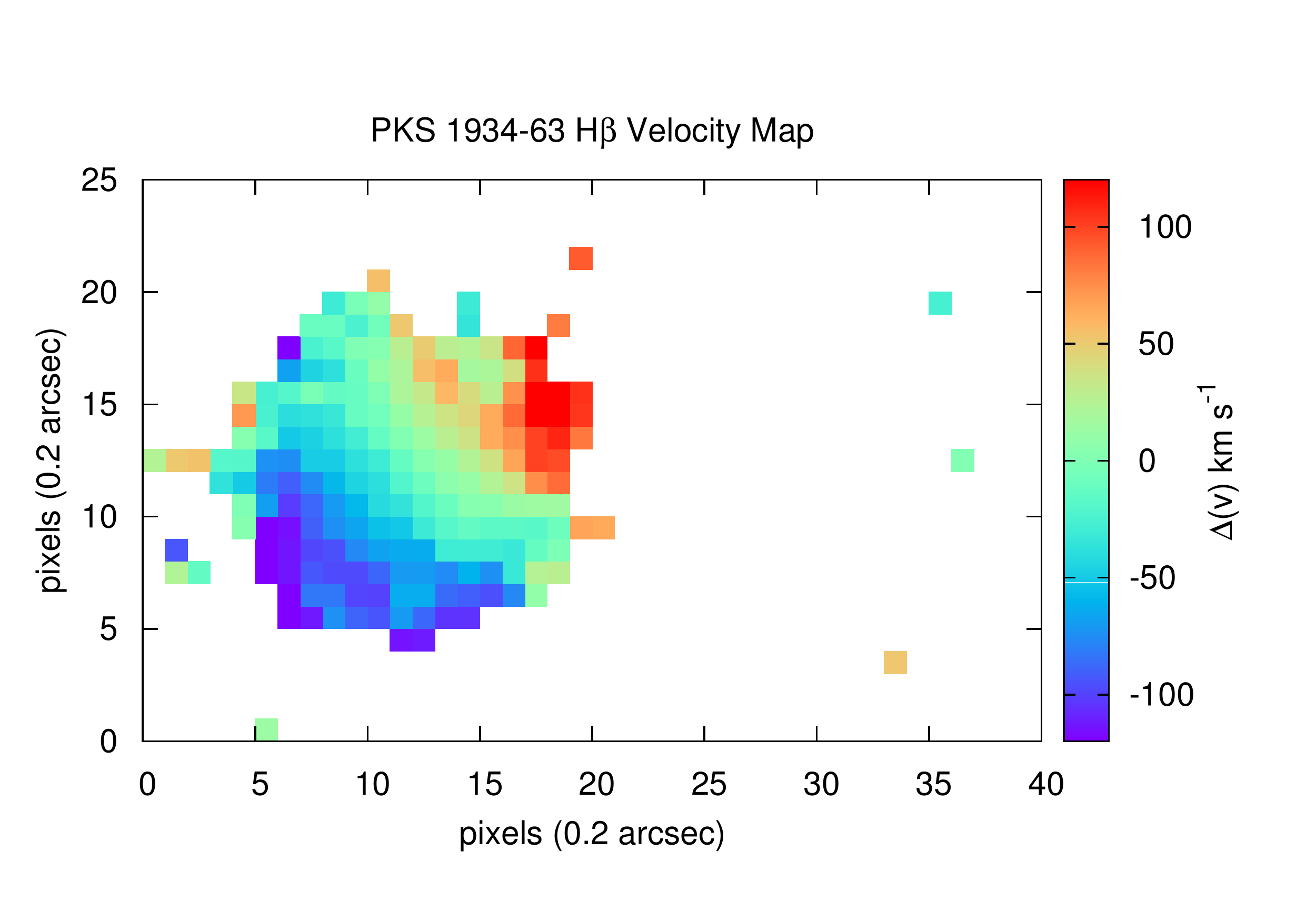}
\caption{Line of sight velocity, $\Delta(v)$ maps obtained by fitting to the [OIII]5007, [OI]6300, $\rm [SII]_{6717}^{6731}$ and $\rm H\beta$ emission lines in the spectra of the individual pixels. Zero-pointed to the $\rm H\alpha$ redshift.}
\end{figure}

The $\rm [OIII]/H\beta$ ratio could be mapped only for the primary (Fig  19), where it is high at $\sim 8$. The low $\rm [SII]6717/6731$ ratio of 0.8--1.0 indicates a high electron density $\rm 800<n_e< 1800~ cm^{-3}$ (again more typical of AGN than star-formation). Neither ratio shows obvious spatial variation. In the companion galaxy  $\rm [OIII]/H\beta$ is almost an order of magnitude lower, and $\rm [SII]6717/6731 \sim1.5$ implying  $\rm n_e<50~ cm^{-3}$.

\begin{figure}
\includegraphics[width=0.8\hsize,angle=-90]{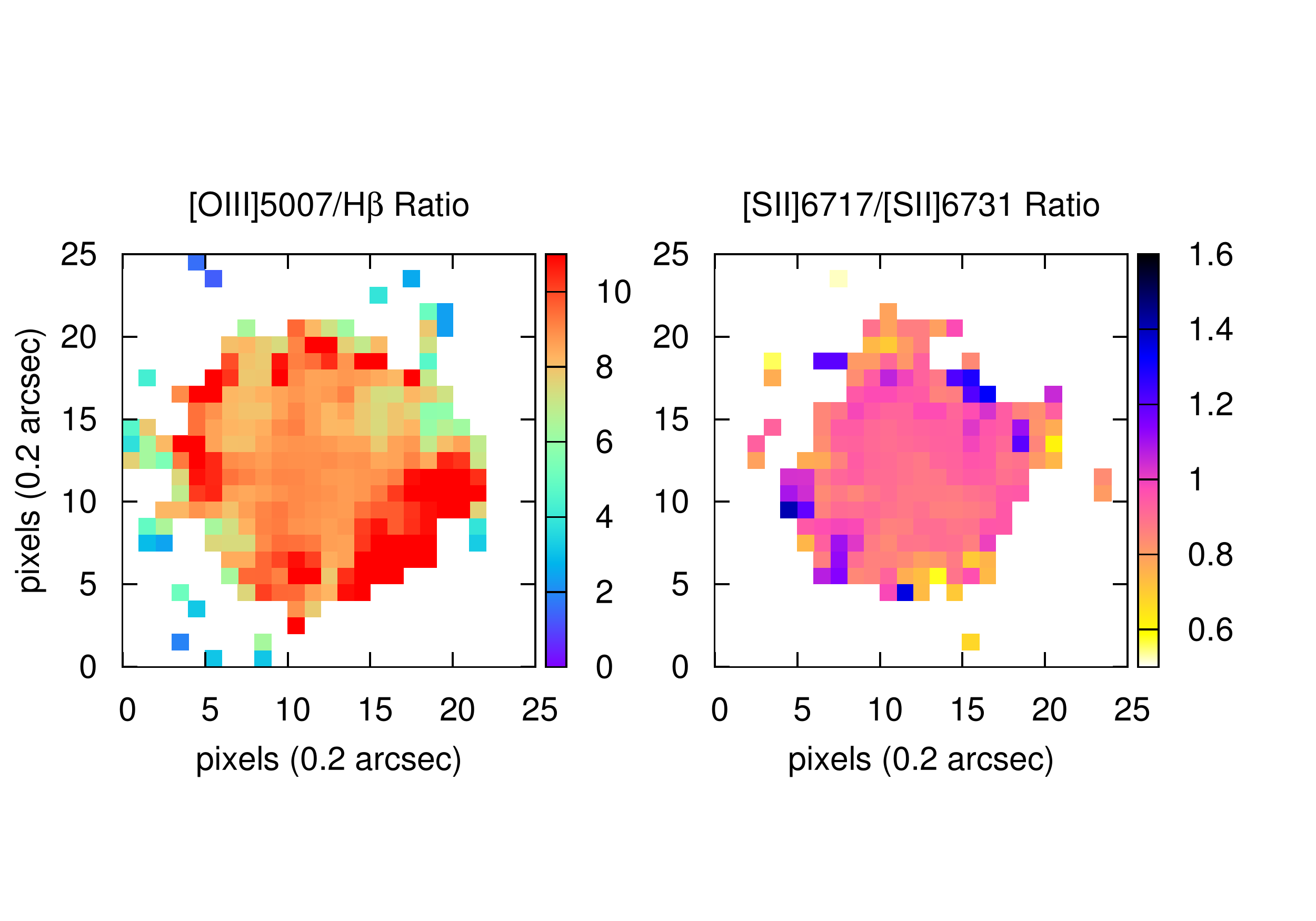}
 \caption{The $\rm [OIII]/H\beta$ and $\rm [SII]6717/6731$ ratios for the primary galaxy, both which have rather uniform values, consistent with AGN dominated emission.}
\end{figure}
\subsection{Rotation Curves from $\Delta(v)$}
To further study the kinematics, we divide the pixels between the two galaxies as described above and extract low-resolution velocity curves from the $\Delta(v)$ maps, for the $\rm H\alpha$ and [NII] fits and for other lines. 

For each galaxy a kinematic axis is defined with position angle $\phi$ through the centre, and  the mean $\Delta(v)$ is determined for pixels within opposite sectors of PA $\phi \pm30^{\circ}$ and $(180^{\circ} +\phi) \pm 30^{\circ}$ as a function of the radial distance along this axis (binned in 0.5 arcsec). In order to estimate the $\phi$ for each galaxy, this calculation is performed for all angles and the difference between the two sides estimated as $\Sigma_{r=0}^{r=2.0"} [v_{rot}(\phi,r)-v_{rot}(\phi,-r)]$. A cosine function was fitted to to the $\phi$ variation to give the phase that maximises this difference; we find $\phi=-45^{\circ}$ for the primary, $\phi=-34^{\circ}$ for the companion. The companion galaxy has a similar long axis, PA $-41^{\circ}$ in the $R$-band, as expected for a disk, although its $\rm H\alpha$ emission is elongated on PA $-75^{\circ}$; the primary is almost round.

Fig 20 shows the  $\rm H\alpha$ and [NII] velocity/rotation curves for the two galaxies. For the primary galaxy this curve is computed in the same way, using the same $\phi$, for [OIII]5007, [OI]6300, [SII] and 
$\rm H\beta$.  The radial component of rotation velocity is estimated as  half the maximum difference between the two sides (amplitude) in $\Delta(v,r)$ at $r\leq 2.0$ arcsec (Table 3).
Holt, Tadhunter \& Morganti (2008) had previously observed the primary galaxy  with the spectrograph WHT-LRIS, slit on PA $-20^{\circ}$, and found a smooth rotation curve  in [OIII]5007 with an amplitude $\simeq 50$ km $\rm s^{-1}$.  Our results are consistent with this but also show marked differences between the emission lines. 

In the primary (AGN) galaxy, $\rm H\alpha$ and $\rm H\beta$ show similar and strong velocity gradients spanning $\sim 200$ km $\rm s^{-1}$. The forbidden line  velocity curves all have  amplitudes about a factor 2 smaller, and may have some sign of a turnover at large radii ($>1.5$ arcsec), where the $\rm H\alpha$ velocity curve only flattens.  The lines [NII] and [OIII] are mildly  ($\simeq 20$ km $\rm s^{-1}$) offset blueward relative to $\rm H\alpha$, but
[OI]6300 is similarly offset redward, which accounts for its redder  colours on Fig 18.  The relative faintness of $\rm H\beta$ only allowed its $\Delta(v)$ to be mapped over the central primary but this is sufficient to measure a large velocity gradient, similar to $\rm H\alpha$ and steeper than the adjacent [OIII]5007.
. These differences can be clearly seen on Fig 21 where the velocities from  [NII] and other lines are plotted against $\rm H\alpha$ $\Delta(v)$ for each individual spatial pixel. 
\begin{figure}
\includegraphics[width=0.75\hsize,angle=-90]{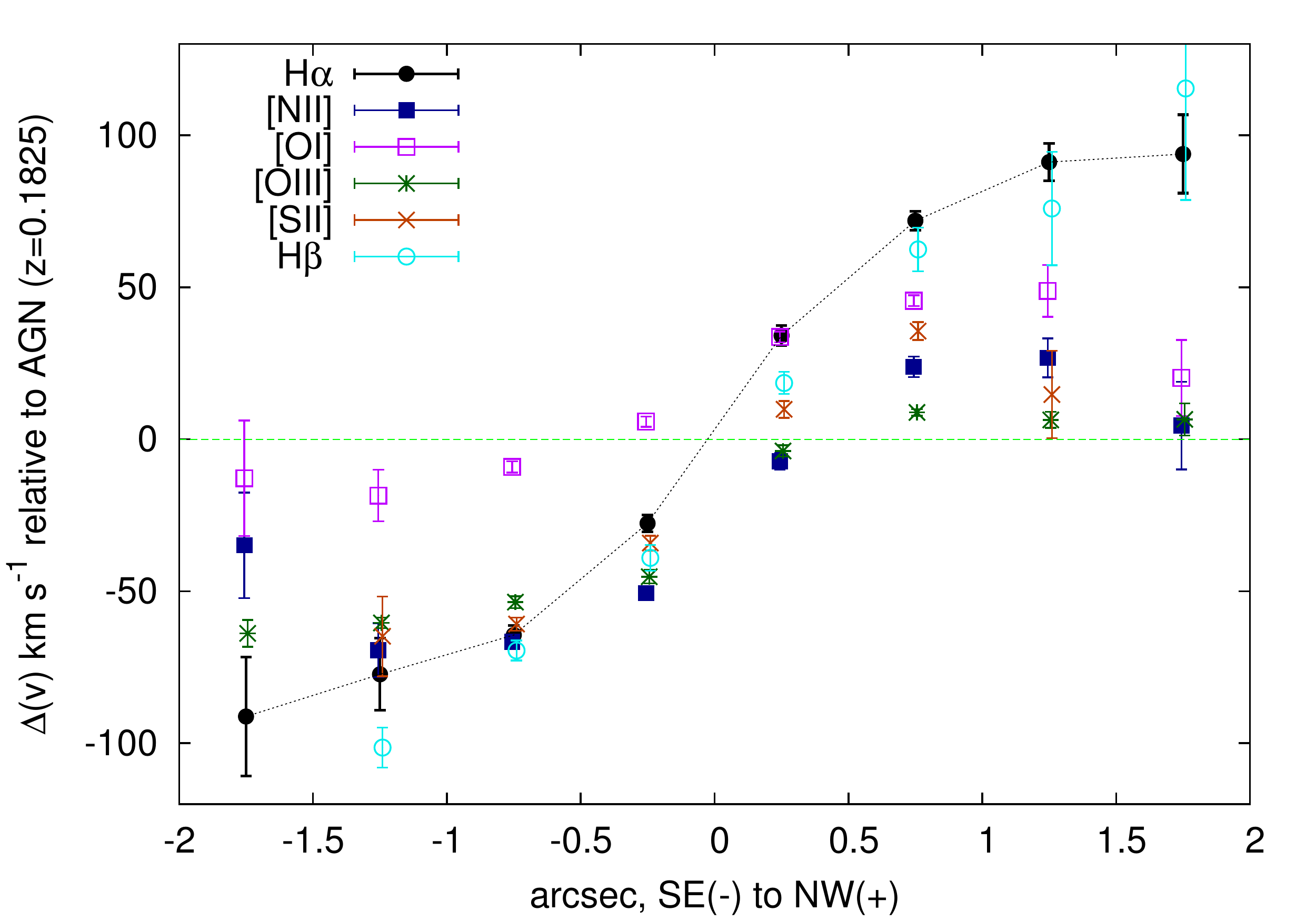}
\includegraphics[width=0.75\hsize,angle=-90]{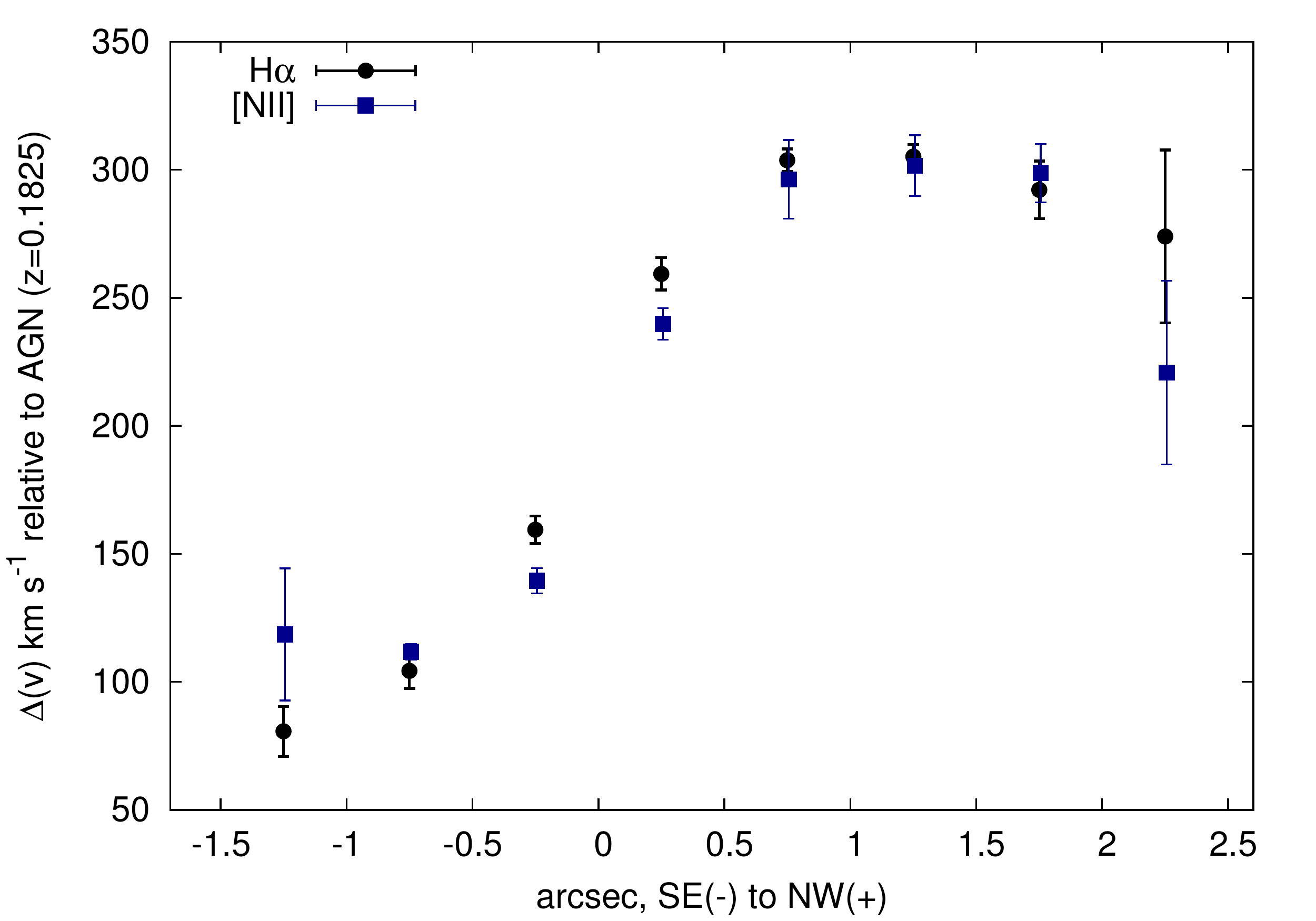}
 \caption{Rotation curves of the the primary (above) and companion (below) galaxies: the line of sight $\Delta(v)$, from $\rm H\alpha$ and other emission lines, as a function of distance $r$ measured along the chosen kinematic axis, which has the fitted PAs of $\phi=-45^{\circ}$ (primary) and $-34^{\circ}$. (secondary).}
\end{figure}

\begin{table}
\caption{Amplitude of the velocity curve estimated for different emission lines, measured from the velocity maps on axes PA $-45^{\circ}$ and $-34^{\circ}$ for the primary and secondary. }
\begin{tabular}{lcc}
\hline
Line & Primary (AGN) & Companion  \\
       & $\Delta(v)$ amplitude km $\rm s^{-1}$\ & $\Delta(v)$ amplitude km $\rm s^{-1}$\\
 \hline
$\rm H\alpha$  & $92.5\pm 11.7$ & $112.2\pm 5.4$  \\

$\rm [NII]6584$  & $48.1\pm 5.5$ & $96.7\pm  10.3$\\

$\rm [OIII]5007$ & $36.4\pm 2.3$ & - \\
$\rm [OI]6300$ & $33.7\pm 6.0$ & - \\
$\rm [SII]_{6717}^{6731}$ & $48.2\pm 1.8$ & - \\
$\rm H\beta$ & $108.4\pm 18.6$ &             -                 \\
\hline
\end{tabular}

\end{table}
\begin{figure}
\includegraphics[width=0.62\hsize,angle=-90]{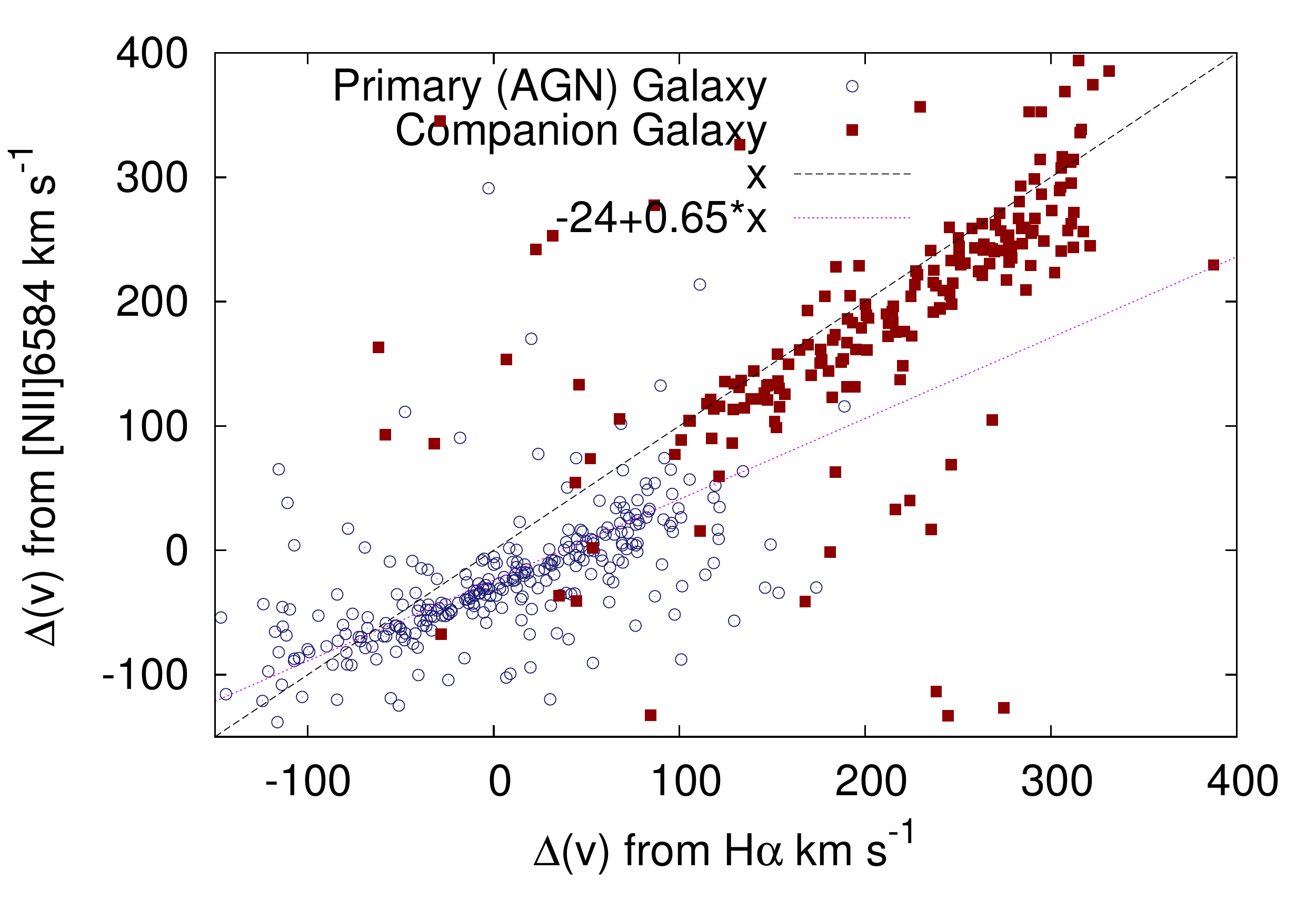}
\includegraphics[width=0.62\hsize,angle=-90]{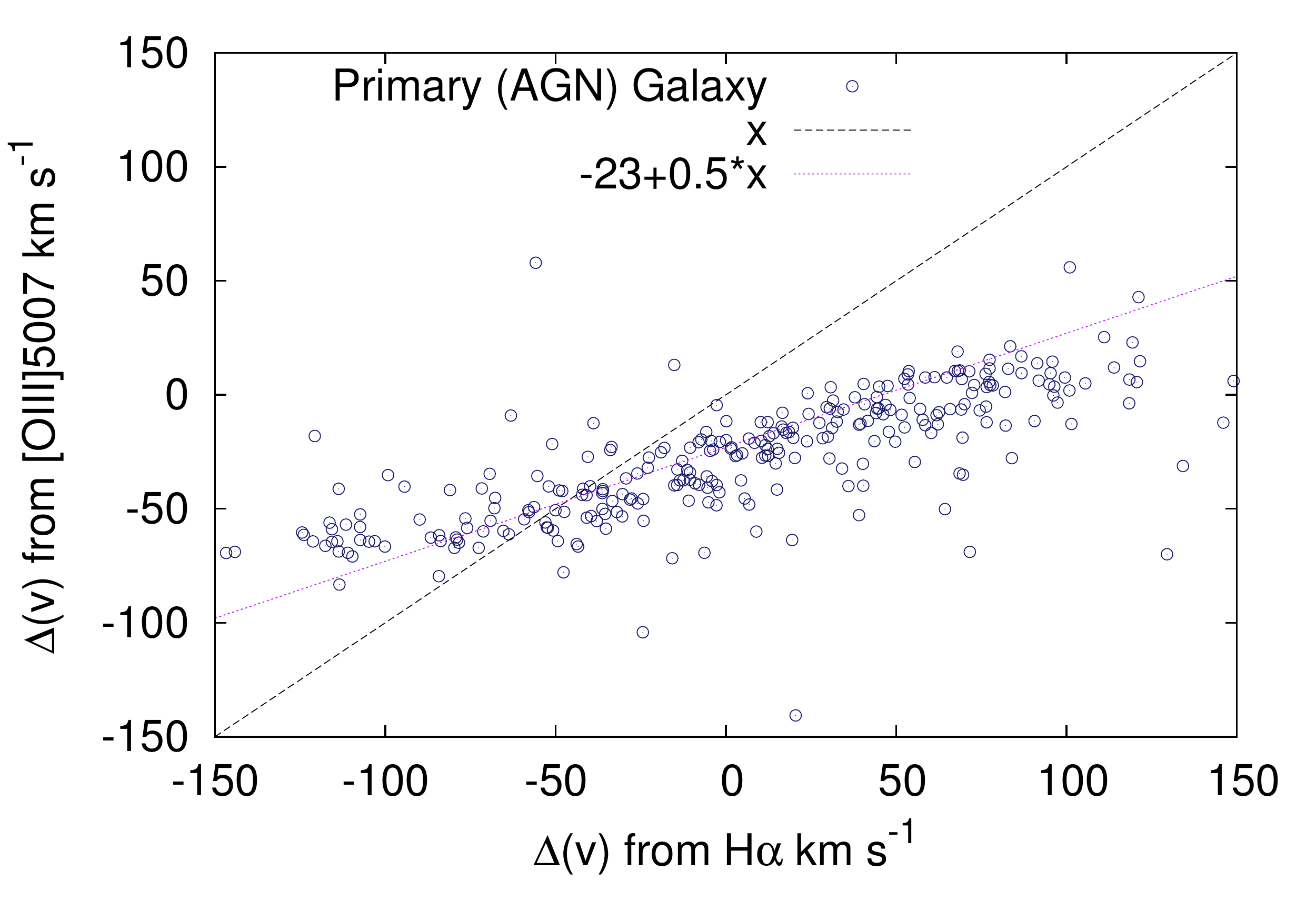}
\includegraphics[width=0.62\hsize,angle=-90]{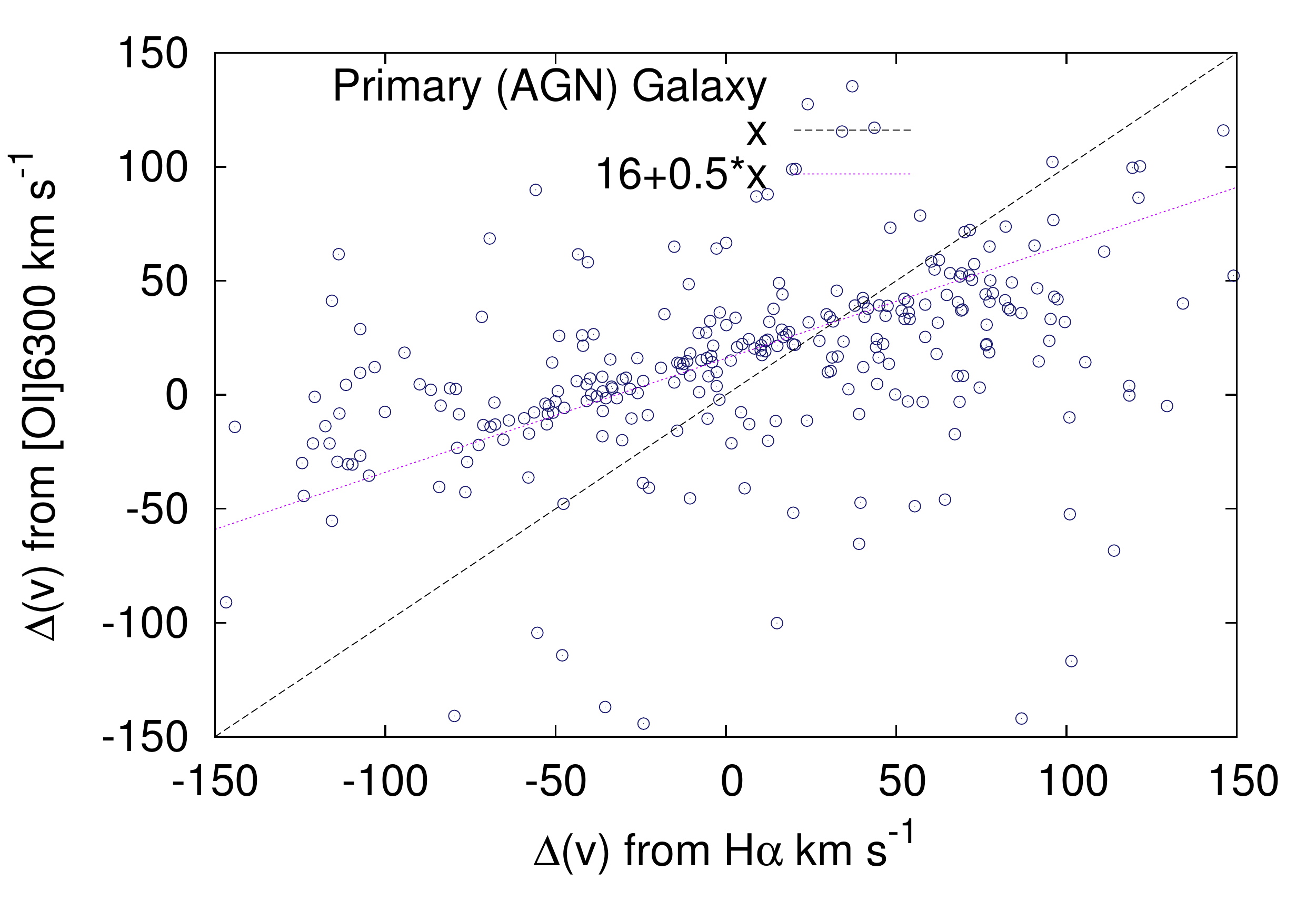}
\includegraphics[width=0.62\hsize,angle=-90]{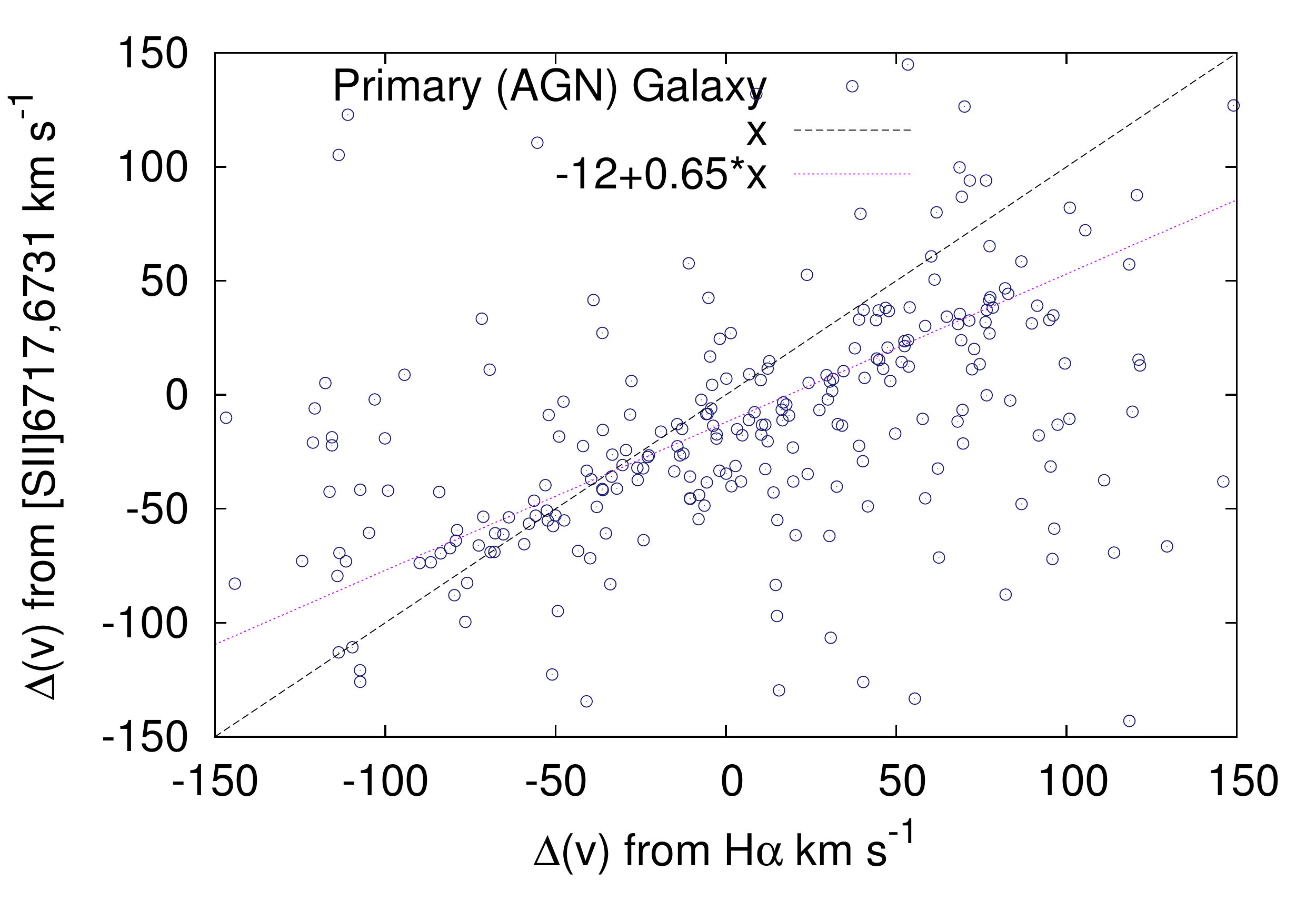}
 \caption{Comparison of single-pixel $\rm H\alpha$, [NII]6584 and [OIII]5007 velocity $\Delta(v)$. For pixels in the companion galaxy the $\rm H\alpha$ and [NII] velocities are very similar and follow the slope of unity (plotted as $y=x$). For the AGN the [NII]  line  shows a reduced velocity gradient (by  $\sim 35\%$) with respect to the $\rm H\alpha$ $\Delta(v)$, as do the [OIII]  and [OI] lines (by $\sim 50\%$), approximately as the plotted functions $y=ax+b$}
\end{figure}

For the companion galaxy the $\rm H\alpha$ and [NII] velocity gradients are similar, and
the interpretation seems clear in that both $\rm H\alpha$ and [NII]  trace the rotation of the HII-regions and stars. The shape and amplitude of the rotation curve, low velocity dispersion ($\sigma<60$ km $\rm s^{-1}$ from the FWHM of single-pixel spectra), and the morphology (S\'ersic $n=1$--2 and arm-like features in the residuals) are all consistent with an inclined (but not edge-on) spiral.

In the case of the primary, the rotation signal could originate from the motion of circumnuclear regions photoionized by the AGN.  From the aperture spectrum of the primary we measured above (section 3.1) a $\rm H\alpha$ FWHM of 476 km $\rm s^{-1}$ or a velocity dispersion $\sigma\simeq 203$ km $\rm s^{-1}$.  If the $\Delta(v)$ amplitude of $\rm H\alpha$  traces, or is a lower limit on, the rotation of the galaxy, then $v_{rot}/\sigma\geq 0.45$ which is already high for a spheroidal and makes this a fast-rotator elliptical (Emsellem et al. 2011).

Shih et al. (2013) studied [OIII]5007 velocity gradients in several CSS radio galaxies (GMOS-IFU observations) and interpreted these as probable outflows (rather than rotation), on the basis that most were aligned with the radio axes, although origin in the ionized interstellar medium was also considered. However, they resolved the [OIII] lines into broad and narrow components with different kinematics, which might allow for both rotation and outflows (as we investigate in 5.3 below). In PKS 1934-63, the velocity gradients in $\rm H\alpha$ and other lines are $\simeq 45^{\circ}$ misaligned with the radio axis
(disfavouring an outflow interpretation) and the velocity maps look like those of a rotating galaxy (e.g.  the `class A'  fast-rotator elliptical models of Naab et al. 2014).

To estimate the rotation required to produce the observed $\Delta(v)$ curves, we compute a simplified `toy model', where the emission is represented as two point sources separated by $\Delta x$ spatially  (on the kinematic axis) and $\Delta v$ in (line-of-sight) velocity space. These could represent the two sides of a structure rotating at $\geq 0.5\Delta v$. Each source is spread spatially by the stellar-fitted Moffat profile and we compute the relative contributions of the redshifted and blueshifted side as a function of distance along the slit axis. Then for each position we model the emission line profile by adding redshifted and blueshifted line profiles in the appropriate ratios (the intrinsic emission line profile is represented using the [OIII] line from the central pixel). Then for each of the model profiles we measure the velocity shift $\Delta(v)$ by fitting a Gaussian, as with the real spectra. Fig 22 shows the modelled velocity curves for 9 combinations of $\Delta v=100, 200, 300$ km $\rm s^{-1}$ and $\Delta x=0.2, 0.3, 0.4$ arcsec.  Although {\it Galfit} fitted the extended $\rm H\alpha$ emission with a larger $r_{hl}$ of 2.38 pixels or 0.476 arcsec, its effective radius is less  because  5/6 of the emission was in the central point-source.
 
  This simplified model, with $\Delta x<$  seeing FWHM, can produce velocity gradients/curves similar to the observed (Fig 20). For $\rm H\alpha$ and $\rm H\beta$, the models with $\Delta v=300$ km $\rm s^{-1}$ and $\Delta x=0.3$--0.4 arcsec are closest to our observation. The velocity curves of the forbidden lines could be reproduced by smaller velocities and/or separations, such as  $\Delta v=200$ km $\rm s^{-1}$ and $\Delta x=0.2$--0.3 arcsec. Note that, at least for $\Delta x<0.4$ arcsec,  the model depends almost as strongly on $\Delta x$ as on $\Delta v$.
The compact dimensions of the line emission mean that the observed velocity shifts  underestimate the true  rotation velocities within the central galaxy; we estimate by a factor $\sim 1.5$ for $\rm H\alpha$ sources, which may orbit the AGN at $\sim 150$ km $\rm s^{-1}$.

 \begin{figure}
\includegraphics[width=0.75\hsize,angle=-90]{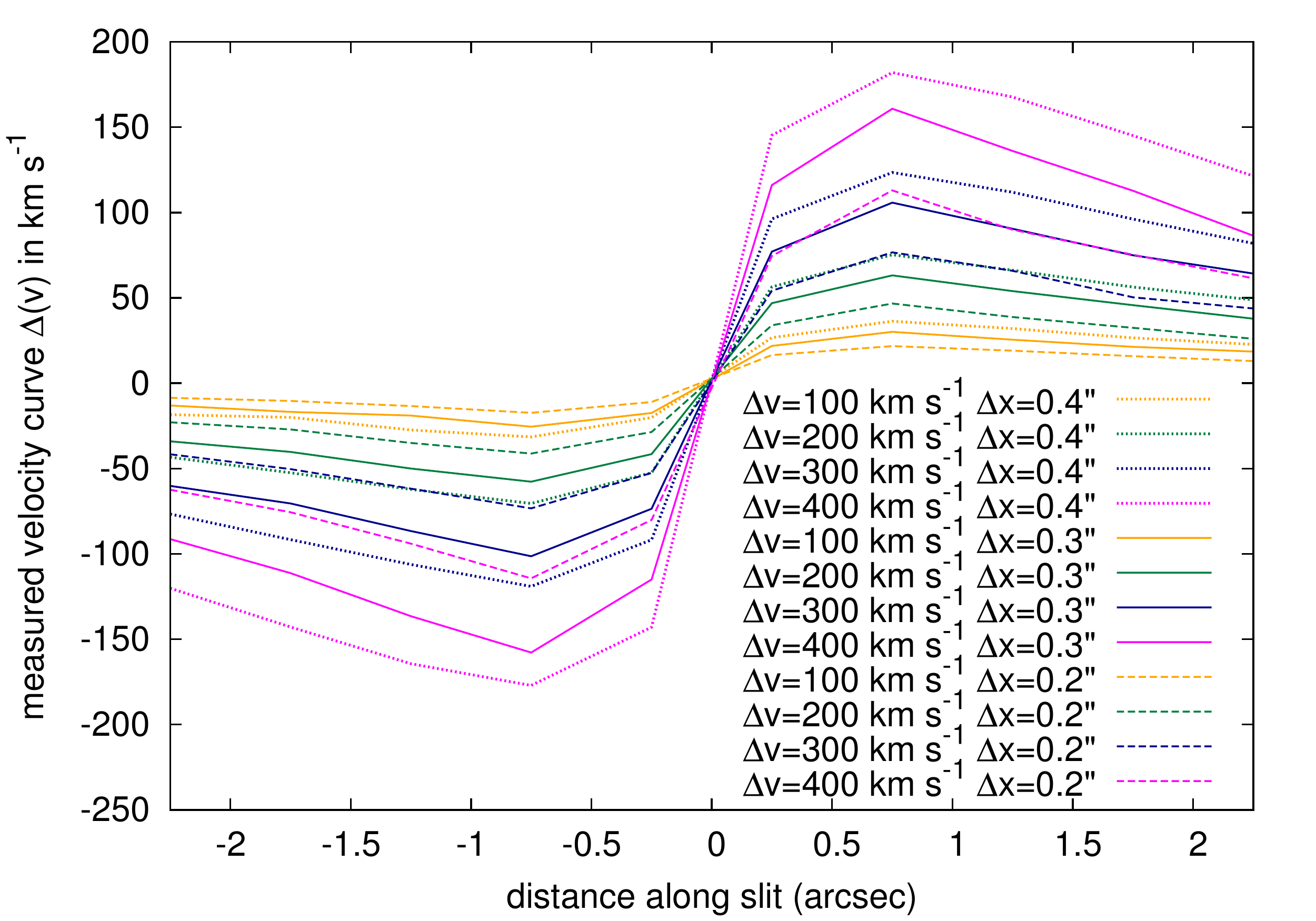}
\caption{Velocity curves along the kinematic axis of the primary galaxy, predicted by our `toy model' of two point-sources separated on the sky (by $\Delta x$) and in velocity space (by $\Delta v$). The models are plotted for combinations of $\Delta x= 0.2$, 0.4 and 0.4 arcsec, and $\Delta v=100$, 200, 300 and 400 km$s^{-1}$; some resemble the observed velocity curve across the primary (Fig 20, above).}
 \end{figure}

\subsection{The Broad Components of [OIII] and [OI]: Different Kinematics}
The above kinematic analysis is based on single Gaussian fits to the emission lines, but the stronger AGN  lines have `winged' profiles which could be more closely fitted as  the sum of narrow and broad components (Gaussians with different FWHM). These might have a different physical origin and kinematic map, e.g  in  WHT-LRIS spectroscopy,  Holt et al. (2008) resolved the [OIII]5007 lines of  compact radio galaxies, including PKS 1934-63 and found broad components blueshifted (relative to the narrow component) by (in most cases) hundreds of km $\rm s^{-1}$, which were interpreted as a signature of fast outflows.

We can perform the two-components analysis in 2D, firstly with [OIII]5007 (even brighter than $\rm H\alpha$ and not blended with other lines).
Again using `fitprof', the [OIII]5007 line is fitted pixel-by-pixel with two Gaussians, with the same initial wavelength and initial FWHM of $6\rm\AA$ and $25\rm\AA$, with both the $\lambda$ and width allowed to very freely. This analysis requires high signal-to-noise and is only successful within $r\simeq 4$--5 pixels of the AGN. If we average the fit results over the  $r\leq 4$ pixel region, the narrow and broad components have mean FWHM of $6.46\pm 0.10\rm\AA$ and $27.22\pm 0.45\rm\AA$, which (subtracting the instrumental width) correspond to $299\pm2$ and $1372\pm 6$ km $\rm s^{-1}$. The ratio of narrow:broad component fluxes is 2.9:1 and  the two have indistinguishable spatial profiles, but they differ kinematically with the mean $\Delta(v)$ (relative to the $\rm H\alpha$ zero-point) of the narrow and broad components $-23.6\pm 3.3$ and $-93.2\pm 7.7$ km $\rm s^{-1}$ respectively. We find the  broad [OIII] component to blueshifted relative to the narrow [OIII] by an average $\Delta(v)=-69.6\pm 8.4$ km $\rm  s^{-1}$, in agreement with Holt et al. (2008), who found $\Delta(v)=-93\pm43$ km $\rm s^{-1}$ for this galaxy, and with IFS we also find a velocity gradient. 

The narrow [OIII] component (on Fig 23) shows essentially the same rotational  velocity map as was obtained with the single Gaussian fit to [OIII] (Fig 18, top) and we similarly fit the PA of the gradient as $-46{^\circ}$. However, the [OIII] broad component has quite different kinematics with a velocity gradient running approximately E-W: we fit a PA of $-85^{\circ}$ (with a large uncertainty of about $\pm 10^{\circ}$), with the W side blueshifted, the opposite sign from the narrow component and thus $\sim 140^{\circ}$ different.
This is near-aligned with the radio axis, on PA $-90.5^{\circ}\pm 1$.

The same analysis can also be performed with the bright, broad and relatively isolated [OI]6300 (Fig 24). Averaging over the pixels with $r<4$ arcsec,
the narrow and broad components have  mean FWHM of $9.30\pm 0.05\rm\AA$ and $33.21\pm 0.2\rm\AA$, which (subtracting the instrumental width) correspond to $359\pm2$ and $1332\pm 6$ km $\rm s^{-1}$. The 
ratio of narrow:broad component fluxes is 1:1.86, so it is not surprising [OI]6300 appears much broader than [OII] (Fig 25). Kinematically,  the mean $\Delta(v)$ (relative to the $\rm H\alpha$ zero-point) of the narrow and broad components are $0.74\pm 3.22$ and $+78.3\pm 2.0$ km $\rm s^{-1}$ respectively. On the single-Gaussian fit [OI] $\Delta(v)$ map and velocity curve (Fig 19 and 21), the [OI] line appeared about 20 km $\rm s^{-1}$ redshifted compared to $\rm H\alpha$, but we see  here this is due to the influence of the strong broad component. The [OI] narrow component is not significantly offset from $\rm H\alpha$ in velocity, and follows the same pattern of galaxy rotation as the narrow [OIII], whereas the [OI] broad component is rather uniformly redshifted without an obvious gradient. 
\begin{figure}
\vskip -0.6cm
 \includegraphics[width=0.75\hsize,angle=-90]{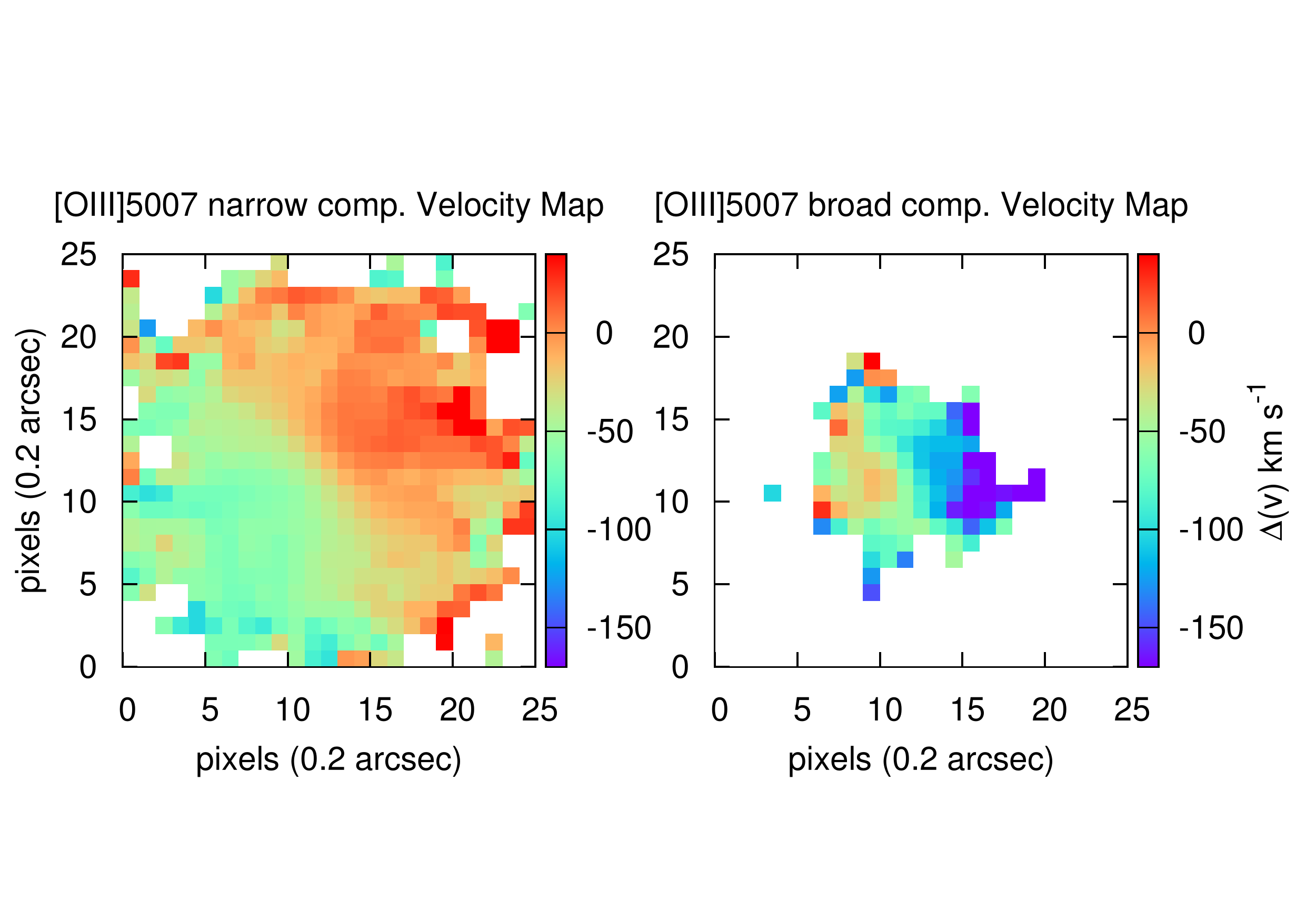}
 
 \caption{Velocity maps for two separately resolved components of [OIII]5007 emission from the primary (AGN) galaxy, given by a double Gaussian fits with a narrow component of mean FWHM $\simeq 6.5\rm\AA$ and a broad component with mean FWHM $\simeq 27.2\rm\AA$.}
\end{figure} 
\begin{figure}
\vskip -0.6cm
 \includegraphics[width=0.75\hsize,angle=-90]{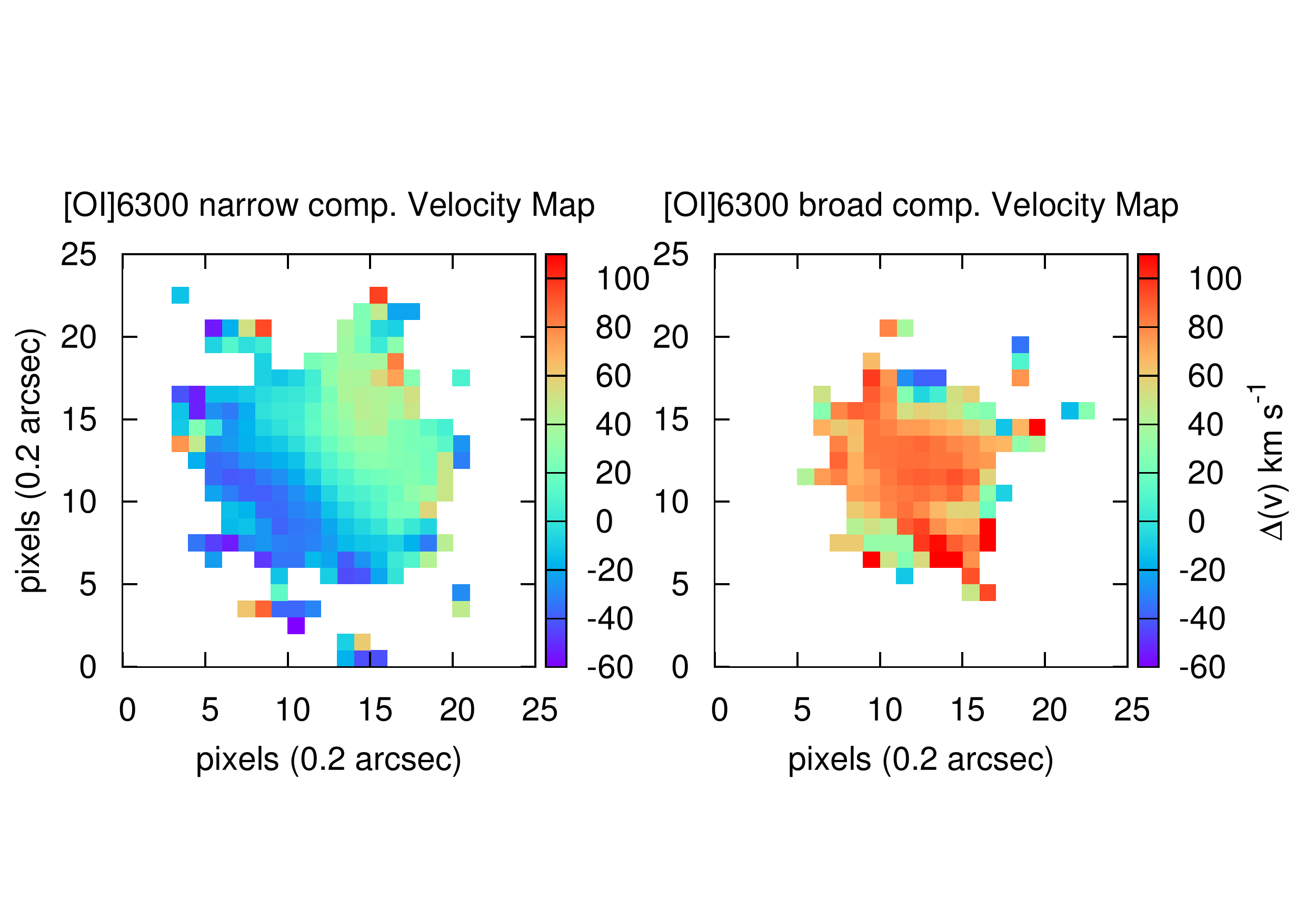}

 \caption{Velocity maps for two separately resolved components of [OI]6300 emission from the primary (AGN) galaxy, given by a double Gaussian fits with a narrow component of mean FWHM $\simeq 9.3\rm\AA$ and a broad component with mean FWHM $\simeq 33.2\rm\AA$.}
\end{figure} 

\begin{figure}
 \includegraphics[width=0.75\hsize,angle=-90]{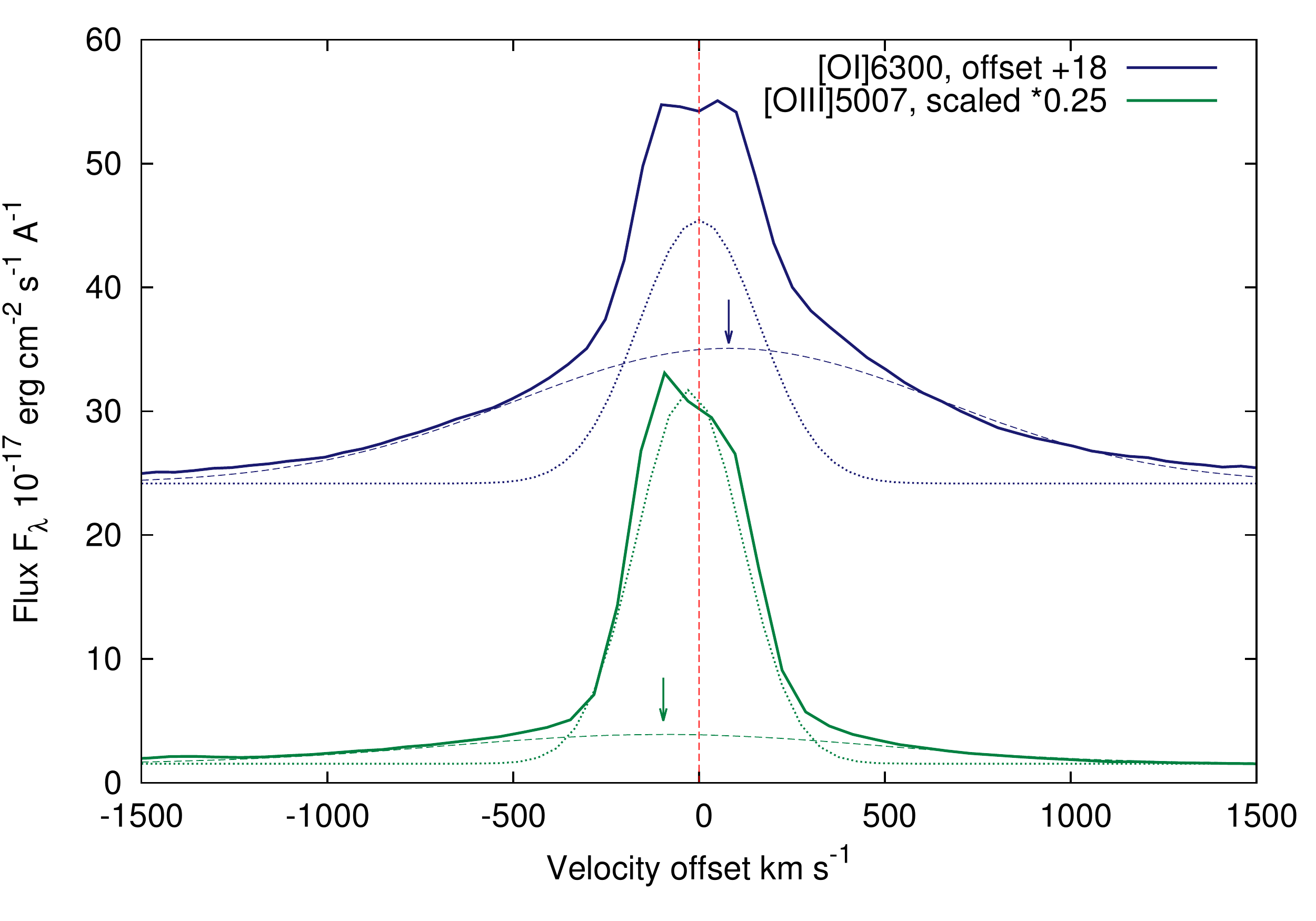}
 \caption{Comparison of the [OIII]5007 and [OI]6300 line profiles for the AGN ($r=5$) aperture spectrum, relative to the $\rm H\alpha$ velocity zero-point. Note the much greater broad component for [OI] and its asymmetry -- excess on the red side, whereas [OIII] has excess on the blue side. The dotted and dashed lines show the Gaussian fits to the narrow and broad components and the arrows show the offsets of the broad components, here -96 km $\rm s^{-1}$ for [OIII] and  +80 km $\rm s^{-1}$ for [OI].}
\end{figure}

\begin{figure}
 \includegraphics[width=0.75\hsize,angle=-90]{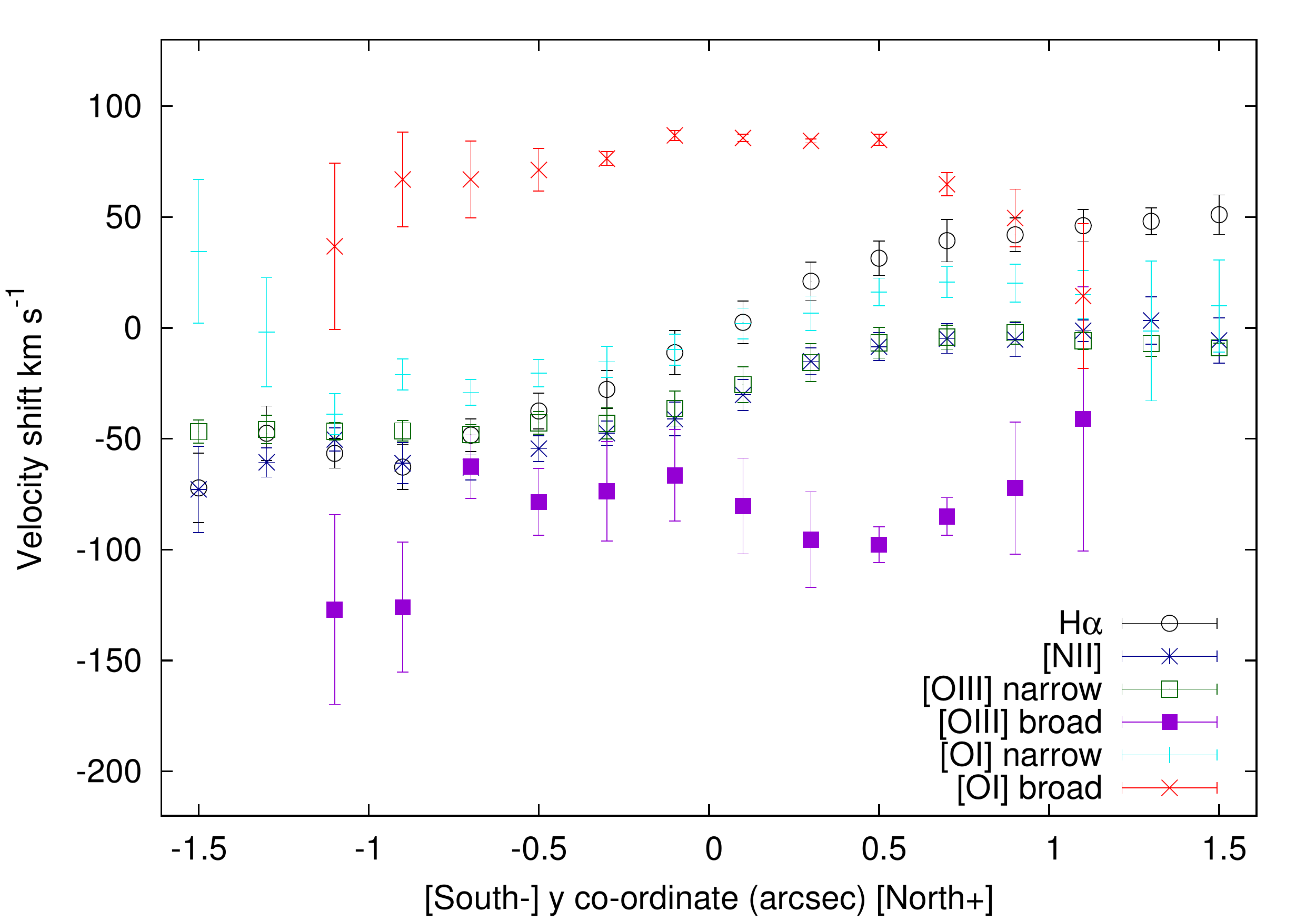}
  \includegraphics[width=0.75\hsize,angle=-90]{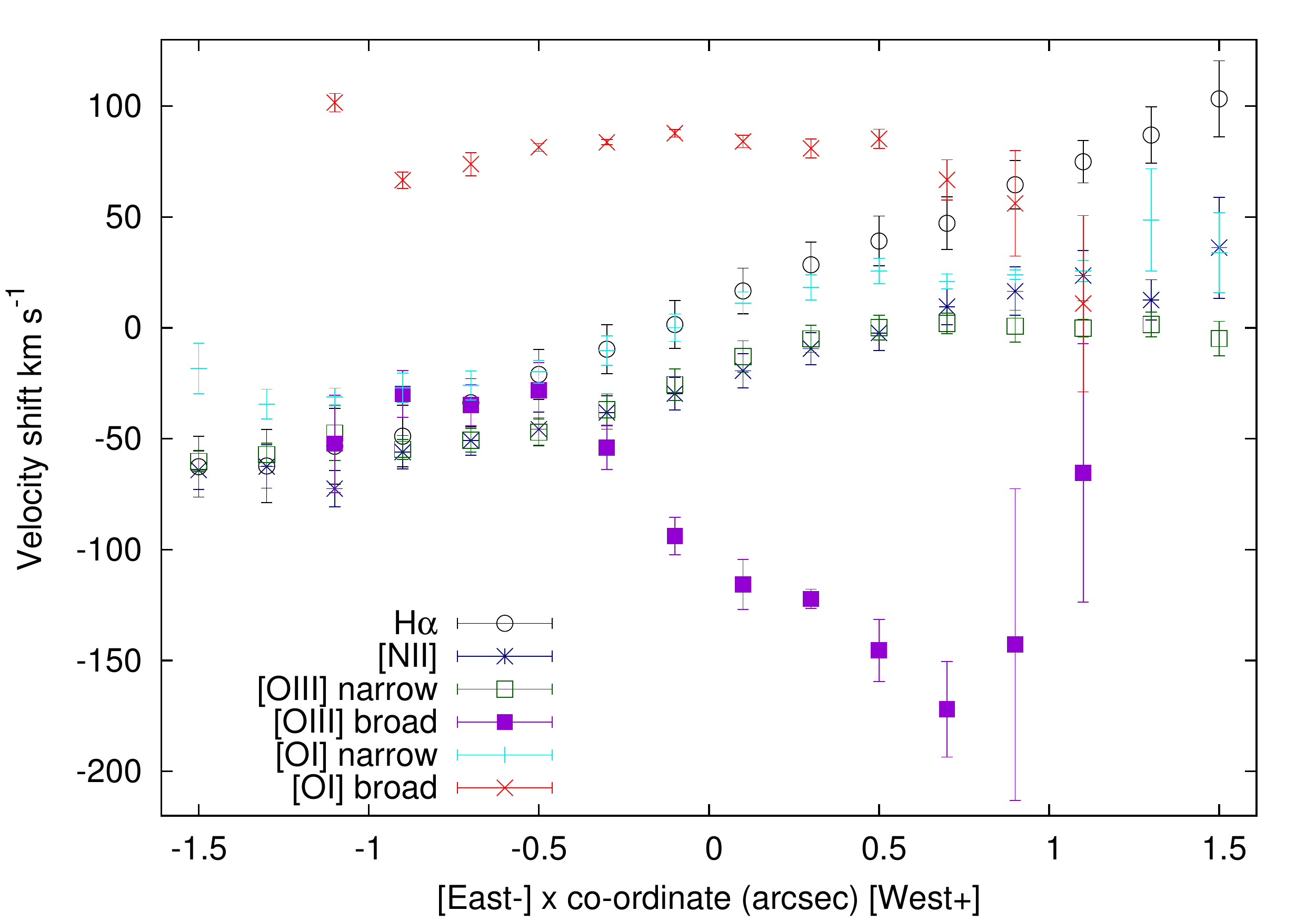}
 \caption{Velocity shifts observed for the [OIII]5007 line broad and narrow components and for $\rm H\alpha$ and [NII], measured from single-Gaussian fits, on E-W (above) and N-S (below) axes running across the AGN centre. At each 1-pixel interval along the chosen axis, the $\Delta(v)$ are averaged over 5 pixels in the perpendicular direction, thus representing the gradient that would be measured on a 1.0 arcsec wide long-slit.}
\end{figure}

We also show this by averaging the $\Delta(v)$  in 5 pixel 
width strips running E-W and N-S across the AGN (i.e. representing 1.0 arcsec width long-slits on the two axes), for the two [OII] components (together with $\rm H\alpha$ and [NII]), on Fig 26. The $\rm H\alpha$ line  shows a strong velocity gradient with the diagonal orientation giving an increase in redshift to both the W and the N, and the [NII] and [OIII]-narrow lines have similar but less steep gradients. In contrast, the [OIII] broad component is blueshifted relative to [OIII]-narrow, with no systematic trend on the N-S axis but a strong E-W gradient. The  E-W velocity shift  over the $r<4$ pixel region amounts to $-144\pm 25$ km $\rm s^{-1}$.

In the AGN/shocks composite model described in Section 3.2, a reasonable fit to observations is given by a high shock velocity 550 km $\rm s^{-1}$ (from $\rm [OI]/H\alpha$) and a ratio of  AGN:shocks and precursor: pure shocks components of  0.5:0.25:0.25 in $\rm H\beta$. In [OIII]5007 these ratios become 0.53:0.36:0.11, whereas in [OI]6300 they are 0.17:0.32:0.51, i.e.  most [OI] emission is from the high velocity shocks. This would explain the much greater prominence of a broad component in the latter line (65\% compared to 26\% of total flux). The widths of the broad components are similar to  the FWHM for velocity dispersions ($\sigma$) matching this shock velocity, i.e.  $\rm FWHM\sim 2.35v_{shock}\simeq 1300$ km $\rm s^{-1}$.  

The alignment with the radio axis, broad FWHM, and the moderate E-W velocity gradient and systematic blueshifting all suggest the broad [OIII] emission originates from a high-velocity outflow on the radio axis, which (as Holt et al. (2008) suggested) is at a low inclination  to the sky plane, tilted towards us on the West side.

\subsection{Alignment of Kinematics with Radio and Ultraviolet Structure}

Labiano et al. (2008) obtained HST ACS imaging of PKS 1934-63 in the ultraviolet F330W filter ($\lambda_{rest}\simeq 2800\rm \AA$) and detected a small central region of UV emission, which was visibly elongated with a long axis given as PA $140^{\circ}$ or $-40^{\circ}$. We obtained this image from http://archive.stsci.edu, it is a 1800s exposure named J96T22010 taken on March 9 2005. Fig 27 shows the primary galaxy region of this image rotated to give N at the top, and showing the elongated UV source (nothing was visible on the F330W image at the position of the companion galaxy). The visible extent of 0.5--0.55 arcsec (1.5--1.7 kpc) is at least $10\times$ the radio hotspot separation.

We found  ellipsoidal fits with {\it s-extractor} and {\it Galfit} on this UV source gave axis ratios  0.65--0.72 and a slightly different PA of $-54^{\circ}$, but it appears asymmetric and may be more complex.  Labiano et al. (2008), with {\it Galfit}  also fitted a point source and two extended components, attributed to `clumps of star-formation.... associated with the  fuelling of the radio activity'. One component was   $r=0.35$ arcsec emission aligned SE-NW, the other very short ($r=0.04$ arcsec) and closer to the E-W (radio) axis.
 It seems the long  axis of the UV source is aligned (within $10^{\circ}$) with the emission-line  velocity gradient from our MUSE data,  which suggests that these two phenomena are connected, e.g. the large component of the UV source might be in the form of a disk centred on the AGN and rotating with the rest of the galaxy.  The type-A rotating elliptical models of Naab et al. (2014), mentioned above, are also described as having `fast rotating central disk-like stellar configurations'. We also found the extended component of $\rm H\alpha $ emission, in two {\it Galfit} fits, to be very similar in PA and extent ( $r_{hl}\simeq 0.4$ arcsec).

The velocity gradient in AGN-dominated lines such as [OI]6300 and [OIII]5007 (the narrow components) suggests that much of the observed line flux comes from this rotating AGN-photoionized  torus or disk, and so has been imprinted with the  rotational Doppler shift pattern. Labiano et al. (2008) give the UV (F330W)  luminosity of the central source as $6.8\times 10^{41}$ ergs $\rm s^{-1}$,  close to the AGN's $\rm H\alpha$ luminosity $8.7\times 10^{41}$ ergs $\rm s^{-1}$. The MgII luminosity could be of the same order and make up a large fraction of the F330W band  flux.  Also some direct scattering of the AGN light is indicated by the Tadhunter et al. (2002) detection of $3.5\pm0.5$ polarization in the $B$-band, almost perpendicular to the radio axis.

 The geometry could be similar to that proposed by Lena et al. (2015) for the Seyfert NGC 1386, where a direct view of the AGN is blocked by an obscuring torus but the radiation emerges on either side in  a wide bicone (which here would be on the E-W radio axis), and the edge of the bicone, on both sides, intersects a rotating disk of gas (in this galaxy aligned SE-NW) at a low angle. The two elongated intersection regions, photoionized by the AGN, would emit in narrow lines and these would show the rotational $\Delta(v)$ gradient of the rotating host galaxy. In PKS 1934-63 the bicone would have a half-opening angle of at least  $45^{\circ}$.  
 This pair of AGN-illuminated intersection regions, to the SE and NW of the AGN,  might correspond to the elongated UV  source (which would explain it being larger than the radio source in that it did not physically originate in the radio burst but is a pre-existing illuminated structure) and explain its $45^{\circ}$ different alignment. In NGC 1386 the intersection regions are seen as two lobes each some 3 arcsec long, but PKS 1934-63 is at a 40$\times$ greater angular-diameter distance and similar features would not be resolved. 
  \begin{figure}

\includegraphics[width=1.0\hsize,angle=0]{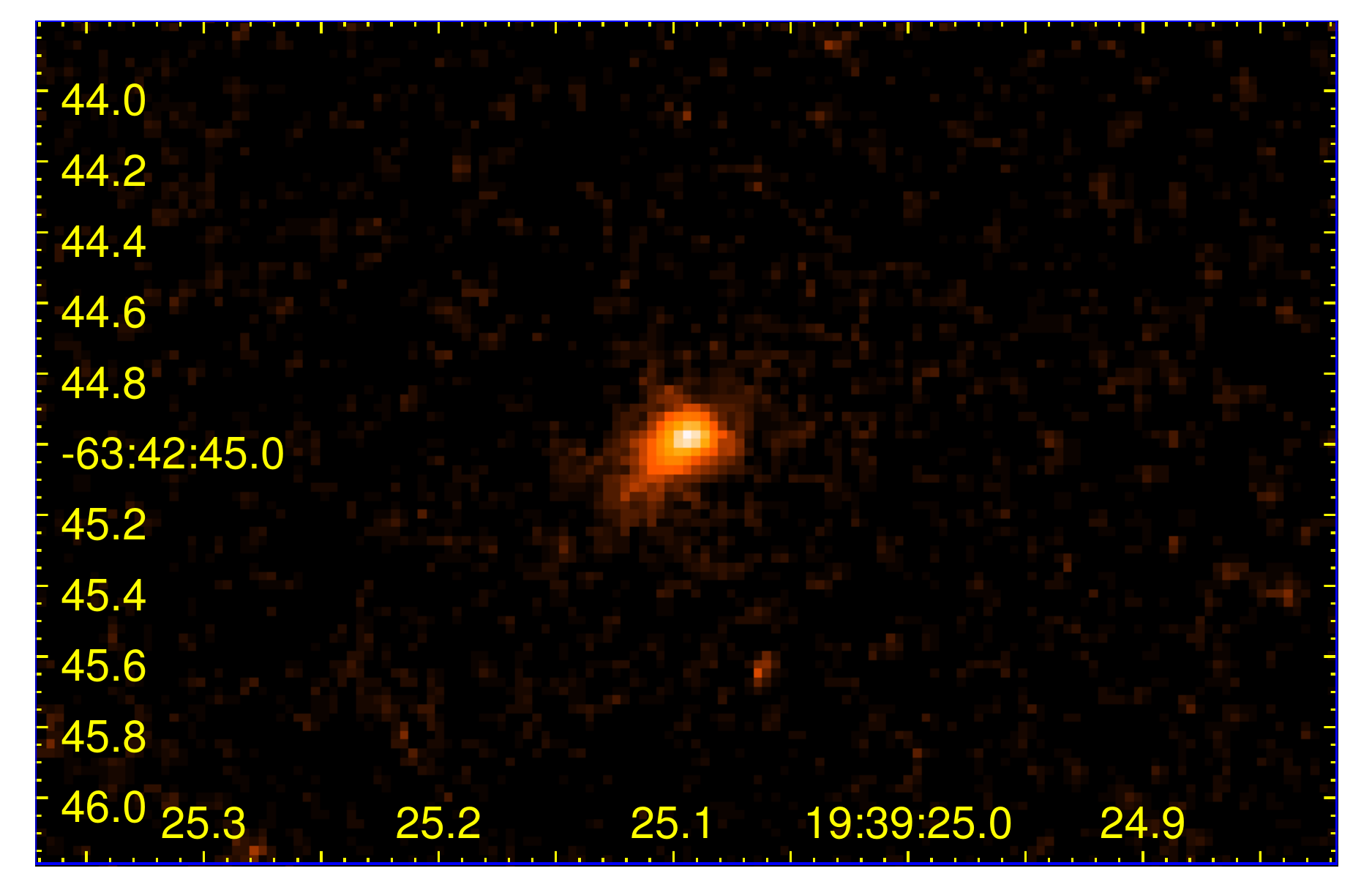}

\caption{HST ACS image of PKS 1934-63 (from the archive) in the ultraviolet F330W band, showing the elongated UV-bright region at the centre of the primary galaxy. Note pixel size is only 0.025 arcsec, and the source is of sub-arcsec diameter. Rotated to give N at the top, E at the left, and showing RA and Dec axes from the HST astrometry.}

\end{figure}

Although (Section 3.2) the diagnostic line ratios of the primary remain consistent with an AGN and shocks, without positive evidence for a starburst, there may be evidence for  star-formation  from the greater amplitude of the rotation signal in $\rm H\alpha$ (and to a lesser degree in [NII] and [SII]) compared to [OI] and [OIII], i.e. in  the lines more associated with star-formation. The AGN-powered $\rm H\alpha$ emission could be supplemented by a similar amount from rapidly orbiting star-formation (HII) regions, which might extend to larger radii (higher up the galaxy's rotation curve), but remain unresolved from the AGN in the MUSE and other ground-based data. Emission coming from two regions on the approaching and receding sides of a fast-rotating disk could explain the slightly double-peaked $\rm H\alpha$ line profiles (Fig 12). The required SFR is quite moderate e.g. the entire $\rm H\alpha$ emission of the primary could be given by a SFR $\rm 3.62~M_{\odot} yr^{-1}$ with no dust, so if a more plausible $\sim 25\%$ of the primary galaxy $\rm H\alpha$ is from star-formation with one magnitude dust extinction, the required SFR is only $\rm \sim 2.3 ~M_{\odot}yr^{-1}$.

Labiano et al. (2008) also suggested that the starburst and AGN are triggered by the same event, with the starburst preceding by $\geq 10$ Myr. The companion galaxy is at PA  $-77.0^{\circ}$, within $14^{\circ}$ of the radio axis, and on the basis of its high relative radial velocity the inter-nuclear vector is probably close to the sky plane, as the radio axis is believed to be (Holt et al. 2008). It has previously been noted that  interacting radio galaxies have some tendency for the companion galaxy position to be aligned with the radio source axis (Roche and Eales 2000), and the near alignment in this example with the `pole' of the AGN might be the most favourable point for the interaction (tidal forces) to start the radio outburst.

\section {Star-Formation History}
In this section the aim is to reconstruct star-formation histories by fitting the observed spectra with stellar population models , using the {\it Starlight} package (Cid Fernandes et al. 2009), with the supplied set of Chabrier-IMF stellar templates. 
 The spectra to be fitted have to be de-redshifted to the rest-frame, and as {\it Starlight} fits only the stellar continuum, strong emission lines need be masked out or otherwise excluded.
The output files give the best-fitting model in terms of tabulated stellar masses assigned to each template (both at formation and present now) which we sum over metallicity to give the mass of stars (present in observed galaxy) as a function of age, and also the fitted model spectrum.

The companion galaxy spectrum was fit with a base of 66 template spectra representing  22 ages from 1 Myr to 10 Gyr (excluding 3 older templates  because the Universe age at observation is only 11.62 Gyr) and three metallicities (Z=0.004, 0.008, 0.02). 

Firstly, Fig 28 shows the {\it Starlight} fit to the companion galaxy spectrum (in the $r=5$ pixel aperture), which estimates the stellar mass $\rm M_{*}=9.98\times 10^9~M_{\odot}$ and the mass-weighted mean age  5.16 Gyr.   Almost all of the stars were formed 1--10 Gyr ago, very few at age $10^7$--$10^8$ yr, and recently there has been a resurgence of star-formation giving rise to $\rm 1.20\times 10^7~M_{\odot}$ at ages $\leq 40$ Myr. This corresponds to $\rm 1.45\times 10^7M_{\odot}$ of young stars at their time of formation  and therefore a recent SFR averaging 0.36 $\rm M_{\odot} yr^{-1}$. On the basis of {\it s-extractor} $R$ magnitudes the aperture correction is 2.25, with which the whole galaxy stellar mass is $\rm 2.25\times 10^{10}M_{\odot}$, and the SFR estimate is similarly increased. With the Tully-Fisher relation of Reyes et al. (2011), the corresponding rotation velocity for this mass is  $163\pm 14$ km $\rm s^{-1}$, consistent with the observed rotation curve if the disk inclination is $\sim 45^{\circ}$ (consistent with the axis ratio of 0.66). Statistical errors estimated by repeatedly running the fit are only $\rm \pm 0.27\times 10^9~M_{\odot}$ for stellar mass and $\pm 0.27$ Gyr for age, and for the young stars $\rm \pm 0.37\times 10^7~M_{\odot}$, meaning there is $>3\sigma$ evidence these are present but the mass is uncertain (fits ranged from 0.63 to $\rm 2.14\times 10^7~M_{\odot}$).

\begin{figure}
\includegraphics[width=0.75\hsize,angle=-90]{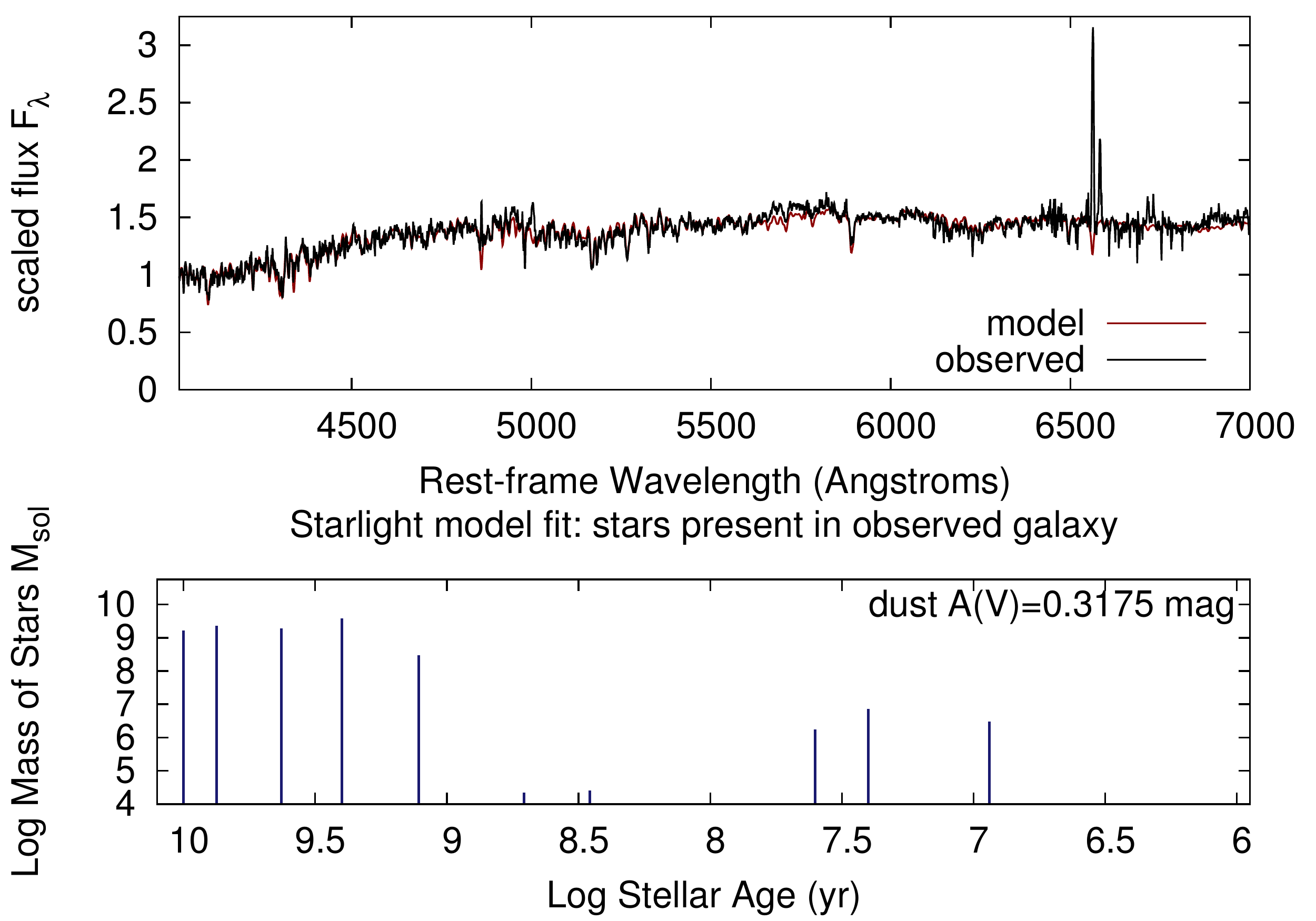}
\caption{Observed spectrum (de-redshifted) and best-fit Starlight model for the companion galaxy in a central $r=5.0$ pixel (radius 1 arcsec) aperture. Below is shown the stellar masses (existing at the time of observation) of each fitted component, against age.}
 \end{figure}
 {\it Starlight} fitted intrinsic dust extinction of $A(V)=0.3175$ mag for the stellar continuum, which is 
 0.26 mag at the wavelength of $\rm H\alpha$. The Balmer decrement of 4.35 corresponds to a much greater 1.19 mag extinction for  the $\rm H\alpha$ line itself, and if this is correct the dust distribution is very inhomogeneous.

 \begin{figure}
\includegraphics[width=0.75\hsize,angle=-90]{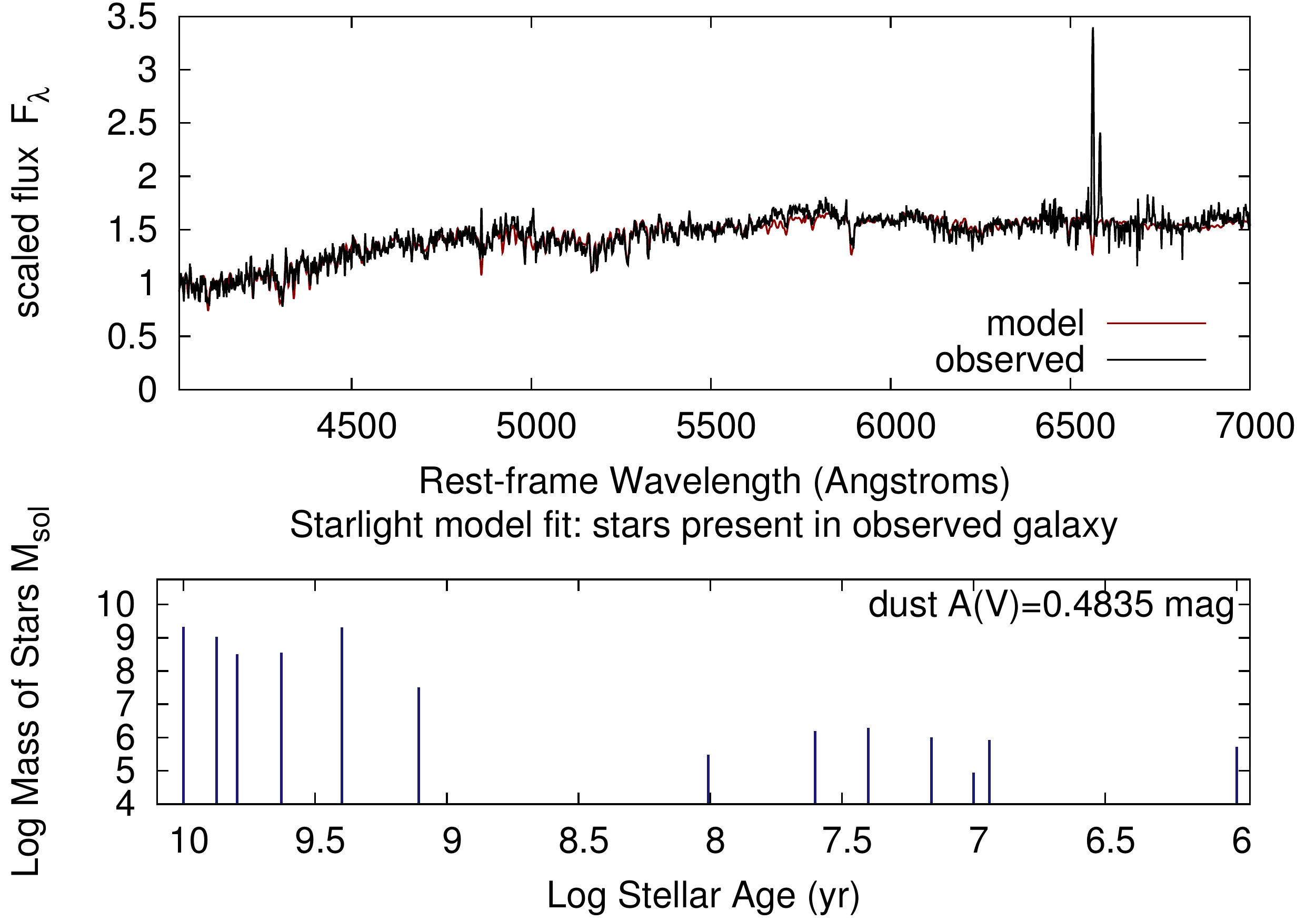}
\caption{As above  for the companion galaxy in a central $r=3.0$ pixel (radius 0.6 arcsec) aperture.}
 \end{figure}
\begin{figure}
\includegraphics[width=0.75\hsize,angle=-90]{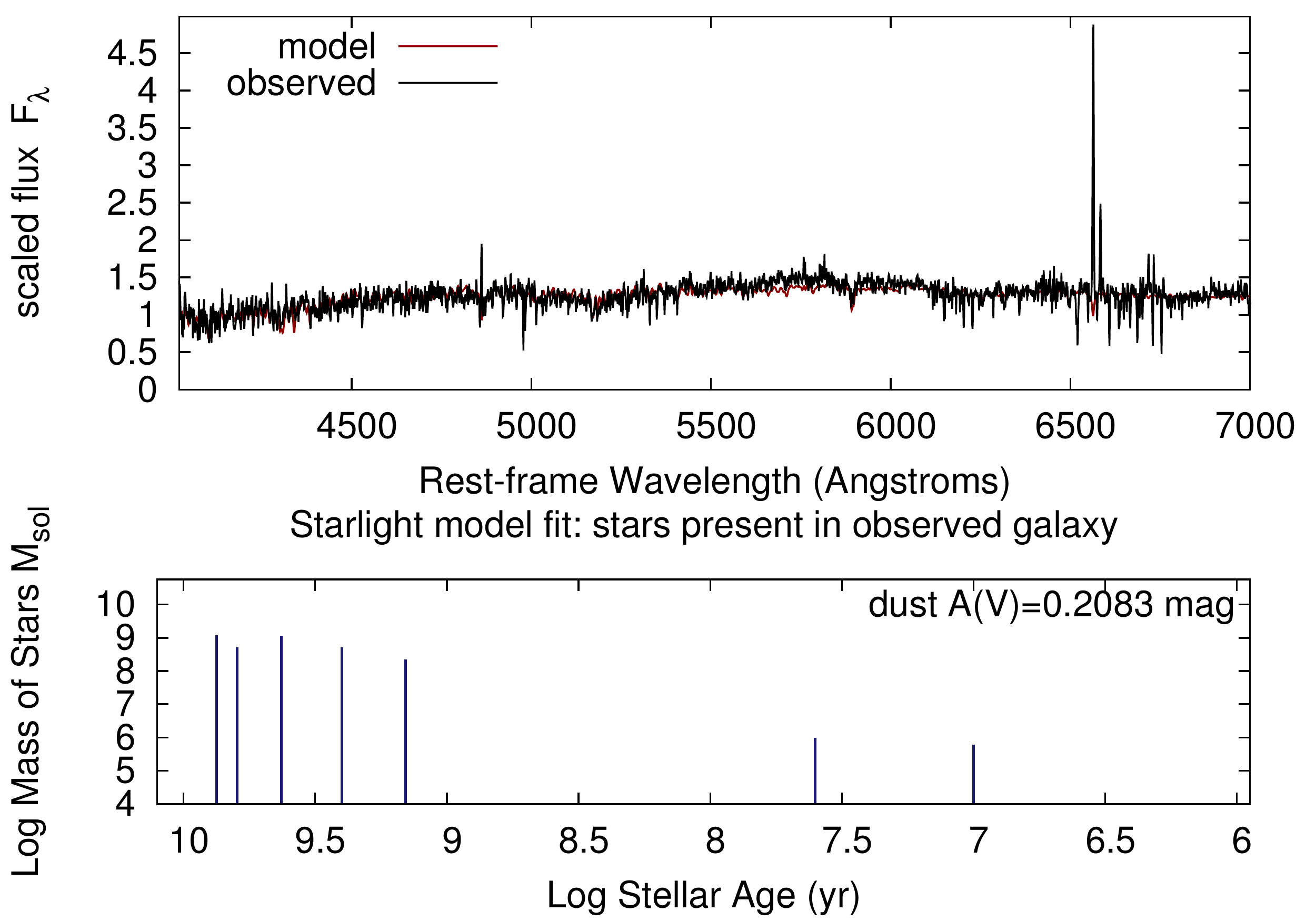}
\caption{As above, spectrum and model for the western arm region of the companion galaxy.}
 \end{figure}
\begin{figure}
\includegraphics[width=0.75\hsize,angle=-90]{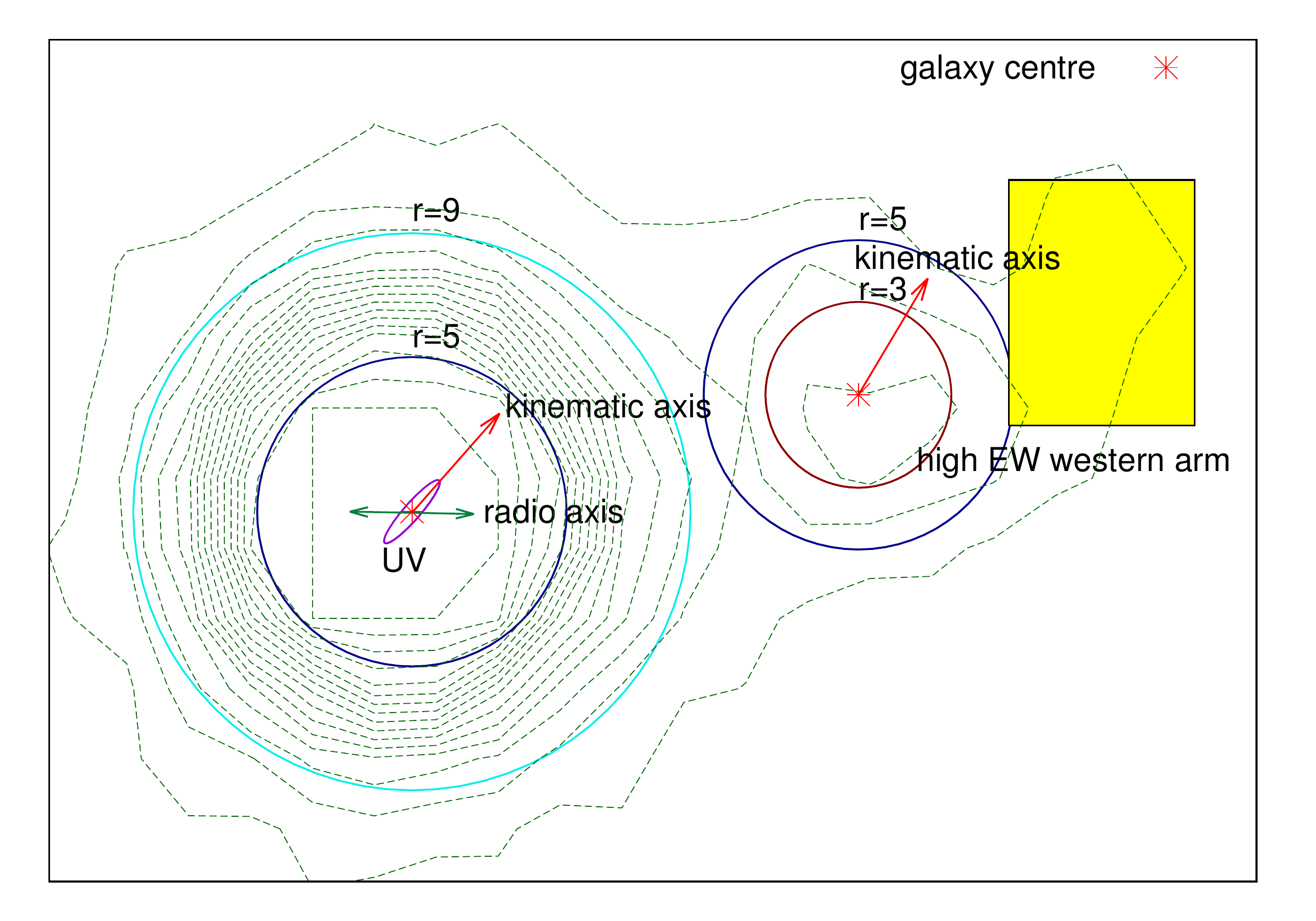}
\caption{Axes of the two galaxies, and radii and regions defined for extracting spectra in this section. The green contours show the $\rm H\alpha$ intensity.}
 \end{figure}

For a spectrum of the central $r=3$ pixel aperture (Fig 29), {\it Starlight} fit finds a heavier dust extinction of $A(V)=0.4835$ mag, with stellar mass  $\rm 5.99\times 10^9~M_{\odot}$, a slightly greater mean age 6.63 Gyr, mostly 1--10 Gyr age stars plus a 0.1\% fraction, $\rm 6.32\times 10^6~M_{\odot}$ formed in the past 40 Myr. This is $\rm 7.61\times 10^6~M_{\odot}$ at formation, giving the mean recent SFR $\rm 0.19~ M_{\odot}yr^{-1}$. The $\rm H\alpha$ EW  $8.9\rm\AA$ and the $\rm [NII]/H\alpha$ of 0.36 are similar to the $r=5$ aperture and the Balmer decrement slightly higher at 4.71.

Thirdly, a spectrum is extracted in a  $6\times7$ pixel box (Fig 30) covering the western arm of the companion galaxy (Fig 31). {\it Starlight} estimates the stellar mass  $\rm 1.93\times 10^9~M_{\odot}$ and the mass-weighted mean age as 5.15 Gyr, with all stars $>1$ Gyr age except for $\rm 1.29\times 10^6~M_{\odot}$ at
 $\leq 40$ Myr. Compared to the central galaxy, the dust extinction is less at $A(V)=0.2083$ mag, the $\rm H\alpha$ $\rm EW_{obs}$ higher at $\rm 14.4\AA$, the $\rm [NII]/H\alpha$ ratio lower at 0.24, but the Balmer decrement is still high at 4.40.

Our models do not find much variation in star-formation history between regions of this galaxy, but the dust seems to be centrally concentrated. The decrease in $\rm [NII]/H\alpha$ from the centre to the western arm may be a metallicity gradient: with the Marino et al. (2013) N2 calibration the $\rm \Delta([NII]/H\alpha)$ corresponds to $\rm \Delta (12+ log O/H)=(8.46-8.54)=-0.08$ dex, over $r\simeq 7.8$ pixels or 4.8 kpc ($\rm \sim 1.2 r_{hl}$), a  typical gradient for a disk galaxy (e.g. Sanchez et al. 2014, Ho et al. 2015).

\begin{figure}
\includegraphics[width=0.75\hsize,angle=-90]{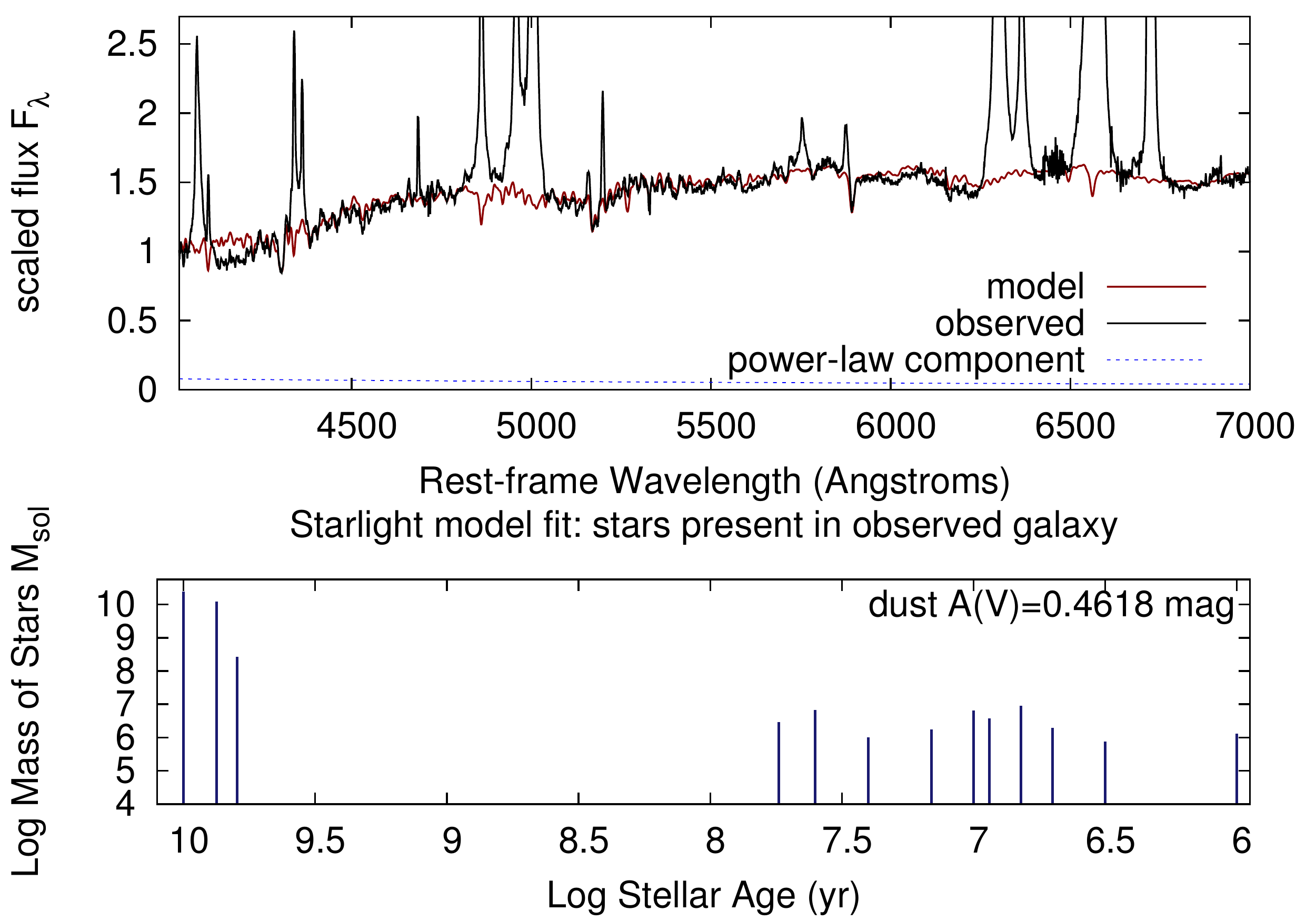}
\caption{Observed spectrum and Starlight model for the primary galaxy extracted in a central $r=5$ pixel (1 arcsec radius) aperture.}
 \end{figure}
\begin{figure}
\includegraphics[width=0.75\hsize,angle=-90]{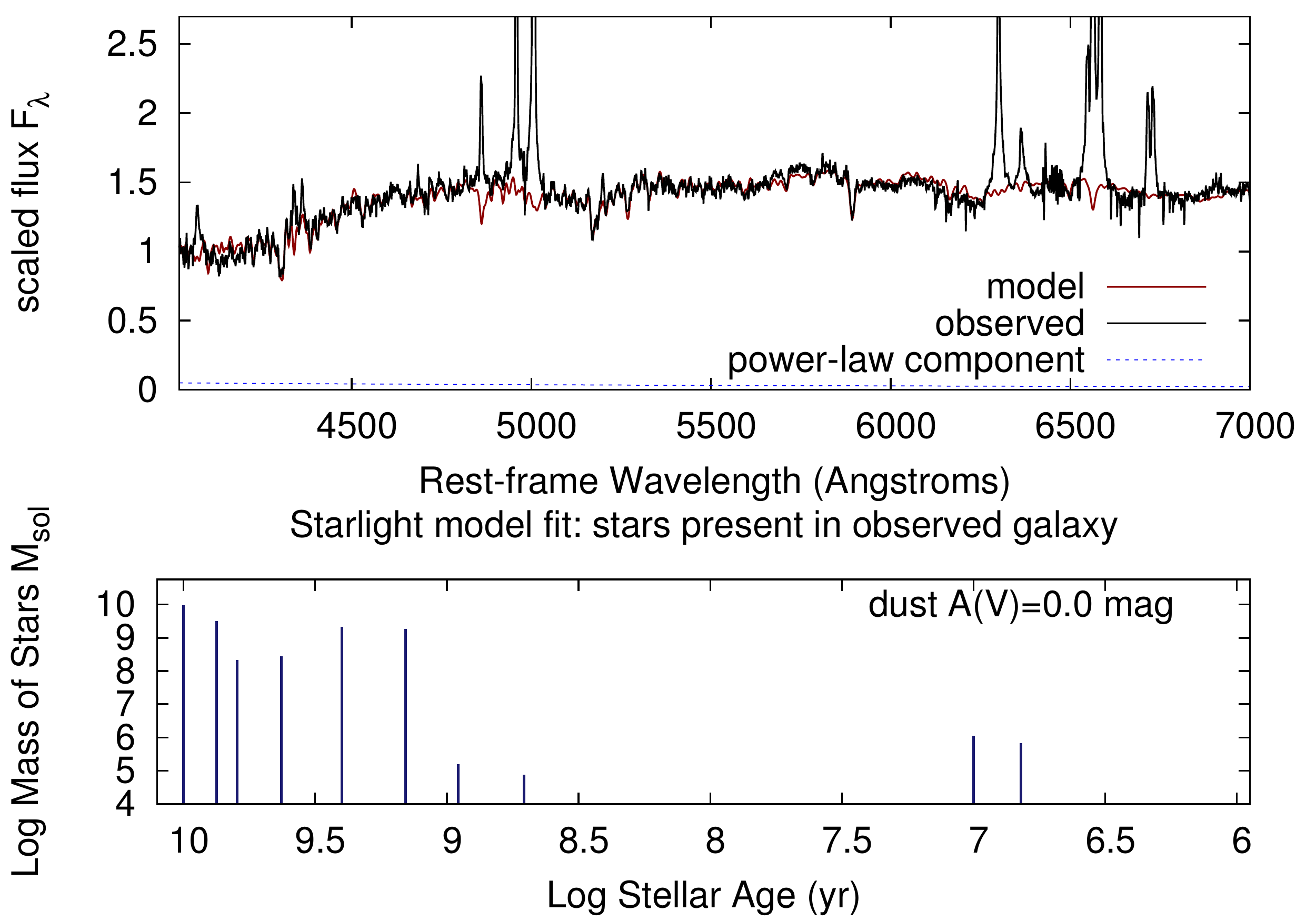}
\caption{Observed spectrum from a $5<r<9$ pixel (1-1.8 arcsec) annulus centred on the AGN (covering the outer part of the primary galaxy), again with the best-fit {\it Starlight} model.}
 \end{figure}

We fitted the primary galaxy spectrum ($r=5$ pixel aperture), this time with a larger base of models including high metallicity ($Z=0.05$) stars and a combination of power-laws $f_{\lambda}\propto \lambda^{-1.5},\lambda^{-1},\lambda^{-0.5}$ and $\rm \lambda^0$ which could represent AGN scattered light and nebular continuum (as in the radio galaxy spectral fits of Tadhunter et al. 2002). 
The strong AGN emission lines can greatly disturb the fit, but if strong ($4\sigma$) clipping is applied in {\it Starlight} they are mostly ignored. Fig 32 shows the continuum best-fitted by an old stellar population with total mass  $3.71\times 10^{10}\rm~ M_{\odot}$ and mass-weighted  mean age 9.14 Gyr, strong dust reddening $A(V)=0.4618$ mag, and a small power-law component, $8\%$ at the blue  end ($4020\rm \AA$), comprised of a mixture $0.039(\lambda/4020)^{-1.5}+0.039(\lambda/4020)^{-1.0}$. On top of this, the model fit a population of $\leq 10^8$ yr age stars amounting to $\rm 3.58\times 10^7 M_{\odot}$, only $0.1\%$ by mass but contributing $22\%$ of the flux at the blue end.
We estimate fit errors $\pm 0.104\times 10^{10}\rm~ M_{\odot}$ and $\pm 0.314$ Gyr age for the old stellar population, and $\pm 0.49\times 10^7 \rm M_{\odot}$ for the young stars mass and $\pm 1.90\%$ for the power-law component, which would mean the latter two are $4\sigma$ detected. 

The $<10^8$ yr component is $\rm 4.23\times 10^7~M_{\odot}$ of stars at time of formation and so gives a mean SFR $\rm 0.42~M_{\odot}yr^{-1}$, but they are concentrated at more recent times, e. g. for age $\leq 10$ Myr (the $\rm H\alpha$ emissions lifetime), stars formed are $\rm 2.63\times 10^7 ~M_{\odot}$, suggesting the SFR increases and could now be $\rm \sim 2.6 ~M_{\odot} yr^{-1}$. 
This seems consistent with  Labiano et al. (2008) accounting for the central UV emission by $\rm \sim 10^7~M_{\odot}$ stars of age $\leq 10$ Myr (allowing for a correction for dust).

As for the stellar mass,  the ratio of total {\it s-extractor} to $r=5$ magnitudes in the $I$-band (less affected by AGN lines than $R$-band) gives the relevant aperture correction as 2.67 in the $I$-band, hence  the total $\rm M_{*}=9.9\times \rm 10^{10}~M_{\odot}$.

In the spectrum from a $5<r<9$ pixel annulus (Fig 33), the AGN line emission is at least a factor 4 less prominent (on the basis of equivalent widths) relative to the host galaxy. The {\it Starlight} fit (again with the $4\sigma$ clipping) finds an old stellar population (mass $\rm 1.70\times 10^{10}~M_{\odot}$) with a more extended formation period giving a mean age of 7.53 Gyr, and zero dust. The power-law component ($4.87\%$) and young stars mass 
($\rm 1.8\times 10^6~M_{\odot}$) are much less than for the centre, and with estimated fit errors  $\pm 3.04\%$ and $\pm 2.38 \times \rm 10^{6}~M_{\odot}$, there is no real evidence for any $<1$ Gyr age stars in the outer host galaxy.

\section{Discussion}
A consistent picture of the central activity and the effects of the interaction may be emerging.
 The AGN host is an elliptical galaxy, with $\rm M_{*}\simeq 1.0\times 10^{11}M_{\odot}$, and $r_{hl}=6.2$ kpc, which makes this one of the smaller radio-AGN hosts. In comparison, Inskip et al. (2010) found a mean $r_{eff}=10.6\pm 1.6$ kpc for narrow-line radio galaxies and Tadhunter et al. (2011) a median mass $\rm 4.7\times 10^{11}M_{\odot}$. Most of its stars are very old, 7--10 Gyr. 
 The interaction is a prograde-prograde type, which has the greatest tidal effects, and the near alignment of the galaxy pair with the radio axis might also be significant. The close passage triggered star-formation, from$<10^8$ yr ago and continuing now.  The tidal inflow of gas could have triggered the radio source  $\sim 10^4$ yr ago, at which point the two radio lobes began to separate at extremely high velocity ($\geq 0.05c$) to  $D=130$ pc today.

The companion galaxy is a rotation-supported disk (barred spiral?) of an estimated  $2.25\times 10^{10}M_{\odot}$ and $r_{hl}=3.9$ kpc, which formed stars over most of its lifetime. The interaction with the primary seems to have triggered new star-formation over the past $\sim 40$ Myr.  From our $r=1$ arcsec aperture spectrum the $\rm H\alpha$ flux  is now $2.412\times 10^{-16}$ erg $\rm cm^{-2}s^{-1}$ (corrected for stellar absorption), giving $\rm L_{H\alpha}=10^{40.36}$ erg $\rm s^{-1}$. Using the AGN-subtracted  narrowband image we estimated a $\rm H\alpha$ aperture correction of 2.045, giving  a whole-galaxy $4.93\times 10^{-16}$ erg $\rm cm^{-2}s^{-1}$
and $\rm L_{H\alpha}=10^{40.67}$ erg $\rm s^{-1}$. The Balmer decrement gave a dust correction factor of 2.99, without much variation, so this might be applied for the whole galaxy. Combining these two corrections gives $\rm L_{H\alpha}=10^{41.14}$ erg $\rm s^{-1}$, 
and the total SFR as $\rm 0.61~M_{\odot}yr^{-1}$. This is consistent with the SFR estimated from the young stars content  in the  {\it Starlight} fit of Section 6, which would sum to $\rm \sim 0.73~M_{\odot}yr^{-1}$ for the whole galaxy. These both give the  specific SFR $\sim 0.03$ $\rm Gyr^{-1}$, less than the inverse age, so this galaxy could not be called starbursting, neither quiescent. 

We agree with Labiano et al. (2008) that the same event (interaction) triggered star-formation (in both galaxies) followed by the  AGN a few $10^7$ yr later.  As (i) neither galaxy has the strong $\rm H\delta$ absorption  characteristic of 0.1--1.0 Gyr age post-starbursts or has many stars in this age range in {\it Starlight} fits, (ii)
 star-formation in the companion galaxy is not concentrated in the nucleus as in late-stage mergers, but extends at least 6 kpc into the western arm, and (iii) the $R$-band residual to the S\'ersic fit (Fig 13) looks like a regular spiral which is not yet strongly disrupted, the galaxies are probably at an early interaction stage, the first perigalacticon. It is possible the SFRs  are still  increasing and the system will become more luminous. 
 The An \& Baan (2012) model predicts GPS types brighten rapidly as $\rm P_{rad}\propto D^{2/3}$ and so the radio luminosity could increase a further factor 4 to $D\simeq 1$ kpc as the source evolves into a CSS type,  and similarly with the evolution trend plotted by Jeyakumar (2016). 
 
Emission lines from the primary galaxy show  velocity gradients on PA $-45^{\circ}$, misaligned with the radio axis PA of $-90.5^{\circ}$. We suggested this gradient was produced by a kpc-scale photoionized disk, viewed with its plane on this axis and rotating with the rest of the galaxy, illuminated by the edges of the E-W oriented bicone of emission from the AGN (somewhat as in the Seyfert described by Lena et al. 2015). One of our key findings is a $\sim 2\times$ stronger velocity gradient in $\rm H\alpha$ compared to the forbidden lines, and we suggested that this disk contained circumnuclear star-formation, orbiting the AGN at 100--150 km $\rm s^{-1}$. This would make the galaxy a very fast rotating elliptical,  but is still consistent with motion under gravity. The central $2800\rm\AA$ emission of Labiano et al. (2008) is elongated on PA $\simeq -45^{\circ}$ as is the extended $\rm H\alpha$ component in our {\it Galfit} fit (Table 2). Yet the host galaxy itself is not visibly elongated in this direction and appears almost round ($b/a=0.92$), which might be investigated with high-resolution imaging. 

 The {\it Starlight} fit found evidence of a young stellar population in the centre of the primary ($r<3$ kpc), but unresolved from the AGN and difficult to distinguish from a blue power-law component. It was estimated there could be 
  $\sim 4\times 10^7~M_{\odot}$ of young stars, with  an increasing SFR which may now reach $\rm \simeq 2.6~M_{\odot} yr^{-1}$. We argued that a SFR of this order could provide  a significant fraction ($\sim 30\%$?) of total  $\rm H\alpha$ flux and therefore affect (i.e. enhance) the $\rm H\alpha$ rotation curve, but still allow the AGN to dominate the diagnostic emission-line ratios. Furthermore the {\it Spitzer} mid-infrared spectroscopy (Dicken et al. 2012) detected  PAH line emission from this galaxy with flux $\rm F(11.3\mu m)=1.3\pm 0.1 \times10^{-14}$ erg $\rm cm^{-2}s^{-1}$, giving luminosity $10^{42.088}$ erg $\rm s^{-1}$. 
  From the conversion derived by Treyer et al. (2010), $\rm log(SFR/L_{11.3})=-41.605\pm 0.183$ (for normal star-forming galaxies assuming Kroupa IMF) this indicates a SFR $3.04
\rm ~M_{\odot}yr^{-1}$ (range 1.98--4.67).

Another kinematic finding is the identification of  a broad ($\rm FWHM\simeq 1350$ km $\rm s^{-1}$) component of [OIII] emission, blueshifted with respect to the narrow component and with a velocity gradient (spanning $144\pm 25$ km $\rm s^{-1}$) aligned with the radio axis rather than the rotation. Its velocity dispersion  also matched the shock velocity $\sim 550$ km $\rm s^{-1}$  estimated from $\rm [OI]/H\alpha=0.8$. This must be the signature of an ionized outflow. An outflow velocity of $\sim 550$ km $\rm s^{-1}$ exceeds by factors $\sim 2$ the velocity of outflows  from most star-forming luminous infra-red galaxies (LIRGs), but is matched by many interacting LIRGs powered by AGN (Arribas et al. 2014). Fast AGN outflows  give rise to broad emission features with FWHM$\simeq 1000$--2000 km $\rm s^{-1}$ and an overall blueshift of anything up to $\sim 850$ km $\rm s^{-1}$ (Holt et al. 2008), probably because the receding side/jet is subject to greater dust extinction (e.g. Humphrey et al. 2006, Villar-Mart\'in et al. 2011). In PKS 1934-63 we found the broad [OIII] is blueshifted relative to narrow [OIII] by an average of  $-69.6\pm 8.4$ km $\rm  s^{-1}$. The overall blueshift and  the velocity gradient, both significant but an order of magnitude less than the FWHM,  can be attributed to the radio axis/jet being oriented close to the sky plane but inclined by several degrees with the west side toward us (the ratio of velocity shift to FWHM might suggest an inclination arcsin $(144/1372)\simeq 6^{\circ}$).

Broad and narrow components with different axes have been seen in a number of other young radio galaxies. As kpc-scale structures and star-formation hotspots have dynamic timescales of a few Myr, if they occur in GPS galaxies they would persist into the CSS stage. Tadhunter et al. (2001) found that in the compact $z=0.15$ radio galaxy PKS 1549-79,  the [OIII]5007 line is blueshifted by 600 km $\rm s^{-1}$ relative to [OI]6300, [OII]3727 and the neutral Hydrogen. In contrast to PKS 1934-63 this radio jet is oriented close to the line of sight, but the explanation is similarly that the broad [OIII] emission comes from an outflow whereas the narrower low-ionization lines originate from extended quiescent gas and possibly star-formation.
The  Shih et al. (2013) GMOS-IFU study of several CSS $D\sim 1$ kpc radio galaxies, found the strong [OIII] lines to have broad (outflow) components with velocity gradients aligned with the radio axes, while the narrow [OIII] components could have similar or significantly different $\Delta(v)$ patterns.

 The Mahony et al. (2016) IFS of the $z=0.045$ radio galaxy 3C293 (again at a slightly later stage with $D\simeq 1.7$ kpc for the inner lobes) similarly found narrow and broad $\rm H\alpha$ components with velocity gradients oriented differently, respectively with the the rotating galaxy disk (long axis) and with the outflowing radio jet. There might be additional insights from measuring the $\rm H\alpha$ and other line velocities separately. 
 
  A comparison might also be made with PKS 2250-41, a  $z=0.31$ radio galaxy in a much later (FRII; $D\sim 70$ kpc) stage with  kinematically distinct narrow and  broad ($\rm FWHM\simeq 500$--900 km $\rm s^{-1}$) emission components  (Villar-Mart\'in et al. 1999). The broad component was attributed to both interaction between the cloud and bowshock ahead of the jet, and entrainment of clouds in the post-shock wind or boundary layers, and is  relatively weak in [OIII] but is the stronger component in [OI]. The narrow emission could be associated with the AGN-photoionized ambient gas in gravitational motion.
In PKS 1934-63 we find that the broad component of [OI] emission, although similar in FWHM to broad [OIII], has a different kinematic pattern with $\sim 80$ km $\rm s^{-1}$ systematic redshift and no obvious $\Delta(v)$ gradient. Perhaps a greater fraction of this originates from `entrained' gas which has been dragged or displaced by the radio jet and is now backflowing towards the galaxy.
PKS 1934-63 may also resemble PKS 2250-41 in that the latter is similarly an $L^*$ elliptical  with strong rotational kinematics (Inskip et al. 2008) and evidence that the AGN was triggered at an early stage (first close passage) of an interaction with a companion disk galaxy.

To understand the evolutionary sequence of radio galaxies it will be important to perform IFS studies of high spatial and spectral resolution at all stages from high-frequency peaked and GPS galaxies to the late and declining (or `relic') FR galaxies, and also to compare high and low ionization radio galaxies. There are radio bursts at earlier stages than PKS 1934-63, such as  the nearby $z=0.014$ PKS 1718-649 (NGC 6238), with $D\sim 2$ pc. This is   a low-ionization galaxy for which the triggering mechanism may be different, but like our target it has broad [OI] lines (Fillipenko 1985), extended $\rm H\alpha$ emission (Keel \& Windhorst 1991), mid-IR evidence of star formation, and a fast-rotating circumnuclear gaseous disk of similar ($r\sim 0.65$ kpc)  dimensions (Maccagni et al. 2014, 2016).
In view of the difficulty resolving AGN and star-forming regions and measuring rotation velocities, it will be of benefit to study radio galaxies such as these with space telescopes and adaptive-optics IFS such as the forthcoming MUSE-GALACSI which may give $<0.05$ arcsec  resolution.

 \section{Summary of Conclusions}
(i) We observed the GigaHertz Peaked Spectrum radio galaxy PKS 1934-63 at optical wavelengths with MUSE.   Our 3D data show PKS 1934-63 is    an interacting pair of galaxies, projected separation 9.12 kpc and $\Delta(v)=216$ km $\rm s^{-1}$. The primary (larger) galaxy hosts the radio AGN (with $\rm H\alpha$ at $\rm 7760.5\AA$ giving $z=0.1825$)  and the companion is an emission-line galaxy at $z=0.1834$.

(ii) The radio galaxy host shows strong emission lines, the brightest [OIII]5007 and $\rm H\alpha$, with
$\rm [OIII]/H\beta$ and $\rm [NII]/H\alpha$ ratios consistent with AGN photoionization.
The  high ratios of   [OI]6300/[OIII]5007 (0.556) and [OIII]4363/[OIII]5007 (0.044)  are characteristic of young radio AGN with high electron temperatures of $T_e\simeq 24000$ K. These ratios and the high values of  $\rm [OII]_{7319}^{7330}/[NII]6584$ and $\rm [SII]_{4069}^{4076}/[SII]_{6717}^{6731}$ favour a  large ($\sim 50\%$ in $\rm H\beta$) contribution from high velocity shocks ($v\simeq 550$ km $\rm s^{-1}$).
The secondary galaxy has lower
$\rm [OIII]/H\beta$ and $\rm [NII]/H\alpha$ ratios indicative of non-AGN star-formation.

(iii) Morphologically the primary galaxy appears spheroidal and almost round: {\it Galfit} in the $R$-band fits a de Vaucouleurs profile with a 14\% point-source and $r_{eff}=6.16$ kpc.
 The $\rm H\alpha$ emission is more centrally concentrated, fit with a point source plus a smaller radius  extended component, with a total luminosity $\rm L_{H\alpha}=10^{41.92}$ erg $\rm s^{-1}$. 

The companion galaxy is fit with a lower S\'ersic index of 1.69 in the red band, $r_{hl}=3.87$ kpc and from the residuals it may be a barred spiral. Its $\rm H\alpha$ emission is extended over the galaxy,  with aperture corrected luminosity $\rm L_{H\alpha}=10^{40.67}$ erg $\rm s^{-1}$. With a further correction of 2.99 for dust extinction its star-formation rate is estimated as 0.61  $\rm M_{\odot}yr^{-1}$. Metallicity is estimated as $\rm 12+log(O/H)\geq 8.53$--8.64 with a moderate gradient (decreasing westwards).

(iv) The $\rm H\alpha$ emission line shows a strong velocity gradient across the two galaxies, running from SE to NW (NW receding), which is made up of the rotation of the two individual galaxies plus their mutual velocity. These lie in similar directions making this a prograde-prograde interaction. The kinematic position angles of the primary and secondary galaxy are estimated as $-45^{\circ}$ and $-34^{\circ}$. The $\rm H\alpha$ line velocity gradient across the primary  gives a minimum rotation velocity $92.5\pm 11.7$ km $\rm s^{-1}$,  making this a fast-rotator elliptical. Considering  the compactness of the $\rm H\alpha$ emission, the true rotation velocity is probably closer to $\sim 150$ km $\rm s^{-1}$. For the companion galaxy the observed rotation, $112\pm 5.4$ km $\rm s^{-1}$ is consistent with an inclined ($\sim 45^{\circ}$) spiral.
 
 (v) In the primary galaxy the velocity gradient is steeper in $\rm H\alpha$ (and $\rm H\beta$) compared to [NII] and especially [OIII]5007 and [OI[6300] i.e. there is a stronger signal of rotation in the lines associated with star-formation rather than AGN. The velocity gradients appear to be aligned with a central ultraviolet-luminous region seen in HST imaging by Labiano et al. (2008). This could be a rotating photoionized disk, emitting AGN lines such as [OIII], with the addition of circumnuclear star-formation giving $\rm H\alpha$ emission out to larger radii and with faster rotation.

(vi) We resolve the strong [OIII]5007 and [OI]6300 emission lines into narrow and broad components, the latter with $\rm FWHM\simeq 1300$--1400 km $\rm s^{-1}$. The [OIII] broad component, making up $26\%$ of the flux is blueshifted by an average of 70 km $\rm s^{-1}$ relative to the narrow component, with a very different velocity gradient, aligned with the radio source; it can be identified with high-velocity ionized outflows from the radio AGN.
The [OI]6300 line has an even more prominent broad component ($65\%$ of  the flux) redshifted by 78 km $\rm s^{-1}$ but with no obvious velocity gradient.

(vii) {\it Starlight} model fits estimate the total stellar masses of the primary and companion as 
as $\rm M_{*}\simeq 1.0\times 10^{11}M_{\odot}$ and
$\rm M_{*}\simeq 2.25\times 10^{10}M_{\odot}$, with predominantly old stellar populations, mean ages about 9 and 5 Gyr. For both galaxies {\it Starlight} also fits a small fraction of very young ($\leq 40$ Myr) stars, but with a lack  of intermediate age ($10^8$--$10^9$ yr) stars. In the companion galaxy this young stellar population, at least $\rm 1.3\times 10^7~M_{\odot}$, is consistent with the SFR estimated from $\rm H\alpha$. For the primary, a  central SFR of  at least $\sim 2\rm ~M_{\odot} yr^{-1}$  could account for the young stars and contribute sufficient $\rm H\alpha$ flux to explain the faster rotation seen in this line. These SFRs would also account for the $11.3\mu m$ PAH flux detected in {\it Spitzer} data.

 (viii) Our interpretation is that the galaxy pair is observed at a stage of first perigalacticon, which tidally induced gas inflow and triggered star-formation, in the centre of the AGN host and extensively in the spiral, from $\sim 40$ Myr ago. After a relatively short delay, and very recently ($\sim 10^4$ yr ago) this started the radio-AGN outburst giving rise to strong emission lines from AGN photoionization and high velocity shocks, and fast outflows of ionized gas.

\section*{Acknowledgments}
This paper is based on observations with the European Southern Observatory Very Large Telescope at Cerro Paranal in Chile, under program  60.A-9335(A). We use data from observations with the NASA/ESA Hubble Space Telescope, obtained from the data archive at the Space Telescope Science Institute. STScI is operated by the Association of Universities for Research in Astronomy, Inc. under NASA contract NAS 5-26555.
We thank Montserrat Villar-Mart\'in for useful input and ideas.

NR, AH, JMG, PP, PL and LC acknowledge Funda\c{c}\~{a}o para a Ci\^{e}ncia e a
Tecnolog\'ia (FCT) support through UID/FIS/04434/2013, and through project
FCOMP-01-0124-FEDER-029170 (Reference FCT PTDC/FIS-AST/3214/2012) funded by
FCT-MEC (PIDDAC) and FEDER (COMPETE), in addition to FP7 project
PIRSES-GA-2013-612701.
We acknowledge support by the exchange programme `Study of Emission-Line Galaxies 
with Integral-Field Spectroscopy' (SELGIFS, FP7-PEOPLE-2013-IRSES-612701), funded by the EU through the IRSES scheme.
NR acknowledges the support of  FCT postdoctoral grant SFRH/BI/52155/2013, and grants CAUP-14/2014-BPD, CAUP-06/2015-BI and CIAAUP-22/2015-BPD.
 AH also acknowledges a Marie Curie Fellowship
co-funded by the FP7 and the FCT (DFRH/WIIA/57/2011) and FP7 / FCT
Complementary Support grant SFRH/BI/52155/2013.
 PL is supported by FCT Postdoctoral grant SFRH/BPD/72308/2010. JMG acknowledges support by the FCT through the Fellowship SFRH/BPD/66958/2009
and POPH/FSE (EC) by FEDER funding through the Programa Operacional
de Factores de Competitividade (COMPETE).
PP is supported by FCT through the Investigador FCT Contract No. IF/01220/2013 and 
POPH/FSE (EC) by FEDER
funding through the programme COMPETE.
LSMC is supported by an Assistantship Grant funded by FCT/MCTES (Portugal) and POPH/FSE (EC). 
MS thanks the Conselho Nacional de Desenvolvimento e Pesquisa (CNPq).

\section*{References} 

\vskip0.15cm \noindent Allen M.G., Groves B., Dopita M., Sutherland R., Kewley L.J., 2008, ApJS 178, 20.

\vskip0.15cm \noindent An Tao,  Baan W.A., 2012, ApJ 760, 77.

\vskip0.15cm \noindent Arribas S., Colina L., Bellocchi E., Maiolino R., Villar-Mart\'in M., 2014, A\&A 568, 14.

\vskip0.15cm \noindent Bacon R., et al., 2015, A\&A, 575, 75.

\vskip0.15cm \noindent Baldwin J.A., Phillips M.M., Terlevich R., 1981, PASP 93, 5.

\vskip0.15cm \noindent Bertin E., Arnouts S., 1996, A\&AS 117, 393.

\vskip0.15cm \noindent Binette L., et al., 2012, A\&A 547, A29.

\vskip0.15cm \noindent Binette L, Dopita M.A., Tuohy I.R., 1985, ApJ 297, 476.

\vskip0.15cm \noindent Calzetti D., Armus L., Bohlin R.C., Kinney A.L., Koornneef J., Storchi-Bergmann T., 2000, ApJ 533, 682.

\vskip0.15cm \noindent Cid Fernandes R., et al.  2009, RMxAC, 35, 127.

\vskip0.15cm \noindent Dicken D., et al., 2012, ApJ 745, 172.

\vskip0.15cm \noindent Emsellem E., et al., 2011, MNRAS 414, 888.

\vskip0.15cm \noindent Fanaroff B.L., Riley J. M., 1974, MNRAS, 167, 31.

\vskip0.15cm \noindent Filippenko A.V., 1985, ApJ 289, 479.

\vskip0.15cm \noindent Fosbury R.A.E.,  Bird M. C., Nicholson W., Wall J.V., 1987, MNRAS 225, 761.

\vskip0.15cm \noindent Groves B., Brinchmann J., Walcher C.J., 2012, MNRAS, 419,1402.

\vskip0.15cm \noindent Heckman T.M., 1980, A\&A, 87, 152.

\vskip0.15cm \noindent Heckman T.M., Smith E.P., Baum S. A., van Breugel W.J.M., Miley G.K., Illingworth G.D., Bothun G.D., Balick B., 1986, ApJ 311, 526.

\vskip0.15cm \noindent Heckman T.M., Lehnert M.D., Strickland D.K., Armus L.,  2000, ApJS, 129, 493.
	
\vskip0.15cm \noindent Henry R.B.C., Edmunds M.G., K\"oppen J., 2000, ApJ 541, 660.

\vskip0.15cm \noindent  Ho I-Ting, Kudritzki R-P, Kewley L.J., Zahid H.J., Dopita M.A.,  Bresolin F., Rupke. D.S.N., 2015, MNRAS 448, 2030.

\vskip0.15cm \noindent Holt J., Tadhunter C.N., Morganti R. 2008, MNRAS 387, 639.

\vskip0.15cm \noindent  Humphrey A., Villar-Mart\'in M., Fosbury R., Vernet J., di Serego Alighieri S.,
2006, MNRAS 369, 1103.

\vskip0.15cm \noindent Humphrey A., Villar-Mart\'in M., Vernet J., Fosbury R., di Serego Alighieri S.,
 Binette L., 2008, MNRAS 383, 11.

\vskip0.15cm \noindent Humphrey A., Binette L., 2014, MNRAS 442, 753.

\vskip0.15cm \noindent Inskip K.J., et al., 2006, MNRAS 370, 1585.

\vskip0.15cm \noindent Inskip K.J., Tadhunter C.N., Dicken D., Holt J., Villar-Mart\'in M., Morganti, R., 2007, MNRAS 382, 95.

\vskip0.15cm \noindent Inskip K. J., Villar-Mart\'in M., Tadhunter C.N., Morganti R., Holt J., Dicken D.,  2008, MNRAS 386, 1797.

\vskip0.15cm \noindent Inskip K. J., Tadhunter C.N., Morganti R., Holt J., Ramos Almeida C., Dicken D.,  2010, MNRAS 407, 1739.

\vskip0.15cm \noindent Jeyakumar S., 2016, MNRAS 458, 3786.

\vskip0.15cm \noindent Kawakatu N., Nagao T., Woo J-H., 2009, ApJ 693, 1686.

\vskip0.15cm \noindent Keel W.C., Windhorst R.A., 1991, ApJ 383, 135.

\vskip0.15cm \noindent Kunert-Bajraszewska M., 2015, Astron. Nachr. 337, 27.

\vskip0.15cm \noindent Labiano A., O'Dea C. P., Barthel P.D., de Vries W. H., Baum S. A., 2008, 
A\&A 477, 491.

\vskip0.15cm \noindent  Lagos P., Telles E., Nigoche Netro A., Carrasco E.R., 2012, MNRAS 427, 740.

\vskip0.15cm \noindent  Lagos P., Telles E.,  Mu\~noz-Tu\~n\'on Casiana, Carrasco E.R., Cuisinier F., Tenorio-Tagle G., 2009, AJ, 137, 5068.

\vskip0.15cm \noindent Lena D., et al., 2015, ApJ, 806, 84.

\vskip0.15cm \noindent Maccagni F., Morganti R., Oosterloo T.A., Mahony, E. K., 2014, A\&A 571, 67.

\vskip0.15cm \noindent Maccagni F.M., Santoro F.., Morganti R., Oosterloo T.A., Oonk J.B.R., Emonts B.H.C., 2016, A\&A preprint arXiV:1602.00701.

\vskip0.15cm \noindent Mahony E.K., Oonk J.B.R., Morganti R., Tadhunter C.,  Bessiere P., Short P.,
Emonts B.H.C.,  Oosterloo T.A., 2016, MNRAS, 455, 2453.

\vskip0.15cm \noindent Marino R.A., et al.,  2013, A\&A 559, 114.

\vskip0.15cm \noindent Morganti R., Tadhunter C.N., Dickson R., Shaw M., 1997, A\&A 326, 130.

\vskip0.15cm \noindent Naab T., et al., 2014, MNRAS 444, 3357.

\vskip0.15cm \noindent Nicholls D.C., Dopita M.A., Sutherland R.S., 2012, ApJ 752, 148.

\vskip0.15cm \noindent Nicholls D.C., Dopita M.A., Sutherland R.S., Kewley L.J., Palay E., 2013, ApJS 207, 21. 

\vskip0.15cm \noindent  O'Dea C.P, Baum S.A., Stanghellini C., 1991 ApJ 380, 66.

\vskip0.15cm \noindent Ojha R., Fey A.L., Johnston K.J., Jauncey D.L., Tzioumis A.K., Reynolds J.E.,
2004, AJ 127, 1977. 

\vskip0.15cm \noindent Peng C.Y., Ho L.C., Impey C.D., Rix H.-W., 2010, AJ 139, 2097.

\vskip0.15cm \noindent Peterson B.A., Bolton J.G., 1972, ApJ 173, L19.

\vskip0.15cm \noindent Pettini M., Pagel B.E.J. 2004, MNRAS, 348, L59.

\vskip0.15cm \noindent Ramos Almeida C., Tadhunter C.N., Inskip K.J., Morganti R., Holt J., Dicken D., 2011, MNRAS 410, 1550.

\vskip0.15cm \noindent Ramos Almeida, C., Bessiere P. S., Tadhunter C. N., Inskip K. J., Morganti R., Dicken D., Gonz\'alez-Serrano J. I., Holt, J., 2013, MNRAS 436, 997.

\vskip0.15cm \noindent Reyes R., Mandelbaum R., Gunn J.E., Pizagno J., Lackner C.N.,
	2011, MNRAS 417, 2347.

\vskip0.15cm \noindent Robinson A., Binette L., Fosbury R.A.E., Tadhunter C.N., 1987, MNRAS 227, 97.

\vskip0.15cm \noindent  Roche N.D., Eales S.A., 2000, MNRAS 317, 120.

\vskip 0.15cm  \noindent S\'anchez S.F., et al. 2014, A\&A 563, 49.

\vskip0.15cm \noindent Shih Hsin-Yi, Stockton A., Kewley L., 2013, ApJ 772, 138.

\vskip0.15cm \noindent Shimmins A.J., 1971, AuJPA 21, 1.

\vskip0.15cm \noindent Shirazi M., Brinchmann J., 2012, MNRAS 421, 1043.

\vskip0.15cm \noindent Sobral D., Best P.N., Smail I., Mobasher B., Stott J., Nisbet D., 2014, 
MNRAS 437, 3516.

\vskip0.15cm \noindent Son D., Woo J-H., Kim S.C., Fu H., Kawakatu N., Bennert V.N., Nagao T., Park D., 2012, ApJ 757, 140.
	
\vskip0.15cm \noindent Tadhunter C., Wills K., Morganti R., Oosterloo T., Dickson R., 
	2001, MNRAS 327, 227.
	
\vskip0.15cm \noindent Tadhunter C., Dickson R., Morganti R., Robinson T. G., Wills K., Villar-Mart\'in M., Hughes M., 2002, MNRAS 330, 977.

\vskip0.15cm \noindent Tadhunter C., Holt J., Gonz\'alez Delgado R., Rodr\'iguez Zaur'n J., 
Villar-Mart\'in M., Morganti R., Emonts B., Ramos Almeida C., Inskip K., 2011, MNRAS 412, 960.

\vskip0.15cm \noindent Treyer M., et al., 2010, ApJ 719, 1191.

\vskip0.15cm \noindent Tzioumis A.K., et al., 1989, AJ 98, 36.

\vskip0.15cm \noindent Tzioumis A.K., et al., 2010, AJ 140,150.

\vskip0.15cm \noindent Villar-Mart\'in M., Tadhunter C., Morganti R., Axon D., Koekemoer A., 1999, MNRAS 307, 24.

\vskip0.15cm \noindent Villar-Mart\'in M., Humphrey A., Delgado R.G., Colina L., Arribas S., 2011, MNRAS.418, 2032.

\vskip0.15cm \noindent de Vries W.H., Barthel P.D., O'Dea C.P., 1997, A\&A 321, 105.

\vskip0.15cm \noindent Wall J.V., Cannon R.D., 1973, AuJPA 31, 1.

\section*{Appendix A: Other Galaxies in the Field of View}

At this time we have measured redshifts for a total of 13 galaxies in the field-of-view (Fig 34), which are catalogued in Table A1 by their detection numbers in the {\it s-extractor} run. These include the two components of PKS 1934-63 (nos. 25 and 26), and the faint, higher-redshift  `third component', which is listed here 
as T3 (not separately detected by {\it s-extractor} because of confusion effects), for which  the spectra  were shown in Section 3. Here we show the MUSE spectra for the other 10 galaxies, extracted in r=4 pixel apertures, which are very diverse and range from $z=0.18$ to $z=1.05$ (Fig 35).

Detections \#24 and \#32 are moderately star-forming with similar spectra (both $\rm EW_{obs}\simeq 21\AA$), separated by only 56.0 kpc (8.9 arcsec on sky plane) and 344 km $\rm s^{-1}$, and may be another interacting pair. No. 32 is visibly a spiral, possible strong $\rm H\delta$ absorption. 

Detections \#20 and \#45 are blue, lower luminosity galaxies,  both with strong star-formation indicated by $\rm EW_{obs}(H\alpha)\simeq 100\AA$, and line ratios indicating low metallicities; $\rm (O3,N2)\simeq(4.03, 0.055)$ and (3.34,0.057), by PP04 calibration giving $\rm 12+log(O/H)\simeq 8.13$ and 8.16. Similar spectra are found for nearby HII or Blue Compact Dwarf galaxies, e.g. as studied by IFS by Lagos et al. (2009, 2012).

Detection \#14 may be physically associated with the AGN, separated by only $\Delta(v)=-250$ km $\rm s^{-1}$ and 98 kpc in the sky plane (31.7 arcsec), but not close enough to be interacting, at least there is no evidence of this. High surface brightness with but no emission lines and with only a few weak absorption features visible, it required  careful matching to other spectra to find the redshift. Detection \#9 is another passive galaxy, with strong CaII lines.

Detection \#31 is faint, star-forming ($\rm EW_{obs}\simeq 40\AA$), and a nearly edge-on disk.
Detection \#17 is unusual; a type 1 QSO (not previously catalogued?) with broad MgII 2800 emission ($\rm FWHM\simeq 73\AA\simeq 4040$ km $\rm s^{-1}$, flux $3.1\times 10^{-16}$ erg $\rm cm^{-2}s^{-1}$) and lines from highly ionized Neon. 

Detections \#2 and \#36 have spectra dominated by a single strong emission line, which from its split appearance and the substantial flux blueward  is almost certainly [OII], meaning redshifts just greater than unity; \#36 is fainter in continuum but a true starburst with $\rm EW_{obs}([OII])\simeq 234\AA$.
\begin{figure}
\hskip-1.0cm
\includegraphics[width=1.25\hsize,angle=0]{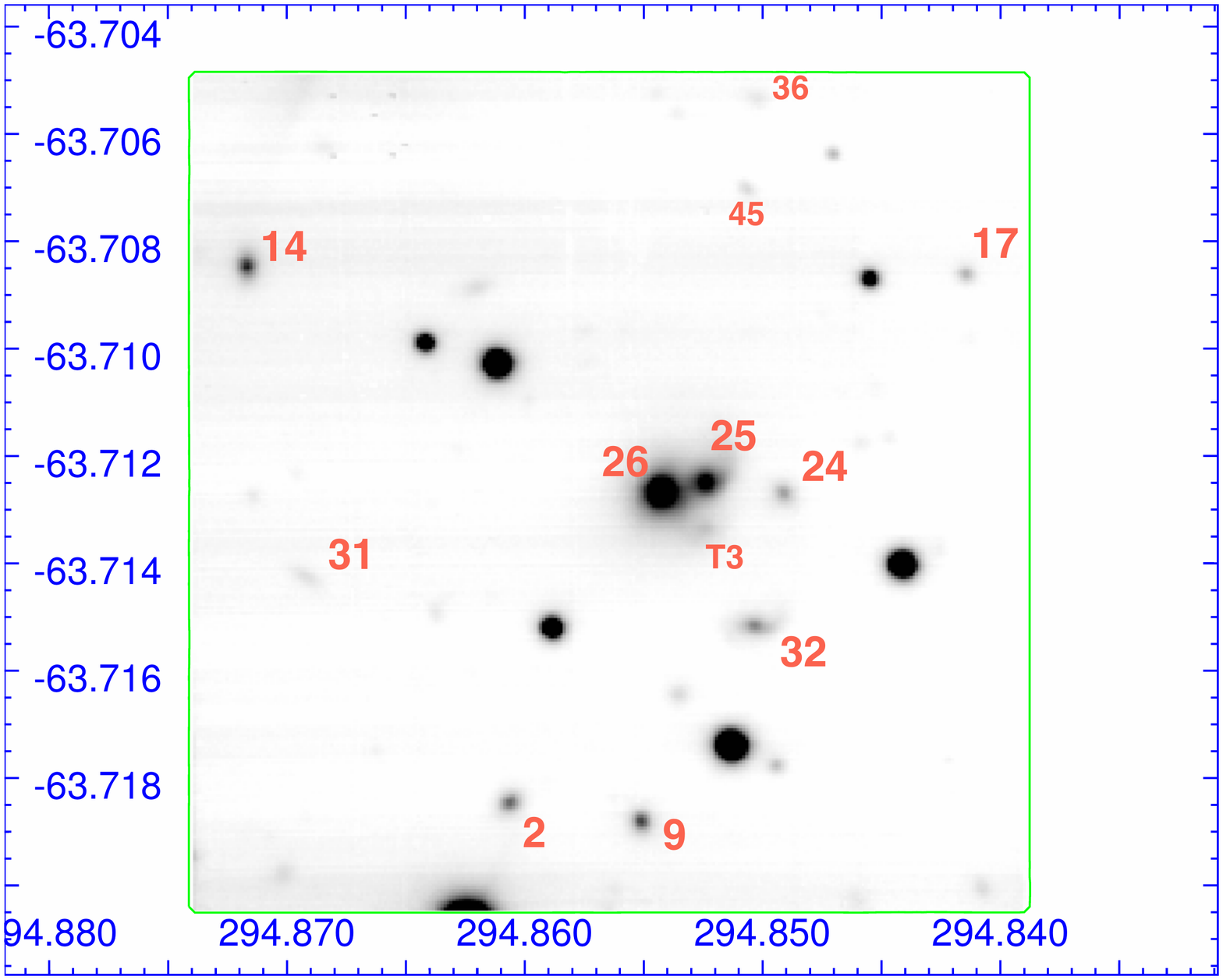}
{{\bf Figure A1.} Positions of the MUSE field of view of galaxies with measured redshifts, numbered as in Table 4. RA and Dec axes are shown, here both in degrees. }
\end{figure}
\begin{figure}
\includegraphics[width=0.75\hsize,angle=-90]{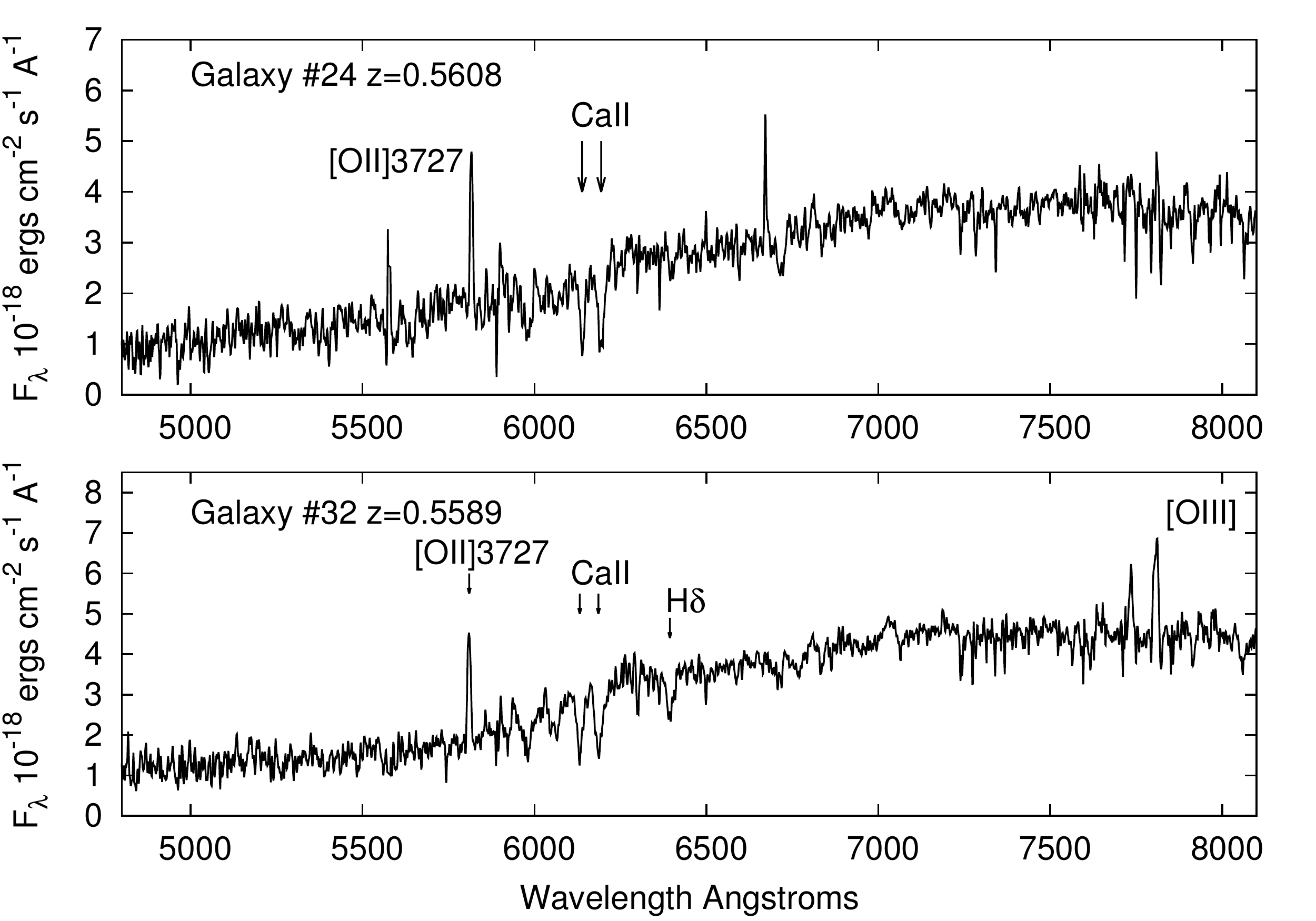}
\end{figure}
\begin{figure}
\includegraphics[width=0.75\hsize,angle=-90]{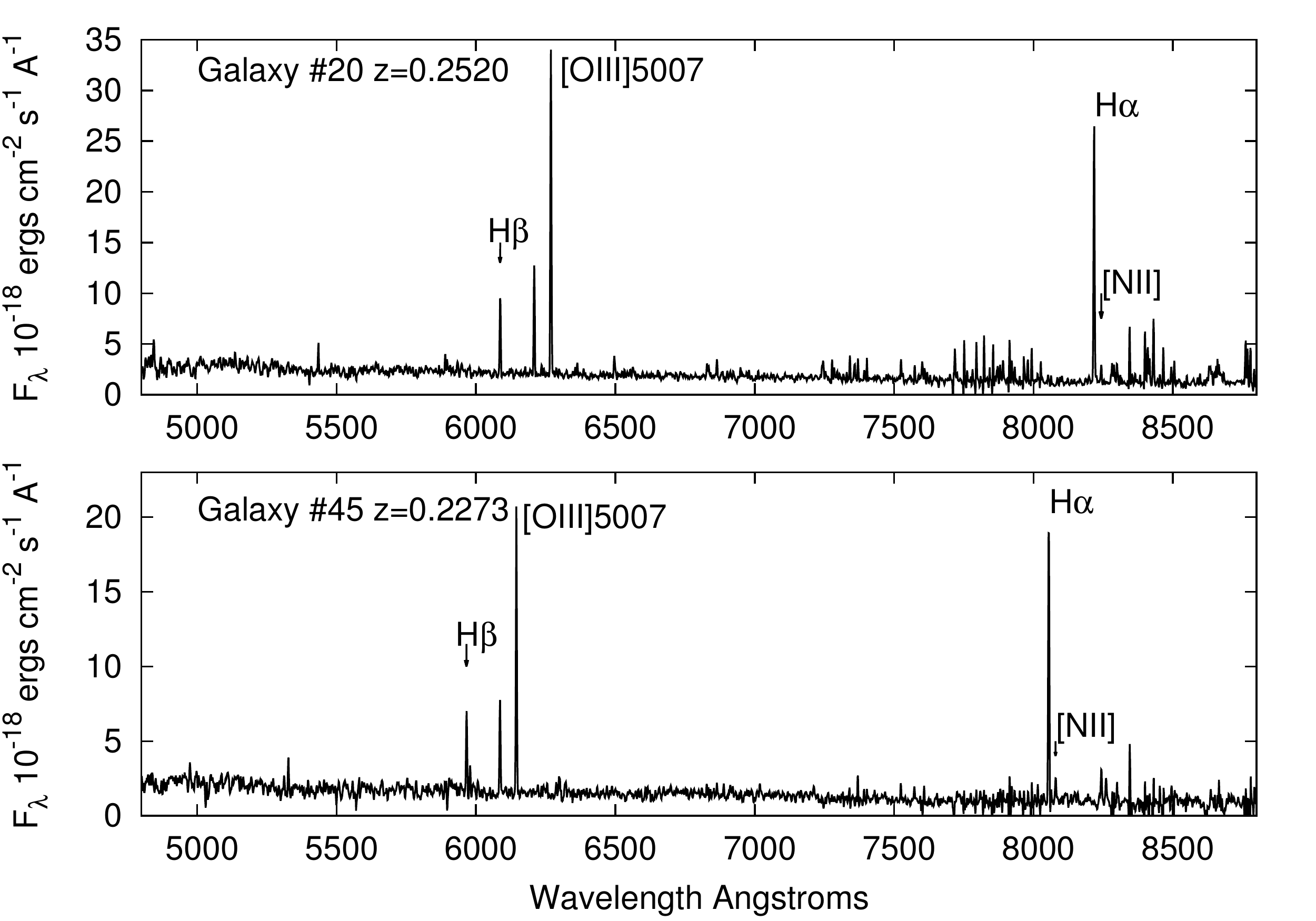}
\end{figure}
\begin{figure}
\includegraphics[width=0.75\hsize,angle=-90]{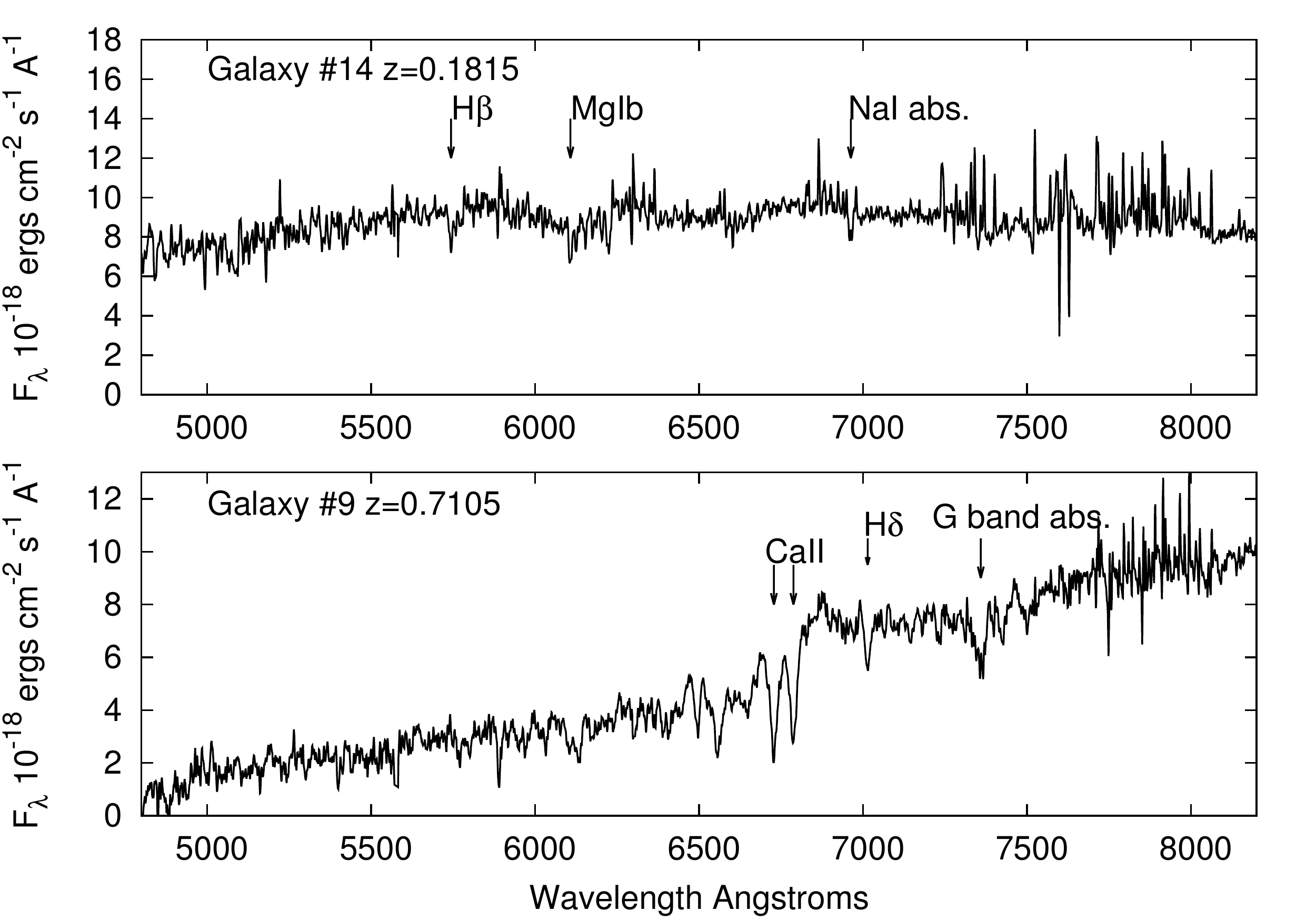}
\end{figure}
\begin{figure}
\includegraphics[width=0.75\hsize,angle=-90]{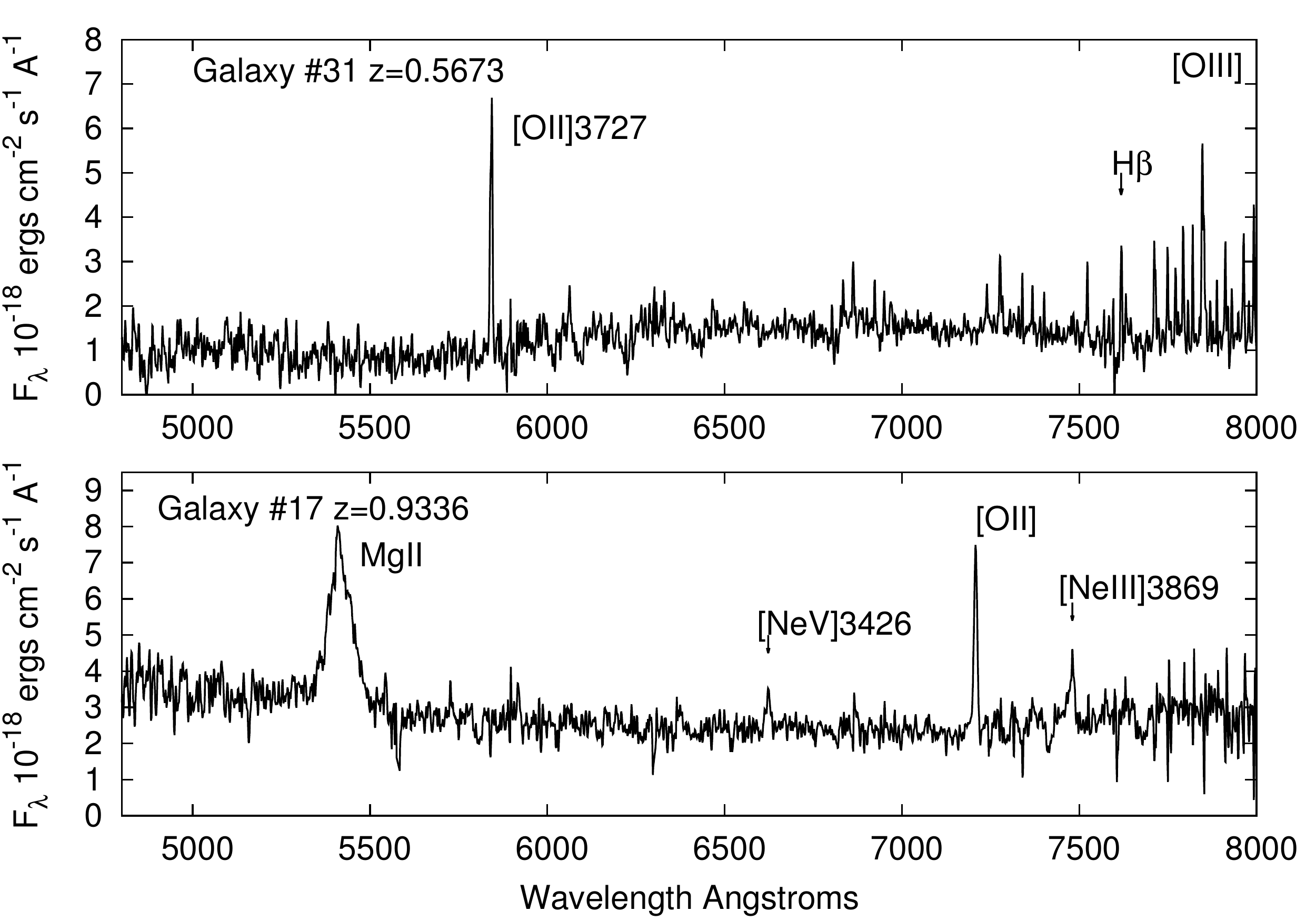}
\end{figure}
\begin{figure}
\includegraphics[width=0.75\hsize,angle=-90]{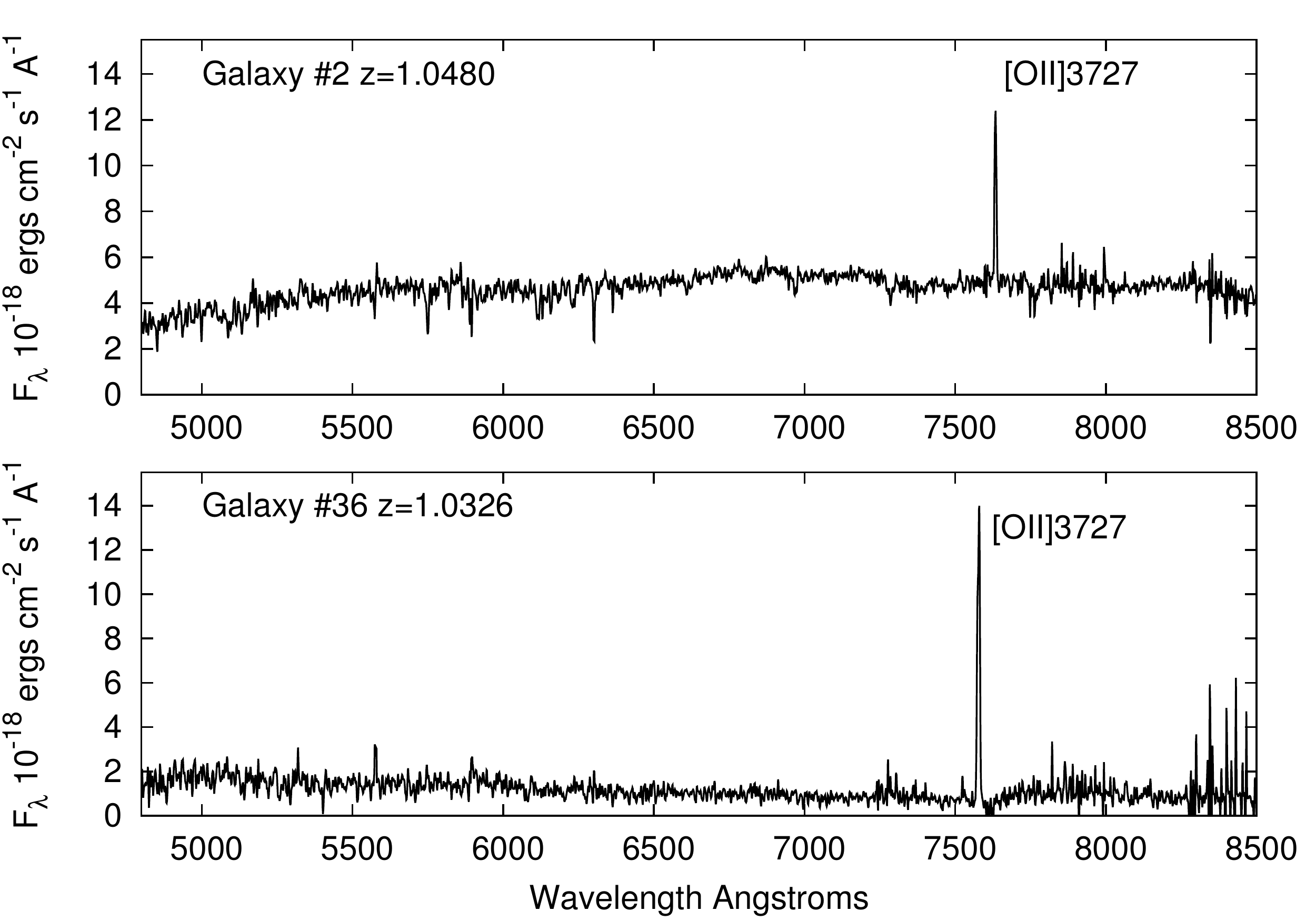}
{{\bf Figure A2.} Spectra for 10 other galaxies on the MUSE field with identified redshifts, listed in Table A1. Extracted in $r=4$ pixel apertures.}
 \end{figure}

 \begin{table*}
{ {\bf Table A1.} List of all galaxies with confidently measured redshifts in the MUSE field-of-view, giving positions, total magnitudes, redshifts and prominent emission(*) and/or absorption lines. Flux is given for either $\rm H\alpha$ (for $z<0.3$) or [OII]3727 for $z>0.3$ (in units $10^{-17}$ erg $\rm cm^{-2}s^{-1}$), and the corresponding luminosity (log L in units erg $\rm s^{-1}$), which can be used as an estimator of SFR. In 25 and 26 only, the flux is corrected for aperture as described in the text.}
\begin{tabular}{lcccccccc}
\hline
No. & RA & Dec & $R$ total & $V-R$ & Redshift &  line flux & log L & features \\
\hline
2   & 294.8607 & -63.7184 & 20.65 & 0.62 & 1.0480 & 7.2 & 41.64 & $\rm [OII]^*$ \\
9   & 294.8551 & -63.7187 & 20.48 &  1.55  & 0.7105 &  0 &   &   CaII abs. \\
14   & 294.8717 & -63.7084 &  19.88 &  0.75  &  0.1815 & 0   &  &  $\rm H\beta$, MgIb, NaI abs.\\
17 & 294.8415 & -63.7085 &  21.49 & 0.36 &  0.9336 & 5.4 & 41.39    & QSO $\rm MgII^*[OII]^*[NeIII/V]^*$ \\
 20 & 294.8621 & -63.7087 &   21.38 &  0.21 & 0.2520 & 12.4 & 40.39 & $\rm H\beta^* [OIII]^* H\alpha^*[NII]^*$\\
24 & 294.8492 &  -63.7127 & 20.29 & 0.64 &  0.5608 & 4.1& 40.73  & $\rm [OII]^*$, CaII \\
25 & 294.8525 & -63.7124 & 18.92 & 0.67 & 0.1834 & 47.3 & 40.65 & companion $\rm H\beta^* [OIII]^*H\alpha^*[NII]^*$\\
26 & 294.8543 &  -63.7126 & 17.64 &  0.73 & 0.1825 & 870 & 41.92 & PKS 1934-63 \\
31 &  294.8692 & -63.7142 & 21.44 & 0.81 &  0.5673 & 4.6 & 40.79 & $\rm [OII]^*H\beta^* [OIII]^*$\\
32 & 294.8493 &  -63.7151 & 20.31 & 1.51 & 0.5589 & 4.8 & 40.79 & spiral $\rm [OII]^*$, CaII \\ 
36  & 294.8503 & -63.7052&  22.29 & 0.47 & 1.0326 & 13.6 & 41.90 & $\rm [OII]^*$ \\
45  & 294.8507 & -63.7070 &  22.65  & 0.04 &  0.2273 &  9.1 & 40.15 & $\rm H\beta^* [OIII]^* H\alpha^*[NII]^*$\\
T3 & 294.8525 & -63.7133  & 20.62 & 0.47 & 0.6461 & 6.8 & 41.10 & $\rm [OII]^* H\delta\gamma\beta^* [OIII]^*$\\ 
\hline
\end{tabular}

\end{table*}

\twocolumn

 \end{document}